\documentclass[a4paper,fleqn]{cas-sc}
\usepackage[numbers,sort&compress]{natbib}

\def\tsc#1{\csdef{#1}{\textsc{\lowercase{#1}}\xspace}}
\tsc{WGM}
\tsc{QE}

\newdefinition{note}{Note}[section]

\usepackage[T1]{fontenc} 
\usepackage{subcaption}
\usepackage{mathtools,amsthm,physics-patch,dsfont,upgreek,pgfplots,subcaption}
\usepackage[shortlabels]{enumitem}
\setlist[enumerate,1]{1),itemsep=2pt,parsep=0pt,topsep=3pt,left=0pt}
\setlist[itemize,1]{left=0pt}
\usepackage{float}
\usepackage{tikz}
\usetikzlibrary{decorations.markings,external,arrows.meta}
\definecolor{regionred}{RGB}{245,176,176}
\definecolor{regionviolet}{RGB}{205,190,238}
\renewcommand{\pi}{\uppi}

\newcommand{\en}{\enspace}

\newcommand{\De}{\Delta}
\renewcommand{\Im}{\operatorname{Im}}

\newcommand{\T}{\mathsf T}
\newcommand{\Th}{\Theta}
\newcommand{\ZZ}{{\mathbb Z}}

\newcommand{\all}{\forall}

\newcommand{\al}{\alpha}
\newcommand{\be}{\beta}
\newcommand{\de}{\delta}
\newcommand{\ka}{\kappa}
\renewcommand{\la}{\lambda}

\newcommand{\e}{\mathrm e}

\renewcommand{\ge}{\geqslant}

\newcommand{\iu}{\mathrm i}
\renewcommand{\le}{\leqslant}

\newcommand{\pd}{\partial}
\newcommand{\sgn}{\operatorname{sgn}}
\newcommand{\si}{\sigma}
\newcommand{\ga}{\gamma}
\newcommand{\Om}{\Omega}
\newcommand{\Ga}{\Gamma}

\renewcommand{\th}{\theta}
\newcommand{\vep}{\varepsilon}

\newcommand{\vp}{\varphi}

\renewcommand{\Re}{\operatorname{Re}}
\newcommand{\CC}{\mathbb C}
\newcommand{\RR}{\mathbb R}
\newcommand{\cB}{\mathcal B}
\newcommand{\cH}{\mathcal H}
\newcommand{\Hb}{H_{\text{bulk}}}
\newcommand{\ha}{\hat a}
\newcommand{\hb}{\hat b}

\newcommand{\bu}{\mathbf u}
\newcommand{\bmu}{\bar\mu}
\newcommand{\dmu}{\de\mu}
\newcommand{\cN}{\mathcal N}
\newcommand{\supp}{\operatorname{supp}}
\newcommand{\myqed}{\tag*{\(\square\)}}

\newcommand{\cD}{\mathcal D}
\newcommand{\wh}{\widehat}

\newcommand{\hc}{\hat c}

\newcommand{\Emax}{E_{\text{max}}}
\newcommand{\Emin}{E_{\text{min}}}
\newcommand{\Hc}{\text{H.c.}}
\newcommand{\eff}{\text{eff}}
\theoremstyle{remark}
\newtheorem{remark}{Remark}[section]
\definecolor{wkbOrbitBlue}{RGB}{18,101,176}
\definecolor{wkbAxisGray}{RGB}{75,88,104}

\newcommand{\BandPlotWidth}{8.2cm}
\newcommand{\BandPlotHeight}{7cm}

\newlength{\BandPanelHeight}
\tikzset{
  wkb axis/.style={draw=wkbAxisGray,line width=.8pt,-{Stealth[length=2.5mm]}},
  wkb tick/.style={draw=wkbAxisGray,line width=.6pt},
  wkb guide/.style={draw=wkbAxisGray!55,densely dashed,line width=.6pt},
  wkb cylinder edge/.style={draw=wkbAxisGray,line width=.6pt},
  wkb rim front/.style={draw=wkbAxisGray,line width=.9pt},
  wkb rim rear/.style={draw=wkbAxisGray!55,densely dashed,line width=.65pt},
  wkb guide front/.style={draw=wkbAxisGray!78,densely dotted,line width=.75pt},
  wkb guide rear/.style={draw=wkbAxisGray!48,densely dashed,line width=.65pt},
  wkb orbit rear/.style={draw=wkbOrbitBlue!48,dashed,line width=1.45pt,
    line cap=round,line join=round},
  wkb halo/.style={draw=white,line width=4.6pt,line cap=round,line join=round},
  wkb curve plain/.style={draw=wkbOrbitBlue,line width=2.25pt,
    line cap=round,line join=round},
  wkb curve/.style={wkb curve plain,
    postaction={decorate},decoration={markings,
      mark=at position .55 with {\arrow{Stealth[length=2.8mm]}}}},
  wkb equivalent/.style={draw=wkbOrbitBlue!55,dashed,line width=1.45pt,
    line cap=round,line join=round,
    postaction={decorate},decoration={markings,
      mark=at position .55 with {\arrow{Stealth[length=2.5mm]}}}},
  wkb turning front/.style={circle,fill=wkbOrbitBlue,draw=white,
    line width=.7pt,inner sep=2.2pt},
  wkb turning rear/.style={circle,fill=white,draw=wkbOrbitBlue!58,
    dashed,line width=1.15pt,inner sep=2pt},
  wkb cut endpoint/.style={circle,fill=white,draw=wkbOrbitBlue,
    line width=1.5pt,inner sep=2pt},
  wkb label/.style={font=\small},
  wkb note/.style={font=\small,align=center,fill=white,
    fill opacity=.9,text opacity=1,inner sep=1.5pt}
}

\newcommand{\WKBCylinderBackground}{%
  \path[shade,top color=blue!1,bottom color=blue!9,opacity=.68]
    (-3.903,-1.213)--(4.097,-1.213)--(3.903,1.213)--(-4.097,1.213)--cycle;
  \path[fill=blue!3,fill opacity=.72]
    plot[domain=0:360,samples=100,variable=\a]
      ({-4-.42*cos(\a)},{1.18*sin(\a)+.28*cos(\a)})--cycle;
  \path[fill=blue!3,fill opacity=.72]
    plot[domain=0:360,samples=100,variable=\a]
      ({4-.42*cos(\a)},{1.18*sin(\a)+.28*cos(\a)})--cycle;

  \draw[wkb rim rear]
    plot[domain=-90:90,samples=60,variable=\a]
      ({-4-.42*cos(\a)},{1.18*sin(\a)+.28*cos(\a)});
  \draw[wkb rim front]
    plot[domain=90:270,samples=60,variable=\a]
      ({-4-.42*cos(\a)},{1.18*sin(\a)+.28*cos(\a)});

  \draw[wkb rim front]
    plot[domain=-90:90,samples=60,variable=\a]
      ({4-.42*cos(\a)},{1.18*sin(\a)+.28*cos(\a)});
  \draw[wkb rim rear]
    plot[domain=90:270,samples=60,variable=\a]
      ({4-.42*cos(\a)},{1.18*sin(\a)+.28*cos(\a)});

  \draw[wkb cylinder edge] (-4,1.18)--(4,1.18);
  \draw[wkb cylinder edge] (-4,-1.18)--(4,-1.18);
  \draw[wkb guide rear] (-4.42,.28)--(4.42,.28);
  \draw[wkb guide front] (-3.58,-.28)--(3.58,-.28);
  \node[wkb note,anchor=west] at (4.48,.28) {$p=0$};
  \node[wkb note,anchor=west] at (3.64,-.28) {$p=\pi$};

  \draw[-{Stealth[length=2.5mm]},wkbAxisGray] (-4.55,-1.62)--(4.62,-1.62)
    node[wkb label,below] {$x$};
  \draw[wkbAxisGray] (-2.8,-1.57)--(-2.8,-1.68)
    node[wkb label,below=1pt] {$x_j^L$};
  \draw[wkbAxisGray] (2.8,-1.57)--(2.8,-1.68)
    node[wkb label,below=1pt] {$x_j^R$};
}

\newcommand{\WKBCylinderCaseI}[1][1]{%
\begin{tikzpicture}[x=1cm,y=1cm,scale=#1]
  \WKBCylinderBackground
  \draw[wkb orbit rear]
    plot[domain=-1:1,samples=150,variable=\t]
      ({2.8*\t-.42*cos(105*sqrt(max(0,1-\t*\t)))},
       {1.18*sin(105*sqrt(max(0,1-\t*\t)))+.28*cos(105*sqrt(max(0,1-\t*\t)))});
  \draw[wkb orbit rear]
    plot[domain=1:-1,samples=150,variable=\t]
      ({2.8*\t-.42*cos(-105*sqrt(max(0,1-\t*\t)))},
       {1.18*sin(-105*sqrt(max(0,1-\t*\t)))+.28*cos(-105*sqrt(max(0,1-\t*\t)))});
  \draw[wkb halo]
    plot[domain=-.515:.515,samples=90,variable=\t]
      ({2.8*\t-.42*cos(105*sqrt(max(0,1-\t*\t)))},
       {1.18*sin(105*sqrt(max(0,1-\t*\t)))+.28*cos(105*sqrt(max(0,1-\t*\t)))});
  \draw[wkb halo]
    plot[domain=.515:-.515,samples=90,variable=\t]
      ({2.8*\t-.42*cos(-105*sqrt(max(0,1-\t*\t)))},
       {1.18*sin(-105*sqrt(max(0,1-\t*\t)))+.28*cos(-105*sqrt(max(0,1-\t*\t)))});
  \draw[wkb curve]
    plot[domain=-.515:.515,samples=90,variable=\t]
      ({2.8*\t-.42*cos(105*sqrt(max(0,1-\t*\t)))},
       {1.18*sin(105*sqrt(max(0,1-\t*\t)))+.28*cos(105*sqrt(max(0,1-\t*\t)))});
  \draw[wkb curve]
    plot[domain=.515:-.515,samples=90,variable=\t]
      ({2.8*\t-.42*cos(-105*sqrt(max(0,1-\t*\t)))},
       {1.18*sin(-105*sqrt(max(0,1-\t*\t)))+.28*cos(-105*sqrt(max(0,1-\t*\t)))});
  \node[wkb turning rear] at (-3.22,.28) {};
  \node[wkb turning rear] at (2.38,.28) {};
\end{tikzpicture}}

\newcommand{\WKBCylinderCaseII}[1][1]{%
\begin{tikzpicture}[x=1cm,y=1cm,scale=#1]
  \WKBCylinderBackground
  \draw[wkb halo]
    plot[domain=-1:1,samples=150,variable=\t]
      ({2.8*\t-.42*cos(180-82*sqrt(max(0,1-\t*\t)))},
       {1.18*sin(180-82*sqrt(max(0,1-\t*\t)))+.28*cos(180-82*sqrt(max(0,1-\t*\t)))});
  \draw[wkb halo]
    plot[domain=1:-1,samples=150,variable=\t]
      ({2.8*\t-.42*cos(360-(180-82*sqrt(max(0,1-\t*\t))))},
       {1.18*sin(360-(180-82*sqrt(max(0,1-\t*\t))))+.28*cos(360-(180-82*sqrt(max(0,1-\t*\t))))});
  \draw[wkb curve]
    plot[domain=-1:1,samples=150,variable=\t]
      ({2.8*\t-.42*cos(180-82*sqrt(max(0,1-\t*\t)))},
       {1.18*sin(180-82*sqrt(max(0,1-\t*\t)))+.28*cos(180-82*sqrt(max(0,1-\t*\t)))});
  \draw[wkb curve]
    plot[domain=1:-1,samples=150,variable=\t]
      ({2.8*\t-.42*cos(360-(180-82*sqrt(max(0,1-\t*\t))))},
       {1.18*sin(360-(180-82*sqrt(max(0,1-\t*\t))))+.28*cos(360-(180-82*sqrt(max(0,1-\t*\t))))});
  \node[wkb turning front] at (-2.38,-.28) {};
  \node[wkb turning front] at (3.22,-.28) {};
\end{tikzpicture}}

\newcommand{\WKBCylinderCaseIII}[1][1]{%
\begin{tikzpicture}[x=1cm,y=1cm,scale=#1]
  \WKBCylinderBackground
  \draw[wkb orbit rear]
    plot[domain=-1:1,samples=150,variable=\t]
      ({2.8*\t-.42*cos(90*(1+\t))},
       {1.18*sin(90*(1+\t))+.28*cos(90*(1+\t))});
  \draw[wkb orbit rear]
    plot[domain=1:-1,samples=150,variable=\t]
      ({2.8*\t-.42*cos(-90*(1+\t))},
       {1.18*sin(-90*(1+\t))+.28*cos(-90*(1+\t))});
  \draw[wkb halo]
    plot[domain=0:1,samples=90,variable=\t]
      ({2.8*\t-.42*cos(90*(1+\t))},
       {1.18*sin(90*(1+\t))+.28*cos(90*(1+\t))});
  \draw[wkb halo]
    plot[domain=1:0,samples=90,variable=\t]
      ({2.8*\t-.42*cos(-90*(1+\t))},
       {1.18*sin(-90*(1+\t))+.28*cos(-90*(1+\t))});
  \draw[wkb curve]
    plot[domain=0:1,samples=90,variable=\t]
      ({2.8*\t-.42*cos(90*(1+\t))},
       {1.18*sin(90*(1+\t))+.28*cos(90*(1+\t))});
  \draw[wkb curve]
    plot[domain=1:0,samples=90,variable=\t]
      ({2.8*\t-.42*cos(-90*(1+\t))},
       {1.18*sin(-90*(1+\t))+.28*cos(-90*(1+\t))});
  \node[wkb turning rear] at (-3.22,.28) {};
  \node[wkb turning front] at (3.22,-.28) {};
\end{tikzpicture}}

\newcommand{\WKBPlaneAxes}[3]{%
  \draw[wkb guide] (-3.35,#1)--(3.35,#1);
  \draw[wkb guide] (-3.35,#2)--(3.35,#2);
  \draw[wkb axis] (-3.55,0)--(3.75,0) node[right,wkb label] {$x$};
  \draw[wkb axis] (-3.35,#1-.13)--(-3.35,#2+.13)
    node[above,wkb label] {$p$};
  #3
  \foreach \xx/\lab in {-2.8/{x^L_j},2.8/{x^R_j}} {
    \draw[wkb tick] (\xx,-.045)--(\xx,.045);
    \node[wkb label,anchor=north] at (\xx,-.08) {$\lab$};
  }
}

\newcommand{\WKBPlaneCaseI}[1][1]{%
  \begin{tikzpicture}[x=1cm,y=1.45cm,scale=#1]
    \node[inner sep=0pt,outer sep=0pt,minimum size=0pt]
      at (-4.5,0) {};
    \node[inner sep=0pt,outer sep=0pt,minimum size=0pt]
    at (5.2,0) {};
  \WKBPlaneAxes{-1}{1}{%
    \foreach \yy/\lab in {-1/{-\pi},0/0,1/{\pi}} {
      \draw[wkb tick] (-3.41,\yy)--(-3.29,\yy);
      \node[wkb label,anchor=east] at (-3.47,\yy) {$\lab$};}}
  \draw[wkb halo] plot[domain=-1:1,samples=150,variable=\t]
    ({2.8*\t},{.62*sqrt(max(0,1-\t*\t))});
  \draw[wkb halo] plot[domain=1:-1,samples=150,variable=\t]
    ({2.8*\t},{-.62*sqrt(max(0,1-\t*\t))});
  \draw[wkb curve] plot[domain=-1:1,samples=150,variable=\t]
    ({2.8*\t},{.62*sqrt(max(0,1-\t*\t))});
  \draw[wkb curve] plot[domain=1:-1,samples=150,variable=\t]
    ({2.8*\t},{-.62*sqrt(max(0,1-\t*\t))});
  \node[wkb turning front] at (-2.8,0) {};
  \node[wkb turning front] at (2.8,0) {};
\end{tikzpicture}}

\newcommand{\WKBPlaneCaseII}[1][1]{%
  \begin{tikzpicture}[x=1cm,y=1.1cm,scale=#1]
    \node[inner sep=0pt,outer sep=0pt,minimum size=0pt]
    at (-4.5,0) {};
    \node[inner sep=0pt,outer sep=0pt,minimum size=0pt]
    at (5.2,0) {};
  \draw[wkb guide] (-3.35,-1)--(3.35,-1);
  \draw[wkb guide] (-3.35,1)--(3.35,1);
  \draw[wkb guide] (-3.35,2)--(3.35,2);
  \draw[wkb axis] (-3.55,0)--(3.75,0) node[right,wkb label] {$x$};
  \draw[wkb axis] (-3.35,-1.16)--(-3.35,2.25)
    node[above,wkb label] {$p$};
  \foreach \yy/\lab in {-1/{-\pi},0/0,1/{\pi},2/{2\pi}} {
    \draw[wkb tick] (-3.41,\yy)--(-3.29,\yy);
    \node[wkb label,anchor=east] at (-3.47,\yy) {$\lab$};
  }
  \foreach \xx/\lab in {-2.8/{x_j^L},2.8/{x_j^R}} {
    \draw[wkb tick] (\xx,-.045)--(\xx,.045);
    \node[wkb label,anchor=north] at (\xx,-.08) {$\lab$};
  }
  \draw[wkb equivalent] plot[domain=1:-1,samples=150,variable=\t]
    ({2.8*\t},{-1+.46*sqrt(max(0,1-\t*\t))});
  \draw[wkb halo] plot[domain=-1:1,samples=150,variable=\t]
    ({2.8*\t},{1-.46*sqrt(max(0,1-\t*\t))});
  \draw[wkb halo] plot[domain=1:-1,samples=150,variable=\t]
    ({2.8*\t},{1+.46*sqrt(max(0,1-\t*\t))});
  \draw[wkb curve] plot[domain=-1:1,samples=150,variable=\t]
    ({2.8*\t},{1-.46*sqrt(max(0,1-\t*\t))});
  \draw[wkb curve] plot[domain=1:-1,samples=150,variable=\t]
    ({2.8*\t},{1+.46*sqrt(max(0,1-\t*\t))});
  \node[wkb turning front] at (-2.8,1) {};
  \node[wkb turning front] at (2.8,1) {};
  \node[wkb cut endpoint] at (-2.8,-1) {};
  \node[wkb cut endpoint] at (2.8,-1) {};
  \node[wkb label,fill=white,inner sep=1pt] at (0,-.34)
    {$p=-\arccos\xi$};
  \node[wkb label,anchor=south] at (0,.49) {$p=\arccos\xi$};
  \node[wkb label,anchor=north] at (0,1.97)
    {$p=2\pi-\arccos\xi$};
\end{tikzpicture}}

\newcommand{\WKBPlaneCaseIII}[1][1]{%
  \begin{tikzpicture}[x=1cm,y=1.45cm,scale=#1]
    \node[inner sep=0pt,outer sep=0pt,minimum size=0pt]
    at (-4.5,0) {};
    \node[inner sep=0pt,outer sep=0pt,minimum size=0pt]
    at (5.2,0) {};
  \draw[wkb guide] (-3.35,-1)--(3.35,-1);
  \draw[wkb guide] (-3.35,1)--(3.35,1);
  \draw[wkb axis] (-3.55,0)--(4.35,0) node[right,wkb label] {$x$};
  \draw[wkb axis] (-3.35,-1.13)--(-3.35,1.13)
    node[above,wkb label] {$p$};
  \foreach \yy/\lab in {-1/{-\pi},0/0,1/{\pi}} {
    \draw[wkb tick] (-3.41,\yy)--(-3.29,\yy);
    \node[wkb label,anchor=east] at (-3.47,\yy) {$\lab$};
  }
  \foreach \xx/\lab in {-2.8/{x_j^L},2.8/{x_j^R}} {
    \draw[wkb tick] (\xx,-.045)--(\xx,.045);
    \node[wkb label,anchor=north] at (\xx,-.08) {$\lab$};
  }
  \draw[wkb halo] plot[domain=-1:1,samples=150,variable=\t]
    ({2.8*\t},{.5+.75*\t-.25*\t*\t*\t});
  \draw[wkb halo] plot[domain=1:-1,samples=150,variable=\t]
    ({2.8*\t},{-(.5+.75*\t-.25*\t*\t*\t)});
  \draw[wkb curve] plot[domain=-1:1,samples=150,variable=\t]
    ({2.8*\t},{.5+.75*\t-.25*\t*\t*\t});
  \draw[wkb curve] plot[domain=1:-1,samples=150,variable=\t]
    ({2.8*\t},{-(.5+.75*\t-.25*\t*\t*\t)});
  \node[wkb turning front] at (-2.8,0) {};
  \node[wkb cut endpoint] at (2.8,1) {};
  \node[wkb cut endpoint] at (2.8,-1) {};
  \draw[wkbAxisGray!60,densely dashed,
    {Stealth[length=2mm]}-{Stealth[length=2mm]}] (3.18,-.96)--(3.18,.96);
  \node[wkb label,align=left,anchor=west,fill=white,inner sep=1pt]
    at (3.32,.47) {$-\pi\equiv\pi$\\on $S^1$};
\end{tikzpicture}}
\ExplSyntaxOn
\cs_set:Npn \__first_footerline:
{
  \group_begin:
  \small\sffamily
  \ifnum\theblind>0\relax
  \else
    \__short_authors:
  \fi
  \group_end:
}
\ExplSyntaxOff
\begin{document}
\let\WriteBookmarks\relax
\def\floatpagepagefraction{1}
\def\textpagefraction{.001}

\pgfplotsset{compat=1.18}


\shorttitle{}    

\shortauthors{Finkel, González-López and Zapata}

\title [mode = title]{Discrete WKB approach to density profiles in inhomogeneous two-band
  free-fermion chains}

\tnotemark[1] 

\tnotetext[1]{This work was partially supported by research grant PID2024-155527NB-I00 from
  Spain's Mi\-nis\-te\-rio de Ciencia, Innovación y Universidades.}

%

\author[1]{F. Finkel}[orcid=0000-0003-0459-7564]


\ead{ffinkel@ucm.es}



\author[1]{A. González-López}[orcid=0000-0001-7620-6097]
\cormark[1]
\ead{artemio@ucm.es}


\cortext[1]{Corresponding author.}

\affiliation[1]{organization={Departamento de Física Teórica, Facultad de Ciencias Físicas,
    Universidad Complutense de Madrid}, addressline={Plaza de las Ciencias~1}, city={Madrid},
  citysep={}, 
  postcode={28040},
  country={Spain}}

\author[2]{M. Zapata}[orcid=0009-0008-0052-7167]



\ead{martin.zapata@mpq.mpg.de}



\affiliation[2]{organization={Theory Division, Max Planck Institute of Quantum Optics},
            city={Garching},
            citysep={}, 
            postcode={85748}, 
            country={Germany}}
          

\begin{abstract}
  We extend the discrete WKB method to one-dimensional bipartite tight-binding chains of free
  fermions with slowly varying hopping amplitudes and on-site potentials. We focus on the
  inhomogeneous Rice--Mele model, suggesting possible experimental realizations thereof in optical
  superlattices. We derive explicit semiclassical expressions for the single-particle wave
  functions and density of states, as well as the sublattice-resolved local particle densities and
  filling fraction at arbitrary Fermi energy. In particular, we show that the WKB density profiles
  agree with those obtained from the local-density approximation. We analyze the spatial
  dependence of the local band edges and its consequences for the global spectral structure, and
  clarify the role played by the phase-space topology in determining the Bohr--Sommerfeld
  quantization conditions. Finally, we illustrate our results with two representative examples: a
  homogeneous Rice--Mele chain subject to a harmonic trap and a fully inhomogeneous chain
  containing a domain wall.
\end{abstract}


\maketitle

\section{Introduction}\label{sec.intro}

Tight-binding lattice Hamiltonians can be realized with a high degree of control in a variety of
experimental platforms, including optical superlattices of ultracold
atoms~\cite{LSZAB16,NTTI16,WZGMR23}. In particular, these platforms allow the hopping amplitudes
and on-site energies of the corresponding tight-binding model to be controlled through the
underlying lattice potentials~\cite{LZMMX22}. A fundamental problem in this context is to
determine local observables---in particular, the particle density---for Hamiltonians whose
parameters vary on a scale much larger than the lattice spacing. To this end, local density
approximations are widely used to relate these quantities in smooth inhomogeneous systems to their
homogeneous counterparts~\cite{As05,BDZ08,GPS08}, particularly for harmonically confined lattice
gases~\cite{BN10}. The situation becomes particularly complex, and therefore richer, for lattice
systems with nontrivial band structure, since a fixed Fermi energy can cross different local bands
and gaps as position varies.

The (homogeneous) Rice--Mele (RM) chain~\cite{RM82} is one of the simplest non-trivial examples of
a tight-binding multiband model. More precisely, the RM chain is a tight-binding model of spinless
fermions in a bipartite one-dimensional lattice, subject to a staggered on-site potential. Its
spectrum is symmetric about zero as a consequence of particle-hole symmetry, and is generically
gapped. Thus the RM chain provides a clean realization of a one-dimensional two-band model.
Although originally formulated as a simple model of a diatomic conjugated polymer, the homogeneous
RM chain has subsequently played the role of a canonical one-dimensional two-band model for the
study of topological phenomena, as witnessed by the vast amount of literature devoted to it. In
particular, its bulk polarization and associated boundary charge, domain-wall excitations, and
edge states, have been extensively investigated, both for the original model and for various
generalizations~\cite{LKPWS20,AGSSM22,Ta24}. The homogeneous RM model is one of the simplest
realizations of a Thouless pump~\cite{Th83}, in which an adiabatic cycle of the hopping and
staggered on-site potential parameters leads to quantized particle transport. As mentioned above,
this phenomenon has been experimentally realized with ultracold atoms in dynamically controlled
optical superlattices~\cite{LSZAB16,NTTI16,WZGMR23}.

Rice--Mele models subject to a spatially varying on-site potential (uniform in each cell) arise
naturally in optical-superlattice realizations, where external confinement produces a scalar shift
of the local energy levels. For instance, in Ref.~\cite{WTD13} a RM-type pump in a harmonic trap
was studied in detail, and the resulting particle-density profile was numerically calculated.
Harmonic confinement also plays an important role in experimental realizations of RM
pumping~\cite{NTTI16}, and its effects on charge transport have been analyzed theoretically in
Ref.~\cite{SC21}. Linearly varying scalar potentials, associated with a static force or electric
field, have also been considered in RM chains, for instance in connection with Wannier--Stark
physics~\cite{CTPF16}. More general inhomogeneous RM (or SSH~\cite{SSH79}) models, associated to
classical families of discrete orthogonal polynomials, were also recently discussed in
Ref.~\cite{CLMPV26}. For the homogeneous non-interacting RM chain, analytical expressions for the
sublattice-resolved bulk density were derived in Ref.~\cite{LKPWS20}, primarily in the context of
boundary-charge physics. However, despite this extensive work, a general analytical description of
the bulk sublattice-resolved particle densities of large (but finite) RM chains with smoothly
varying, position-dependent hopping amplitudes and on-site potentials, valid at arbitrary filling,
appears to be lacking. The aim of this paper is precisely to fill this gap, obtaining explicit
analytic expressions for these quantities in the general setting described above. The method we
shall apply, which to the best of our knowledge has not been used in this context, is based on a
discrete WKB approximation for multi-component recurrence relations.

Discrete WKB-type methods for three-term recurrence or difference equations, although not as well
known as their continuous counterpart, are not new in the mathematics~\cite{DM67} and
physics~\cite{Br93} literature. In fact, a discrete WKB approximation was successfully applied by
the present authors to evaluate the particle density in inhomogeneous fermionic chains of XX type
at arbitrary filling~\cite{ZFG26}, explaining saturation/depletion phenomena which result in
entanglement entropy suppression~\cite{FG21,MSSSR22}.

It should be stressed, however, that in the references mentioned above the discrete WKB method was
applied to \emph{scalar} recurrence relations. Extensions of discrete WKB methods to matrix
difference equations have also been considered in the mathematical literature~(see, e.g.,
\cite{CS17}), but their physical applications appear to be largely unexplored. In this work we
extend this framework to the \emph{vector} three-term recurrence relation satisfied by the
two-component single-particle eigenfunctions of the RM chain. More precisely, we apply a WKB-like
approximation directly to the discrete eigenvalue problem in the slowly-varying, large-system
limit, without first reducing this problem to a system of Schrödinger-like differential equations.
This approach allows us to capture the essential bulk behavior of the rapidly oscillating
single-particle eigenfunctions throughout the spectrum, which we then use to derive closed-form
expressions for the sublattice-resolved particle density at arbitrary fillings for general
slowly-varying inhomogeneities. In fact, in the inhomogeneous case the \emph{local}
single-particle spectrum of the RM chain is bounded by four smooth position-dependent functions
defining the local band edges. At a fixed Fermi energy $\vep_F$ (i.e., at arbitrary filling), the
same chain may therefore consist of spatial regions that are empty, partially filled (metallic),
locally insulating, or fully occupied. Our formulas reproduce with remarkable accuracy the
particle-density profiles across these regions and, in particular, determine the position of their
boundaries, as illustrated by several concrete examples. Our examples include situations both with
and without a global gap, as well as cases featuring a Dirac point at which the dimerization
changes sign.

The paper is organized as follows. In Section~\ref{sec.model} we set up our notation and discuss
possible experimental realizations of inhomogeneous RM chains in the framework of optical
superlattices of ultracold atoms. Section~\ref{sec.homogeneous} reviews the computation of the
single-particle spectrum of the homogeneous RM model, showing that when the number of sites is odd
the energies and their corresponding eigenfunctions can be determined in closed form.
Section~\ref{sec.LDA} outlines the derivation of the local-density approximation to the
sublattice-resolved particle density of the inhomogeneous RM chain, using the corresponding
results for its homogeneous counterpart presented in one of the appendices. In
Section~\ref{sec.consWKB} we develop the discrete WKB treatment of the single-particle wave
functions, which is used in Section~\ref{sec.WKBdens} to determine the density of states and the
particle-density profiles, including the average density, filling fraction, and sublattice
imbalance. We illustrate our general results in Section~\ref{sec.examples} by means of two
representative examples, namely the homogeneous Rice--Mele chain in a harmonic trap and an
inhomogeneous chain containing a domain wall. Finally, in Section~\ref{sec.conc} we present our
conclusions and outline possible future developments suggested by the present work. Several
technical details and auxiliary results are collected in the Appendices.

\section{The model}\label{sec.model}

We shall take the Hamiltonian of the inhomogeneous RM model~\cite{RM82} with unit cells of size
$Z=2$ as
\begin{equation}
  \label{HRM}
  H = \sum_{n=0}^{L-1}(v_na_n^\dagger b_n+\Hc)+\sum_{n=0}^{L^*-2}(w_nb_n^\dagger
  a_{n+1}+\Hc)
  +\sum_{n=0}^{L^*-1}\mu_n^Aa_n^\dagger a_n+\sum_{n=0}^{L-1}\mu_n^Bb_n^\dagger b_n,
\end{equation}
where:
\begin{itemize}
\item $L$ is the number of cells, labeled from $0$ to $L-1$
\item $L^*=L$ for the ``even'' chain (with $L$ cells), whereas $L^*=L+1$ for the ``odd'' chain
  (with $L$ cells plus a single site at the end)
\item $v_n$, $w_n$, $\mu_n^A$, and $\mu_n^B$ are real parameters, with $v_n,w_n\ne0$.
\end{itemize}
We shall also use in what follows the notation
\begin{equation}
  \label{mude}
  \bar\mu_n=\frac12(\mu_n^A+\mu_n^B),\qquad \dmu_n=\frac12(\mu_n^A-\mu_n^B).
\end{equation}
Note that the $2L$ (or
$2L+1$) sites can alternatively be numbered either by a single integer $m$ or by a pair $(n,\si)$,
where $\si=0$ for the A sublattice and $\si=1$ for the B sublattice. Obviously, $m=2n+\si$; more
generally, for a polymer with $Z>2$ sites per cell we would have $m=nZ+\si$, with
$\si=0,\dots,Z-1$.

In fact, the condition $v_n,w_n\neq0$ allows us to take these coefficients real and positive
without loss of generality by a suitable local $\mathrm U(1)$ gauge transformation of the fermion operators, since
there are no gauge-invariant phases on an open chain. This also holds true for polymer models with
arbitrary sublattice size $Z$ (see Appendix~\ref{app.gauge}).

In the last decade, advances in ultracold-atom optical superlattices have enabled the experimental
realizations of a variety of theoretical models in condensed-matter physics. For instance, in
ref.~\cite{LZMMX22} the authors propose a simple experimental set-up to simulate a wide class of
Aubry--André--Harper~\cite{Ha55,AA80} models, which include as a particular case the homogeneous
RM chain. In fact, it is possible to extend the ideas of the previous reference to experimentally
realize a class of inhomogeneous RM chains. More precisely, we propose to use an optical
superlattice with potential
\begin{equation}\label{Voptlatt}
  V(x)=V_0\sin^2\left(\frac{2\pi x}{a}\right) +V_1(x)\sin^2\left(\frac{\pi
      x}{a}-\frac{\phi(x)}2\right),
\end{equation}
where $a/2$ is the optical lattice spacing, ${V_1(x)}\ll V_0$, and both $V_1(x)$ and the phase
$\phi(x)$ are slowly varying in the lattice, namely
\[
  \abs{\frac{V_1'(x)}{V_1(x)}}\ll\frac1a,\qquad \abs{\phi'(x)}\ll\frac1a\,.
\]
Note that the potential~\eqref{Voptlatt} differs from that used in ref.~\cite{LZMMX22} in the
modulation of $V_1$ and $\phi$. We shall assume throughout that the additional potentials are
sufficiently weak that the relevant low-energy single-particle states can be accurately expanded
in terms of the lowest-band Wannier functions of the unperturbed periodic Hamiltonian
\begin{equation}\label{h0}
  h_0=-\frac{\hbar^2}{2M}\pd_x^2+V_0\sin^2\left(\frac{2\pi x}{a}\right),
\end{equation}
where $M$ is the atomic mass\footnote{In the ensuing discussion we restore Planck's constant
  \(\hbar\) to make the connection with the physical parameters of the optical superlattice
  explicit.}. The tight-binding Hamiltonian obtained from the above potential is then
approximately
\[
  H=\sum_{m}\left(h_{m,m+1}c_m^\dagger c_{m+1}+\Hc\right)+\sum_{m}h_{mm}c^\dagger_mc_m\,,
  \qquad h_{ij}\coloneqq\mel*{W_i}{h}{W_j},
\]
where
\[
  h=-\frac{\hbar^2}{2M}\pd_x^2+V(x),
\]
and the $W_j$'s are the lowest-band Wannier functions of the periodic Hamiltonian $h_0$. It can be
shown (see Appendix~\ref{app.AAH}) that the matrix elements $h_{m,m+1}$ and $h_{mm}$ are of the
form
\begin{equation}\label{hij}
h_{m,m+1}\simeq A_1+V_0B_1-(-1)^{m} S\left(\frac{am}2\right)t_S,\qquad
h_{mm}\simeq A_0+V_0B_0+\frac12V_1\left(\frac{am}2\right)-(-1)^{m} C\left(\frac{am}2\right)t_C,
\end{equation}
where $A_i$, $B_i$, $t_S$ and $t_C$ are constant parameters, and
\begin{equation}\label{CSx}
  C(x)=\frac12 V_1(x)\cos\phi(x),\qquad S(x)=\frac12 V_1(x)\sin\phi(x).
\end{equation}
Setting $a_n=c_{2n}$, $b_{n}=c_{2n+1}$, we see that $H$ is the Hamiltonian of an inhomogeneous RM
chain with parameters
\begin{alignat*}{2}
  v_n&\simeq A_1+V_0B_1-S(an)t_S,\qquad\quad
       & w_n&\simeq A_1+V_0B_1+S(an)t_S,\\
  \bar\mu_n&\simeq A_0+V_0B_0+\frac12 V_1(an),
             & \dmu_n&\simeq-C(an)t_C.
\end{alignat*}
Thus the slowly varying amplitude $V_1$ controls the position dependence of $\bmu_n$, while the
functions $C(x)$ and $S(x)$ (involving both the amplitude $V_1$ and the slowly varying phase
$\phi$) respectively determine that of $\dmu_n$ and $w_n-v_n$.

Although the chain parameters $v_n$, $w_n$, $\bar\mu_n$, and $\de\mu_n$ in the previous equation
are all position-dependent, the sum $v_n+w_n$ is not. We propose two possible ways of generating a
(small) dependence of $v_n+w_n$ on position, to obtain a more general inhomogeneous RM model. The
first one consists in adding to the constant amplitude $V_0$ of the main lattice potential
$V_0\sin^2(2\pi x/a)$ a slowly varying term $\de V_0(x)$. If $\delta V_0(x)$ is sufficiently small
with respect to $V_0$, we can again assume that the relevant low-lying single-particle states can
be approximately expanded in terms of the localized lowest-band Wannier functions of the periodic
Hamiltonian $h_0$. To leading order in the spatial variation, the previous formulas therefore
remain valid with $V_0$ replaced by $V_0+\delta V_0(an)$, making $v_n+w_n$ position-dependent. An
alternative way of generating position dependence in the sum $v_n+w_n$ is to add to the
Hamiltonian $h$ an array of barriers centered at the intracell bonds,
\[
  V_2(x)=\sum_mU(b_m)g(x-b_m),\qquad b_m\coloneqq ma+\frac{a}4,
\]
with $\abs{U(x)}\ll V_0$ slowly varying on the scale of the lattice spacing and $g(s)$ even and
sharply localized about $s=0$ (for instance, $g(s)=e^{-s^2/(2\si^2)}$, with $\si\ll a$).
Proceeding as in Appendix~\ref{app.AAH} it can be shown that, under the above assumptions on $U$
and $g$, the parameters $v_n$ and $\bmu_n$ become
\[
  v_n\simeq A_1+V_0B_1-S(an)t_S+U(na)D,\qquad \bar\mu_n\simeq A_0+V_0B_0+\frac12 V_1(an)+U(na)F,
\]
for suitable constants $D$ and $F$ determined by overlap integrals involving the Wannier functions
and $g$, while $w_n$ and $\dmu_n$ remain approximately unchanged. In particular, $v_n+w_n$
acquires the (small) position-dependent contribution $U(na)D$, as claimed.

\section{The homogeneous Rice--Mele model}\label{sec.homogeneous}

Let us briefly recall the solution of the homogeneous Rice--Mele model, for which $v_n=v$,
$w_n=w$, and $\de\mu_n=\de\mu$ for all $n$, with $v\ne w$. We shall assume without loss of
generality that $\mu_n=\bar\mu=0$, since the general case $\bar\mu\ne0$ can be recovered by
replacing $\vep$ by $\vep-\bar\mu$. We shall also denote single-particle wave functions by
\[
  \ket\psi=\sum_{n=0}^{L^*-1}\al_n\ket{n}_A+\sum_{n=0}^{L-1}\be_n
  \ket{n}_B\eqqcolon\ket\al+\ket\be,\qquad
  \text{with}\quad\ket{n}_A=a_n^\dagger\ket0,\quad\ket{n}_B=b_n^\dagger\ket0,
\]
where $\ket0$ is the vacuum. The eigenvalue equation $H\!\ket\psi=\vep\ket\psi$ is thus equivalent
to the system of linear recurrence relations
\begin{subequations}\label{rr}
  \begin{align}
    \label{rr1}
    (\vep-\dmu)\al_n&=v\be_n+w\be_{n-1},\qquad 0\le n\le L^*-1,\\
    \label{rr2}
    (\vep+\dmu)\be_n&=v\al_n+w\al_{n+1},\qquad 0\le n\le L-1,
  \end{align}
\end{subequations}
with boundary conditions depending on the parity of the number of sites (see below).

Let us start by finding a three-term recurrence relation for the coefficients $\be_n$ by
multiplying eq.~\eqref{rr2} by $\vep-\dmu$ and using eq.~\eqref{rr1} to eliminate $\al_n$ and
$\al_{n+1}$, namely
\begin{equation}\label{rraln}
  (\vep^2-\dmu^2-v^2-w^2)\be_n=vw(\be_{n-1}+\be_{n+1}),\qquad 0\le n\le L-1.
\end{equation}
This recurrence relation with constant coefficients has the two linearly independent
solutions\footnote{More precisely, these solutions are linearly independent unless $q=0$ or
  $q=\pi$. It can be shown that in the former case $v/w=-L/(L+1)$ and $\be_n=A(n+1)$, while in the
  latter $v/w=L/(L+1)$ and $\be_n=A(-1)^n(n+1)$ for some constant $A$. However, the first of these
  solutions must be discarded, as we are assuming that $v,w>0$, whereas for fixed $v\ne w$ the
  second one cannot occur in the thermodynamic limit $L\to\infty$. For this reason, in what
  follows we shall assume that $q\ne0,\pi$.} $\be_n=\e^{\pm\iu nq}$ provided that
\begin{equation}\label{disprel}
  \vep^2-\dmu^2-v^2-w^2=2vw\cos q.
\end{equation}
This equation relates the single-particle energy $\vep$ with the quasimomentum $q$. More
precisely, a set of quasi-momenta $\{q_k\}$ corresponds to twice as many energies, through the
dispersion relations
\begin{equation}\label{disprel-ens}
  \vep_\pm(q_k)=\pm\sqrt{\dmu^2+v^2+w^2+2vw\cos q_k}\eqqcolon\pm E(q_k).
\end{equation}
The general solution of eq.~\eqref{rraln} is a linear combination with
constant coefficients of the two independent solutions $\e^{\pm\iu nq}$.

When the number of sites is even we have $L^*=L$, and the boundary conditions are
\begin{equation}
  \label{bdrye}
  \be_{-1}=\al_L=0.
\end{equation}
Imposing the condition $\be_{-1}=0$ we conclude that
\begin{equation}\label{ben}
  \be_n=A\sin\bigl((n+1)q\bigr)
\end{equation}
for some non-zero constant $A$. The coefficient $\al_n$ is then obtained from eq.~\eqref{rr1}
provided that $\vep-\dmu\ne0$. This condition certainly holds, since otherwise eqs.~\eqref{rr} are
easily seen to imply that $\al_n=\be_n=0$ for all $n$. We therefore have
\begin{equation}\label{aln}
  \al_n=\frac{A}{\vep-\dmu}\left[v\sin\bigl((n+1)q\bigr)+ w\sin\bigl(nq\bigr)\right].
\end{equation}
Finally, imposing the remaining boundary condition $\al_{L}=0$ yields the transcendental equation
determining the allowed quasimomentum $q$, and hence the single-particle spectrum:
\begin{equation}\label{qeigeq}
  v\sin\bigl((L+1)q\bigr)+ w\sin\bigl(Lq\bigr)=0.
\end{equation}
\begin{remark}
  Since, as explained in the previous footnote, we can assume that $q\ne0,\pi$, this equation can
  be equivalently written as
\[
  vU_{L}(x)+wU_{L-1}(x)=0,\qquad x=\cos q,
\]
where $U_m$ is the Chebyshev polynomial of the second kind and degree $m$. Using the properties of
these polynomials, it can be shown that for $v/w>L/(L+1)$ there are $L$ real roots $q_1,\dots,q_L$
of eq.~\eqref{qeigeq} in the interval $(0,\pi)$, whose corresponding $2L$ eigenvalues are obtained
from eq.~\eqref{disprel-ens}. It can also be shown that for $v/w< L/(L+1)$ there are only $L-1$
such roots. The additional root of the eigenvalue equation~\eqref{qeigeq} is of the form
$q=\pi+\iu\ka$, giving rise to the exceptional eigenvalues\footnote{It can be shown that these
  eigenvalues are real, as expected.} (edge modes)
\[
  \pm\sqrt{\dmu^2+v^2+w^2-2vw\cosh\ka}.
\]
  Substituting $q=\pi+\iu\ka$ into eq.~\eqref{qeigeq} we obtain the relation
  \[
    v\sinh\bigl((L+1)\ka\bigr)-w\sinh\bigl(L\ka\bigr)=0,
  \]
  which in the thermodynamic limit yields
  \[
    \e^{-\ka}\simeq\frac{v}{w},
  \]
  whence
  \[
    v^2+w^2-2vw\cosh\ka\simeq v^2+w^2-vw\left(\frac{v}{w}
      +\frac{w}{v}\right)=0.
  \]
  Therefore the exceptional eigenvalues are approximately equal to $\pm\abs{\dmu}$ in the
  thermodynamic limit.\qed
\end{remark}

The transcendental equation~\eqref{qeigeq} determining the allowed momenta of the even chain
cannot be solved in closed form, and its solution depends on the chain parameters $(v,w)$. As we
shall see below, both difficulties are avoided by considering the odd chain (i.e, with an
additional site at the right end), obtained by setting $L^*=L+1$ in eq.~\eqref{HRM}. This has no
effect on bulk quantities in the thermodynamic limit, but as we shall see below it considerably
simplifies the calculation of the particle densities.

Indeed, in the case of an odd number of sites the boundary conditions read
\begin{equation}
  \label{bceven}
  \be_{-1}=\be_L=0,
\end{equation}
where the last equality comes from eq.~\eqref{rr1} with $L^*=L+1$. From the condition $\be_{-1}=0$
and eq.~\eqref{rr1} we still obtain eqs.~\eqref{ben} and~\eqref{aln} for the components of the
wave function. However, in this case the boundary condition $\be_L=0$ yields the $L$ allowed
momenta in closed form:
\begin{equation}\label{qkeven}
  \sin\bigl((L+1)q\bigr)=0\en\implies\en q=\frac{k\pi}{L+1}\eqqcolon q_k,\qquad k=1,\dots,L,
\end{equation}
with corresponding $2L$ eigenvalues given by eq.~\eqref{disprel}. Finally, if $\vep=\dmu$, from
eq.~\eqref{rr1} and the boundary conditions~\eqref{bceven} we deduce that $\be_n=0$ for all $n$
and
\[
  v\al_n+w\al_{n+1}=0\en\implies\en\al_n=A\left(-\frac{v}{w}\right)^n.
\]
Normalizing we obtain
\[
  \al_n=
    \sqrt{\frac{1-v^2/w^2}{1-\left(v^2/w^2\right)^{L+1}}}\,\left(-\frac{v}{w}\right)^n;
\]
this is an \emph{edge mode}, concentrated on the left end if $v<w$ or on the right one if $v>w$.

To normalize the wave functions with energies $\vep\ne\dmu$, we note that the recurrence
relations~\eqref{rr1}-\eqref{rr2} can be written as
\[
  (\vep-\dmu)\ket\al=T\ket\be,\qquad  (\vep+\dmu)\ket\be=T^\dagger\ket\al,
\]
where $T:\CC^L\to\CC^{L+1}$ is the operator defined by
\[
  T=\sum_{n=0}^{L-1}\big(v\ket{n}_A+w\ket{n+1}_A\big)\!\bra{n}_B.
\]
We thus have
\[
  (\vep-\dmu)\braket{\al}{\al}=\braket{\al}{T\be}=\braket{\smash{T^\dagger}\al}{\be}=(\vep+\dmu)
  \braket{\be}{\be}\implies\braket{\al}{\al}=\frac{\vep+\dmu}{\vep-\dmu}\braket{\be}{\be},
\]
and therefore
\[
  \braket{\psi}{\psi}=\braket{\al}{\al}+\braket{\be}{\be}
  =\frac{2\vep}{\vep-\dmu}\braket{\be}{\be}.
\]
Hence it suffices to compute $\braket{\be}{\be}$, which is straightforward:
\[
  \abs{A}^{-2}\braket{\be}{\be}=\sum_{m=1}^L\sin^2(mq_k)
  =\sum_{m=1}^L\sin^2\biggl(\frac{mk\pi}{L+1}\biggr)=\frac{L+1}2\,.
\]
Thus the normalized eigenfunctions have components
\begin{equation}
  \label{normeig}
  \al_n=\frac{\sgn(\vep-\dmu)}{\sqrt{(L+1)\vep(\vep-\dmu)}}
  \left(v\sin\bigl((n+1)q_k\bigr)+w\sin\bigl(nq_k\bigr)\right),\qquad
   \be_n=\sqrt{\frac{\vep-\dmu}{(L+1)\vep}}\,\sin\bigl((n+1)q_k\bigr)
\end{equation}
up to an inessential overall phase.

The single-particle energies of the RM chain with an odd number of sites and $\bmu=0$ are the
numbers
\begin{equation}\label{sp-odd}
  \vep_{-}(q_1)<\cdots<\vep_-(q_L)<\dmu<\vep_+(q_L)<\cdots<\vep_+(q_1),
\end{equation}
where $\vep_\pm$ and $q_k$ are given by eqs.~\eqref{disprel-ens} and~\eqref{qkeven}. Thus, in the
thermodynamic limit $L\to\infty$ the single-particle spectrum consists of the two bands
$[-\Emax,-\Emin]$ and $[\Emin,\Emax]$, with
\[
  E_{\text{min}}=\sqrt{\dmu^2+(v-w)^2},\qquad
  E_{\text{max}}=\sqrt{\dmu^2+(v+w)^2},
\]
together with the isolated point $\de\mu$. In particular, the energy gap $2\Emin$ is
non-vanishing, since we are assuming that $v\ne w$.

\section{Local density approximation}\label{sec.LDA}

In this section we derive the local density approximation (LDA) to the sublattice-resolved
particle densities of the inhomogeneous RM model for arbitrary fillings (i.e., Fermi energies).
The key idea is to replace the parameters in the corresponding quantities for the homogeneous
model by their position-dependent counterparts. To this end, we first need to express the
sublattice-averaged particle density (or average density, for short) and the sublattice imbalance
of the homogeneous model,
\[
  \bar\rho\coloneqq\frac1{2a}\big(\expval*{a_n^\dagger a_n}+\expval*{b_n^\dagger b_n}\big),\qquad
  \de\rho\coloneqq\frac1{2a}\big(\expval*{b_n^\dagger b_n}-\expval*{a_n^\dagger a_n}\big),
\]
computed in Appendix~\ref{sec:homdens}, in terms of the Fermi energy $\vep_F$.

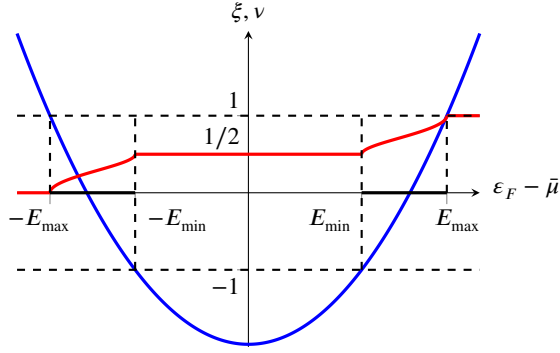
\begin{figure}[pos=t]
  \centering
  \begin{tikzpicture}
\pgfmathsetmacro{\aa}{sqrt(29)} 
\pgfmathsetmacro{\bb}{sqrt(89)} 

\begin{axis}[
  width=7.7cm,
  height=5.76cm,
  axis lines=middle,
  xmin=-11, xmax=11,
  ymin=-2, ymax=2.1,
  samples=500,
  no markers,
  tick align=outside,
  xtick={-\bb,-\aa,\aa,\bb},
  xticklabels={,,,},
  ytick={-1,0.5,1},
  yticklabels={,,},
  xlabel={$\vep_F-\bmu$},
  ylabel={$\xi,\nu$},
  xlabel style={at={(axis description cs:1.01,0.50)},anchor=west},
  ylabel style={at={(axis description cs:0.50,1.00)},anchor=south},
  clip=false,
]

\addplot[blue, very thick, domain=-11:11] {(x^2 - 59)/30};

\addplot[red, very thick, domain=-11:-\bb] {0};
\addplot[red, very thick, domain=-\bb:-\aa] {acos((x^2 - 59)/30)/360};
\addplot[red, very thick, domain=-\aa:\aa] {0.5};
\addplot[red, very thick, domain=\aa:\bb] {1 - acos((x^2 - 59)/30)/360};
\addplot[red, very thick, domain=\bb:11] {1};

\addplot[black, dashed, thick] coordinates {(-11,1) (11,1)};
\addplot[black, dashed, thick] coordinates {(-11,-1) (11,-1)};

\node[anchor=east, yshift=6pt] at (axis cs:0,1) {$1$};
\node[anchor=east, yshift=6pt] at (axis cs:0,0.5) {$\textstyle 1/2$};
\node[anchor=east, yshift=-6pt] at (axis cs:0,-1) {$-1$};

\addplot[black, dashed, thick] coordinates {(-\bb,0) (-\bb,1)};
\addplot[black, dashed, thick] coordinates {(-\aa,-1) (-\aa,1)};
\addplot[black, dashed, thick] coordinates {(\aa,-1) (\aa,1)};
\addplot[black, dashed, thick] coordinates {(\bb,0) (\bb,1)};

\node[anchor=north, yshift=-3pt, xshift=-4pt] at (axis cs:-\bb,0) {$-E_{\max}$};
\node[anchor=north, yshift=-3pt, xshift=16pt] at (axis cs:-\aa,0) {$-E_{\min}$};
\node[anchor=north, yshift=-3pt, xshift=-12pt] at (axis cs:\aa,0) {$E_{\min}$};
\node[anchor=north, yshift=-3pt, xshift=4pt] at (axis cs:\bb,0) {$E_{\max}$};

\addplot[black, very thick] coordinates {(-\bb,0) (-\aa,0)};
\addplot[black, very thick] coordinates {(\aa,0) (\bb,0)};

\end{axis}

\end{tikzpicture}
\caption{schematic plot of $\xi(\vep_F)$ (blue) and $\nu(\vep_F)$ (red) as functions of the Fermi
  energy $\vep_F$. The two metallic bands $[\bmu-\Emax,\bmu-\Emin]$ and $[\bmu+\Emin,\bmu+\Emax]$
  are represented by thick lines on the horizontal axis.}
  \label{fig.xinu}
\end{figure}

Suppose, to begin with, that $\vep_F$ is in the first band, i.e., that
\[
  -E_{\text{max}}\le \vep_F-\bmu\le-E_{\text{min}},
\]
where we have restored the constant on-site energy $\bmu$ for reasons that shall be clear in the
sequel. Since $\nu=M/(2L+1)$ and
\[
  (\vep_F-\bmu)^2=(\vep_M-\bmu)^2=\dmu^2+v^2+w^2+2vw\cos\biggl(\frac{M\pi}{L+1}\biggr)
  \simeq\dmu^2+v^2+w^2+2vw\cos(2\pi\nu),
\]
we can write
\begin{equation}\label{xidefh}
  \nu(\vep_F)=\frac1{2\pi}\arccos\xi(\vep_F),\qquad\text{with}\quad
  \xi(\vep_F)=\frac{(\vep_F-\bmu)^2-\dmu^2-v^2-w^2}{2vw}.
\end{equation}
Thus for $-E_{\text{max}}\le \vep_F-\bmu\le -E_{\text{min}}$ we have
\begin{equation}\label{bdrhocons}
  \bar\rho(\vep_F)=\frac1{2\pi a}\arccos\xi(\vep_F),\qquad
  \de\rho(\vep_F)=\frac{\ka\dmu}{2\pi a\sqrt{vw}}\,
  F\biggl(\frac12\arccos(\xi(\vep_F)),\ka\biggr),
\end{equation}
with $\ka$ given by eq.~\eqref{ka}. Obviously, for $\vep_F<\bmu-E_{\text{max}}$ we have
$\nu(\vep_F)=0$ and therefore $\bar\rho(\vep_F)=\de\rho(\vep_F)=0$. Since $\xi(\vep_F)>1$, and
hence $\Re\arccos\xi(\vep_F)=0$, for $\vep_F<\bmu-\Emax$ (cf.~Fig.~\ref{fig.xinu}), we can extend
eq.~\eqref{bdrhocons} to Fermi energies $\vep_F<\bmu-E_{\text{max}}$ by replacing
$\arccos\xi(\vep_F)$ by its real part. In other words,
\begin{equation}
  \label{densB1}
  \bar\rho(\vep_F)=\frac1{2\pi a}\Re\arccos\xi(\vep_F),\qquad
  \de\rho(\vep_F)=\frac{\ka\dmu}{2\pi a\sqrt{vw}}\,
  F\biggl(\frac12\Re\arccos(\xi(\vep_F)),\ka\biggr),
\end{equation}
for Fermi energies $\vep_F\le\bmu-\Emin$.

Suppose next that $\vep_F$ is in the insulating gap $(\bmu-\Emin,\bmu+\Emin)$. In this region
$\nu(\vep_F)=1/2$ and $\xi(\vep_F)<-1$ (cf.~Fig.~\ref{fig.xinu}), so that
$\Re\arccos\xi(\vep_F)=\pi$. We conclude that eq.~\eqref{densB1} is still valid for these
energies.

Finally, suppose that $\vep_F$ lies in the second metallic band $[\bmu+\Emin,\bmu+\Emax]$. In this
case
\[
  \vep_F=\bmu+E_{2L-M+2}\implies
  (\vep_F-\bmu)^2\simeq\dmu^2+v^2+w^2+2vw\cos\bigl(2\pi(1-\nu)\bigr)
  \implies 1-\nu(\vep_F)=\frac1{2\pi}\arccos\xi(\vep_F).
\]
Since in this region $-1\le\xi(\vep_F)\le1$, comparing with eq.~\eqref{barrhoderhoup} we conclude
that for $\vep_F\in[\bmu+\Emin,\bmu+\Emax]$ we have
\begin{equation}
  \label{densB12}
  \bar\rho(\vep_F)=\frac1a-\frac1{2\pi a}\Re\arccos\xi(\vep_F),\qquad
  \de\rho(\vep_F)=\frac{\ka\dmu}{2\pi a\sqrt{vw}}\,
  F\biggl(\frac12\Re\arccos(\xi(\vep_F)),\ka\biggr).
\end{equation}
This equation actually holds for $\vep_F>\bmu+\Emax$, since for these energies $\xi(\vep_F)>1$
(cf.~Fig.~\ref{fig.xinu}), and therefore
\[
  1-\nu=0=\frac1{2\pi}\Re\arccos\xi(\vep_F).
\]
Summarizing, if we define the Fermi quasimomentum $p_F(\vep)\in[0,\pi]$ by
\begin{equation}
  \label{Fermimom}
  p_F(\vep)=\Re\arccos\xi(\vep),
\end{equation}
the formulas
\begin{subequations}
  \label{Fermidensh}
  \begin{align}
    \bar\rho(\vep_F)
    &=
    \begin{cases}
      \frac{p_F(\vep_F)}{2\pi a},&\vep_F\le\bmu\\[1ex]
      \frac1{2\pi a}\big(2\pi-p_F(\vep_F)\big),&\vep_F>\bmu,
    \end{cases}
      \label{Fermidenshb}
    \\
    \de\rho(\vep_F)&=\frac{\ka\dmu}{2\pi a\sqrt{vw}}\,
                     F\Bigl(\frac12p_F(\vep_F),\ka\Bigr)
                     \label{Fermidenshd}
  \end{align}
\end{subequations}
hold for all $\vep_F\in\RR$.

In the spirit of the local density approximation, we now approximate the average density and
sublattice imbalance of a general inhomogeneous Rice-Mele model by replacing in the previous
formulas the constant parameters $v,w,\bmu,\de\mu$ by the continuous versions
$v(x),w(x),\bmu(x),\de\mu(x)$ of the model's parameters $v_n,w_n,\bmu_n,\de\mu_n$. More precisely,
we define the local quasimomentum $p_F(x,\vep)$ at $x=na\in[0,La]\eqqcolon[0,\ell]$ by
\begin{equation}
  \label{pF}
  p_F(x,\vep)=\Re\arccos\xi(x,\vep)=
  \begin{cases}
    \arccos\xi(x,\vep), &\abs{\xi(x,\vep)}\le1,\\
    \pi, &\xi(x,\vep)\le-1,\\
    0, &\xi(x,\vep)\ge1,
  \end{cases}
\end{equation}
where
\begin{equation}
  \label{xix}
  \xi(x,\vep)=\frac{\big(\vep-\bar\mu(x)\big)^2-\dmu(x)^2-v(x)^2-w(x)^2}{2v(x)w(x)},
  \qquad
  \dmu(x)=\frac12\big(\mu^A(x)-\mu^B(x)\big).
\end{equation}
The two metallic bands at $x$ are the two intervals
\begin{equation}\label{cBpm}
\cB_-(x)=[\bar\mu(x)-\Emax(x),\bar\mu(x)-\Emin(x)]\qquad
\cB_+(x)=[\bar\mu(x)+\Emin(x),\bar\mu(x)+\Emax(x)],
\end{equation}
with
\begin{equation}\label{Eminmax}
  \Emin(x)=\sqrt{\dmu(x)^2+\big(v(x)-w(x)\big)^2}\,,\qquad
  \Emax(x)=\sqrt{\dmu(x)^2+\big(v(x)+w(x)\big)^2}\,,
\end{equation}
on which $-1\le\xi(x,\vep_F)\le 1$ and consequently $\Im\arccos\xi(x,\vep_F)=0$. Note that the
Fermi quasimomentum $p_F(x,\vep_F)$ is equal to $0$ for $\vep_F<\bar\mu(x)-\Emax(x)$, increases
monotonically from $0$ to $\pi$ as $\vep_F$ sweeps through the lower local band $\cB_-(x)$, stays
equal to $\pi$ in the local insulating gap~$\big(\bar\mu(x)-\Emin(x),\bar\mu(x)+\Emin(x)\big)$,
decreases monotonically from $\pi$ to $0$ as $\vep_F$ traverses the upper local band $\cB_+(x)$,
and remains equal to $0$ for $\vep_F>\bar\mu(x)+\Emax(x)$. The LDA for the average density and
sublattice imbalance is then the local analogue of eq.~\eqref{Fermidensh}, namely
\begin{subequations}\label{LDArhos}
  \begin{align}
    \label{LDA1}
    \bar\rho(x,\vep_F)&=
    \begin{cases}
      \frac{p_F(x,\vep_F)}{2\pi a},&\vep_F\le\bar\mu(x)\\[1ex]
      \frac1{2\pi a}\big(2\pi-p_F(x,\vep_F)\big),\en&\vep_F>\bar\mu(x),
    \end{cases}
    \\ \label{LDA2}
    \de\rho(x,\vep_F)&=\frac{\dmu(x)}{\pi a\Emax(x)}\,
    F\Bigl(\frac12p_F(x,\vep_F),\ka(x)\Bigr),
  \end{align}
  with
  \begin{equation}
    \label{kaLDA}
    \ka(x)=
    \frac{2\sqrt{v(x)w(x)}}{\Emax(x)}=\sqrt{1-\frac{\Emin^2(x)}{\Emax^2(x)}}\,.
  \end{equation}
\end{subequations}
Note that for $\Emin(x)\lesssim\Emax(x)$ (i.e., for narrow local metallic bands) the previous
formula implies that
\[
  \de\rho(x,\vep_F)\simeq\frac{\dmu(x)}{2\pi a\Emax(x)}\,p_F(x,\vep_F).
\]
Likewise, for $\Emin(x)\ll\Emax(x)$ (i.e., for a narrow local insulator gap) we have\footnote{More
  precisely, in the first line of the following formula we must require that
  $\pi-p_F(x,\vep_F)\gg\Emin(x)/\Emax(x)$.}
\[
  \de\rho(x,\vep_F)\simeq
  \frac{\dmu(x)}{\pi
    a\Emax(x)}\,\begin{cases}
    \log\tan\Bigl(\frac14p_F(x,\vep_F)+\frac\pi4\Bigr),\quad
    &0\le p_F(x,\vep_F)<\pi\\[2ex]
     \log\left(\frac{4\Emax(x)}{\Emin(x)}\right),& p_F(x,\vep_F)=\pi.
  \end{cases}
\]

\section{WKB wave function}\label{sec.consWKB}

In this section we shall apply a generalization of the discrete WKB method developed in
Ref.~\cite{ZFG26} to derive a semiclassical approximation of the single-particle eigenfunctions of
a general inhomogeneous RM model. This approximation will be used in the next section to obtain
closed formulas for the model's average density and sublattice imbalance in the thermodynamic
limit.

Let us set $x=na$, where $a$ is the lattice spacing, and take $a\to0+$ and $L\to\infty$ with
$La\to\ell$ finite. We shall assume that the model's parameters $v_n,w_n,\mu_n^A,\mu_n^B$ vary
slowly and, as in the previous section, denote by $v(x),w(x),\mu^A(x),\mu^B(x)$ their continuum
limits. The coefficients $(\al_n,\be_n)$ of an eigenfunction of the single-particle Hamiltonian
with energy $\vep$,
\[
  \ket\psi=\sum_{n=0}^{L-1}(\al_na_n^\dagger+\be_n b_n^\dagger)\ket0,
\]
satisfy the system of recurrence equations
\begin{equation}\label{contsysdisc}
  (\vep-\mu_n^A)\al_n=v_n\be_n+w_{n-1}\be_{n-1},\qquad (\vep-\mu_n^B)\be_n=v_n\al_n+w_{n}\al_{n+1},
\end{equation}
with $n=0,\dots,L-1$ and boundary conditions $\be_{-1}=\al_L=0$. The continuum version of this
system is
\begin{equation}
  \label{contsys}
  \big(\vep-\mu^A(x)\big)\al(x)=v(x)\be(x)+w\Bigl(x-\frac a2\Bigr)\be(x-a),\quad
  \big(\vep-\mu^B(x)\big)\be(x)=v(x)\al(x)+w\Bigl(x+\frac a2\Bigr)\al(x+a).
\end{equation}
Note that, since $x$ is a cell coordinate, we have taken the intracell parameters $v(x)$,
$\mu^A(x)$, and $\mu^B(x)$ to be cell-centered and the intercell parameter $w(x)$ to be
bond-centered, i.e.,
\[
  v_n=v(x_n),\qquad \mu^{A,B}_n=\mu^{A,B}(x_n),\qquad w_n=w\bigl(x_{n+\frac12}\bigr),
\]
where $x_m=ma$. Introducing the two-component spinor
\[
  \Psi(x)=
  \begin{pmatrix}
    \al(x)\\ \be(x)
  \end{pmatrix},
\]
we can write eqs.~\eqref{contsys} in matrix form as
\begin{equation}\label{contsysmat}
  \big(\vep-\bmu(x)-\dmu(x)\si^z\big)\Psi(x)=v(x)\si^x\Psi(x)+w\Bigl(x-\frac
  a2\Bigr)\si^+\Psi(x-a)+w\Bigl(x+\frac a2\Bigr)\si^-\Psi(x+a),
\end{equation}
where $\si^{x,y,z}$ are the three Pauli matrices and $\si^\pm=\frac12(\si^x\pm\iu\si^y)$. In the
spirit of the WKB approximation, let us express the wave function as\footnote{In fact, the
  eigenfunction $\Psi(x)$ depends on the eigenvalue $\vep$. We shall, however, temporarily
  suppress the dependence of $\Psi$ and related quantities on $\vep$ for the sake of conciseness.}
\[
  \Psi(x)=u(x,a)\,\e^{\frac\iu{a}S(x,a)},
\]
with
\[
  u(x,a)=\sum_{n\ge0}a^nu_n(x),\qquad S(x,a)=\sum_{n\ge0}a^n S_n(x).
\]
We then have
\begin{align*}
  u(x\pm a,a)&=\sum_{m,n\ge0}(\pm1)^m\frac{a^{m+n}}{m!}u_n^{(m)}(x)=
  u_0(x)+a\big(u_1(x)\pm u_0'(x)\big)+O(a^2),\\
  S(x\pm a,a)&=S(x,a)+\sum_{m\ge1,n\ge0}(\pm1)^m\frac{a^{m+n}}{m!}S_n^{(m)}(x)
  =S(x,a)\pm a S_0'(x)+a^2\left(\frac12S_0''(x)\pm S_1'(x)\right)+O(a^3)
\end{align*}
and
\[
  w\Bigl(x\pm\frac a2\Bigr)=w(x)\pm\frac a2w'(x)+O(a^2),
\]
where the prime denotes derivative with respect to $x$. Inserting these expansions into
eq.~\eqref{contsysmat} and setting
\begin{equation}\label{pdef}
  q=S_0'(x)
\end{equation}
we obtain, to zeroth order in the lattice spacing $a$, the equation
\[
  \big(\vep-\bmu(x)-\dmu(x)\si^z\big)u_0(x)=\left(v(x)\si^x+\e^{-\iu q}w(x)\si^++\e^{\iu
      q}w(x)\si^-\right)u_0(x),
\]
or equivalently
\[
  \big(\vep-\bmu(x)-h(x,q)\big)u_0(x)=0,
\]
with
\[
  h(x,q):=
  \begin{pmatrix}
    \dmu(x)& v(x)+w(x)\e^{-\iu q}\\
    v(x)+w(x)\e^{\iu q}& -\dmu(x)
  \end{pmatrix}.
\]
Since the spinor $u_0(x)$ cannot be identically zero, it must be an eigenvector of the matrix
$h(x,q)$ with eigenvalue $\vep-\bmu(x)$. Hence
\begin{equation}\label{cosp}
  \det\bigl(\vep-\bmu(x)-h(x,q)\bigr)=\big(\vep-\bmu(x)\big)^2-\dmu(x)^2-v(x)^2-w(x)^2-2v(x)w(x)\cos
  q=0,
\end{equation}
or equivalently (restoring, for the time being, the dependence in $\vep$)
\begin{equation}\label{cospx}
  \cos q=\frac{\big(\vep-\bmu(x)\big)^2-\dmu(x)^2-v(x)^2-w(x)^2}{2v(x)w(x)}=\xi(x,\vep).
\end{equation}

Setting, for definiteness,
\begin{equation}
  \label{pfull}
  p(x,\vep)\coloneqq
  \begin{cases}
    \arccos\xi(x,\vep)\in[0,\pi],& \abs{\xi(x,\vep)}\le1\\
    \iu\arccosh\xi(x,\vep), &\xi(x,\vep)>1\\
    \pi+\iu\arccosh\abs{\xi(x,\vep)},\quad &\xi(x,\vep)<-1,
  \end{cases}
\end{equation}
we have
\begin{equation}\label{qpmp}
  q=\pm p(x,\vep)\quad\bmod{2\pi},
\end{equation}
and therefore
\begin{equation}\label{S0p}
  S_0(x,\vep)=\pm\int^x\dd x\,p(x,\vep).
\end{equation}
Note that $p(x,\vep)$ is real in the \emph{classically allowed region}
$\big\{x\in[0,\ell]:\abs{\xi(x,\vep)}\le 1\big\}$. Let us define the set $\Om_\vep$ as the
classically allowed region minus the set of \emph{turning points} (or band crossings)
$\big\{x\in[0,\ell]:\abs{\xi(x,\vep)}=1\big\}$, namely
\begin{equation}
  \Om_\vep=\big\{x\in[0,\ell]:\abs{\xi(x,\vep)}< 1\big\}.
\end{equation}
In general, this region will be the disjoint union
\[
  \Om_\vep=\bigsqcup_{j=1}^{g(\vep)}I_j(\vep),\qquad
  I_j(\vep)=\bigl(x_j^L(\vep),x_j^R(\vep)\bigr),
\]
of $g(\vep)$ disjoint intervals. A similar decomposition holds for the \emph{classically forbidden
  region}:
\[
  \big\{x\in[0,\ell]:\abs{\xi(x,\vep)}>1\big\}=\bigsqcup_{j=1}^{\tilde g(\vep)}\tilde I_j(\vep),
\]
with $\tilde I_j(\vep)$ an open interval.

\subsection{Classically allowed intervals ($\abs{\xi(x,\vep)}<1$)}

Let us suppose, to begin with, that $x\in I_j(\vep)$, so that $\abs{\xi(x,\vep)}<1$ and
$p(x,\vep)$ is real. Consider, first, the solution with $q=p(x,\vep)$ (i.e., we take the ``$+$''
sign in eq.~\eqref{S0p}). From the $O(a)$ terms in eq.~\eqref{contsysmat} we obtain\footnote{Here
  and in what follows,$\frac{\pd h}{\pd q}$ is evaluated at $q=p(x,\vep)$.}
\[
  (\vep-\bmu-h)u_1=\pdv{h}{q}\left[-\iu\, u_0'+\left(\frac12p'\sigma^z-\frac{\iu w'}{2w}+S_1'\right)
    u_0\right],
\]
where for convenience we have suppressed the $x$ dependence and used the identity
\[
  \pdv{h}{q}(x,q) =\iu w(x)\left(- \e^{-\iu q}\sigma^++\e^{\iu q}\sigma^-\right).
\]
Since the matrix $\vep-\bmu(x)-h(x,p(x))$ is singular and self-adjoint, by Fredholm's alternative
the r.h.s.\ of the above system must be orthogonal to $u_0$, i.e.
\begin{equation}
  \label{Fredh}
  u_0^\dagger\pdv{h}{q}\left[-\iu\,u_0'+\left(\frac12p'\sigma^z-\frac{\iu w'}{2w}+S_1'\right) u_0\right]=0.
\end{equation}
To further simplify this condition, let us write the spinor $u_0$ as follows:
\[
  u_0(x)=A(x)\eta(x),
\]
with $\eta(x)\coloneqq(\eta_A(x)\;\eta_B(x))^\T$ and $\eta(x)^\dagger\eta(x)=1$. Since $\eta(x)$ is
an eigenvector of $h(x,p(x))$ with eigenvalue $\vep-\bmu(x)$, solving the eigenvalue equation we
obtain
\[
  \eta_A\propto v+w\e^{-\iu p},\qquad \eta_B\propto \vep-\bmu-\dmu=\vep-\mu^A.
\]
Moreover, from the identity
\[
  \abs{v+w\e^{-\iu p}}^2=\left(\vep-\bar\mu\right)^2-\dmu^2=\left(\vep-\mu^A\right)\left(\vep-\mu^B\right)
\]
it follows that we can take
\[
  \eta_A=\frac{v+w\e^{-\iu
      p}}{\sqrt{2(\vep-\bmu)(\vep-\mu^A)}},\quad
  \eta_B=\sgn(\vep-\bmu)\sqrt{\frac{\vep-\mu^A}{2(\vep-\bmu)}}\,,
\]
or, more symmetrically and in an appropriate gauge,
\begin{equation}\label{etaAB}
  \eta_A=\e^{\iu\frac\phi2}\sqrt{\frac{\vep-\mu^B}{2(\vep-\bmu)}},\quad
  \eta_B=\sgn(\vep-\bmu)\,\e^{-\iu\frac\phi2}\sqrt{\frac{\vep-\mu^A}{2(\vep-\bmu)}}\,,
\end{equation}
where
\begin{equation}
  \phi(x,\vep)\coloneqq\arg(v(x)+w(x)\,\e^{-\iu p(x,\vep)})\,.
\end{equation}
(Note, in passing, that $\vep-\bmu$, $\vep-\mu^A$, and $\vep-\mu^B$ have the same sign.) We shall
also absorb the phase of $A(x)$ in $\Re S_1(x)$ and $\e^{-\Im S_1(x)}$ into $A(x)$, so that we can
take $A(x)>0$ and $S_1(x)$ real without loss of generality.
\begin{remark}
  Of course, eq.~\eqref{etaAB} is fully consistent with eq.~\eqref{uAB}. Indeed, defining
\begin{equation}\label{Exq}
  E(x,q)=\sqrt{\dmu(x)^2+v(x)^2+w(x)^2+2v(x)w(x)\cos q}\,,
\end{equation}
we have
\[
E(x,p(x,\vep))=\pm(\vep-\bmu(x))=\abs{\vep-\bmu(x)},
\]
with $\pm=\sgn(\vep-\bmu(x))$, and therefore
\begin{align*}
  E\mp\dmu&=\pm(\vep-\bmu-\dmu)=\pm(\vep-\mu^A)=\abs*{\vep-\mu^A},\\
  E\pm\dmu&=\pm(\vep-\bmu+\dmu)=\pm(\vep-\mu^B)=\abs*{\vep-\mu^B}.
            \myqed
\end{align*}
\end{remark}

With the previous definitions, the compatibility condition~\eqref{Fredh} becomes
\begin{equation}
  \label{Fredh2}
  \eta^\dagger\pdv{h}{q}\left[\eta'+\left(\frac{w'}{2w}
      +\frac{A'}{A}+\frac{\iu p'}2\sigma^z+\iu S_1'\right)
    \eta\right]=0.
\end{equation}
Let us define the Fermi velocity $v_F(x)$ by
\begin{align}
  v_F(x,\vep)&\coloneqq\frac{\pd \vep}{\pd q}(x,p(x,\vep))=\pm\frac{\pd E}{\pd q}(x,p(x,\vep))
  =\frac{\mp v(x)w(x)\sin p(x,\vep)}{E(x,p(x,\vep))}
  =\frac{v(x)w(x)\sqrt{1-\xi(x,\vep)^2}}{\bmu(x)-\vep}.
          \label{vF}
\end{align}
Since, by the Feynman--Hellmann theorem,
\begin{equation}\label{vFintro}
  \eta^\dagger\frac{\pd h}{\pd q}\eta
  =\frac{\pd\vep}{\pd q}=v_F,
\end{equation}
we can rewrite eq.~\eqref{Fredh2} as
\begin{equation}
  \label{Fredh3}
  \eta^\dagger\pdv{h}{q}\eta'+\frac{\iu p'}2\eta^\dagger\left(\pdv{h}{q}\sigma^z\right)\eta
  +\left(\frac{w'}{2w}+\frac{A'}{A}
      +\iu S_1'\right)v_F=0.
\end{equation}
Taking the total derivative of eq.~\eqref{vFintro} with respect to $x$ we obtain
\[
  v_F'\coloneqq\frac{\dd v_F}{\dd x}=2\Re\left(\eta^\dagger\pdv{h}{q}\eta'\right)+\iu
  p'\eta^\dagger\left(\pdv{h}{q}\sigma^z\right)\eta+\frac{w'}{w}v_F,
\]
which substituted into eq.~\eqref{Fredh3} yields
\begin{subequations}
  \begin{equation}\label{FredhR}
    \frac12v_F'+\frac{A'}{A}v_F +\iu\left(v_FS_1'+R\right)=0,
  \end{equation}\label{R}
  with
  \begin{equation}
    R=\Im\left(\eta^\dagger\pdv{h}{q}\eta'\right).
  \end{equation}
\end{subequations}
Taking the real part of eq.~\eqref{FredhR} and integrating we obtain
\begin{equation}
  \label{Ax}
  A(x)=\frac1{\sqrt{\abs{v_F(x)}}}
  =\sqrt{\frac{\abs{\vep-\bmu(x)}}{v(x)w(x)\abs{1-\xi(x)^2}^{1/2}}}\,,
\end{equation}
where we have used eq.~\eqref{cospx} and suppressed the $\vep$ dependence on $A$, $v_F$ and $\xi$
for convenience. The $O(1)$ correction to the phase $S_1(x)$ can be obtained from the imaginary
part of eq.~\eqref{FredhR}, namely
\[
  S_1=-\int\dd x\,\frac{R}{v_F}\,.
\]
Putting everything together, we obtain the following expression for the WKB wave function
corresponding to the solution (branch) $q=p(x,\vep)$ in eq.~\eqref{qpmp} on the interval
$I_j(\vep)$:
\[
  \Psi_j^{(+)}(x,\vep)=\frac{\eta(x,\vep)}%
  {\sqrt{\abs{v_F(x,\vep)}}}\,\e^{\iu\wh\th_j(x,\vep)},
\]
with
\[
  \wh\th_j(x,\vep)=
  \int^x_{x_j^L(\vep)}\frac{\dd s}a\arccos\xi(s,\vep)+S_1(x).
\]
It can be readily ascertained that the solution $\Psi_j^{(-)}$ corresponding to the branch
$q=-p(x,\vep)$ is given by the complex conjugate of $\Psi_j^{(+)}$, i.e.,
\[
\Psi_j^{(-)}(x,\vep)=\frac{\eta^*(x,\vep)}%
{\sqrt{\abs{v_F(x,\vep)}}}\,\e^{-\iu\wh\th_j(x,\vep)}
\]
The general solution of eq.~\eqref{contsys} on the interval $I_j(\vep)$ is a linear combination of
the solutions $\Psi_j^{(\pm)}$ with complex coefficients. Since $\Psi(x,\vep)$ can be chosen to be
real, we conclude that the restriction of the WKB wave function to the interval
$I_j(\vep)\subset\Om_\vep$ is proportional to\footnote{Obviously, the choice of the real or
  imaginary part in the following equation is merely a matter of convenience, due to the arbitrary
  constant phase $\de_j(\vep)$.}
\begin{equation}\label{PsijWKB}
  \Psi_j(x,\vep)=\frac{\Im\left(\eta(x,\vep)\,\e^{\iu\th_j(x,\vep)}\right)}%
  {\sqrt{\abs{v_F(x,\vep)}}}\,,
\end{equation}
where
\begin{equation}\label{Th}
  \th_j(x,\vep)=\int^x_{x_j^L(\vep)}\frac{\dd s}a\arccos\xi(s,\vep)+S_1(x)+\de_j(\vep),
\end{equation}
with $\de_j(\vep)\in\RR$ a constant local phase\footnote{The local phases $\de_j(\vep)$ can be
  determined imposing appropriate matching conditions between the different branches of the wave
  function. However, as will be apparent shortly, this will not be needed for our purposes.}, and
the normalized spinor~$\eta(x,\vep)$ is defined in eq.~\eqref{etaAB}. By eqs.~\eqref{etaAB},
\eqref{Ax}, and \eqref{PsijWKB}, the components of $\Psi_j$ in the
interval $I_j(\vep)$ are given by
\begin{subequations}\label{etaAetaB}
  \begin{align}
    \al_j(x,\vep)
    &=\sqrt{\frac{\vep-\mu^B(x)}{2\big(\vep-\bmu(x)\big)\abs{v_F(x,\vep)}}}\,
      \sin\left(\th_j(x,\vep)+\frac{\phi(x,\vep)}2\right),
      \label{etaAf}\\
    \be_j(x,\vep)
    &=\sgn\bigl(\vep-\bmu(x)\bigr)
      \sqrt{\frac{\vep-\mu^A(x)}{2\big(\vep-\bmu(x)\big)\abs{v_F(x,\vep)}}}\,
      \sin\left(\th_j(x,\vep)-\frac{\phi(x,\vep)}2\right).
      \label{etaBf}
  \end{align}
\end{subequations}
Note that
\begin{equation}\label{vphi}
  \vp_j(x,\vep)=\int^x_{x_j^L(\vep)}\frac{\dd s}a\arccos\xi(s,\vep)+\de_j(\vep)
\end{equation}
satisfies
\[
  \vp_j'(x,\vep)=\frac1a\arccos\xi(x,\vep)=\frac{p(x,\vep)}a=\pm\frac{S_0'(x,\vep)}a.
\]
Thus $\pm\vp_j(x,\vep)$ differs from the eikonal $S_0(x,\vep)/a$ by a constant.

\subsection{Classically forbidden intervals ($\abs{\xi(x,\vep)}>1$)}

Consider next the case in which the point $x$ lies on one of the disjoint intervals
$\tilde I_j(\vep)$ inside the classically forbidden region $\abs{\xi(x,\vep)}>1$. We now have
\[
  p(x)=\iu\arccosh\abs{\xi(x)}+\pi\,\Th\bigl(-\xi(x)-1\bigr),\qquad x\in\tilde I_j(\vep)
  =\big(\tilde x_j^L(\vep),\tilde x_j^R(\vep)\big),
\]
where $\Th$ is Heaviside's theta function. In this case
\[
  \pm\frac{\iu\,S_0(x)}a=-\int^x\frac{\dd x}a\,\arccosh{\abs{\xi(x)}}+\frac{\iu\pi
    x}a\,\Th\bigl(-\xi(x)-1\bigr),
\]
where $\Th\bigl(-\xi(x)-1\bigr)$ is constant over the whole interval $\tilde I_j(\vep)$ and on
chain sites
\[
  \e^{\pm\frac{\iu\pi x}a}=\e^{\pm\iu n\pi}=(-1)^n
\]
is a staggered sign. Hence to lowest exponential order in $a$ the WKB wave function is of the form
\[
  \Psi(x)\simeq\left(c_j^L(\vep)\e^{-\ka_L(x,\vep)}\eta_+(x,\vep)+c_j^R(\vep)\e^{-\ka_R(x,\vep)}
  \eta_-(x,\vep)\right)
  \e^{\frac{\iu\pi
      x}a\Th(-\xi(x,\vep)-1)},\qquad x\in\tilde I_j(\vep),
\]
where the spinors $\eta_{\pm}(x,\vep)$ are (appropriately normalized) linearly independent
eigenvectors of $h(x,\pm p(x,\vep))$ with eigenvalue $\vep-\bmu(x)$, and
\[
  \ka_L(x,\vep)=\int_{\tilde x_j^L(\vep)}^x\frac{\dd s}a\,\arccosh\abs{\xi(s,\vep)},
  \qquad
  \ka_R(x,\vep)=\int_x^{\tilde x_j^R(\vep)}\frac{\dd s}a\,\arccosh\abs{\xi(s,\vep)}
\]
are both positive. The coefficients $c_j^{L,R}$ are typically $O(1)$, so that the WKB wave
function is exponentially decaying in the classically forbidden region
$\{x\in[0,\ell]:\abs{\xi(x,\vep)}>1\}$. Thus this function is supported in the classically allowed
region on which $\abs{\xi}\le1$ and $p(x,\vep)$ is real, up to exponentially small terms.
Neglecting these terms, we can write
\begin{equation}\label{PsiWKB}
  \Psi(x,\vep)=\sum_{j=1}^{g(\vep)}C_j(\vep)\Psi_j(x,\vep),
\end{equation}
for suitable complex constants $C_j(\vep)$. As we shall show below, however, it is \emph{not}
necessary to determine these constants nor the phases $\de_j(\vep)$ to compute the model's bulk
particle-density profiles in the thermodynamic limit.

By eq.~\eqref{PsiWKB}, for each $x\in[0,\ell]$ the range of single-particle energies $\vep$ for
which $x\in\supp\Psi$ (up to exponentially small tails) is the union of the disjoint intervals
(local metallic bands) $\cB_-(x)$ and $\cB_+(x)$ given by eq.~\eqref{cBpm}. Hence the global
metallic bands $\cB_\pm$ should be approximately given by
\[
  \cB_-=\Bigl[\vep^-_{\mathrm{min}},\vep^-_{\mathrm{max}}\Bigr],\qquad
  \cB_+=\Bigl[\vep^+_{\mathrm{min}},\vep^+_{\mathrm{max}}\Bigr],
\]
where
\begin{alignat*}{2}
  \vep^-_{\mathrm{min}}
  &=\min\limits_{0\le x\le\ell} \big(\bmu(x)-\Emax(x)\big),\qquad
    \vep^-_{\mathrm{max}}
  &=\max\limits_{0\le x\le\ell}\big(\bmu(x)-\Emin(x)\big),\\
  \vep^+_{\mathrm{min}}
  &=\min\limits_{0\le x\le\ell} \big(\bmu(x)+\Emin(x)\big),\qquad
    \vep^+_{\mathrm{max}}&=\max\limits_{0\le x\le\ell} \big(\bmu(x)+\Emax(x)\big)
\end{alignat*}
and $E_{\text{min},\text{max}}(x)$ are defined in eq.~\eqref{Eminmax}. Note that, although
$\cB_-(x)\cap\cB_+(x)=\emptyset$ for all\footnote{This holds true except at points $x$ where
  $\dmu(x)=0$ and $v(x)=w(x)$ simultaneously, so that $E_{\min}(x)=0$.} $x\in[0,\ell]$, the global
bands $\cB_-$ and $\cB_+$ may overlap. If this is the case, at fillings for which the Fermi energy
lies in the overlap, the chain is not a global band insulator and is metallic in the spectral
sense, even though it may contain locally insulating regions.

\section{WKB approximation of the particle-density profiles}\label{sec.WKBdens}

With the previous formulas for the components of the single-particle semiclassical wave function
$\Psi(x)$, it is straightforward to obtain an approximation for the particle densities. To this
end, we need first to compute the density of states. 

\subsection{Density of states}

\begin{figure}[pos=t]
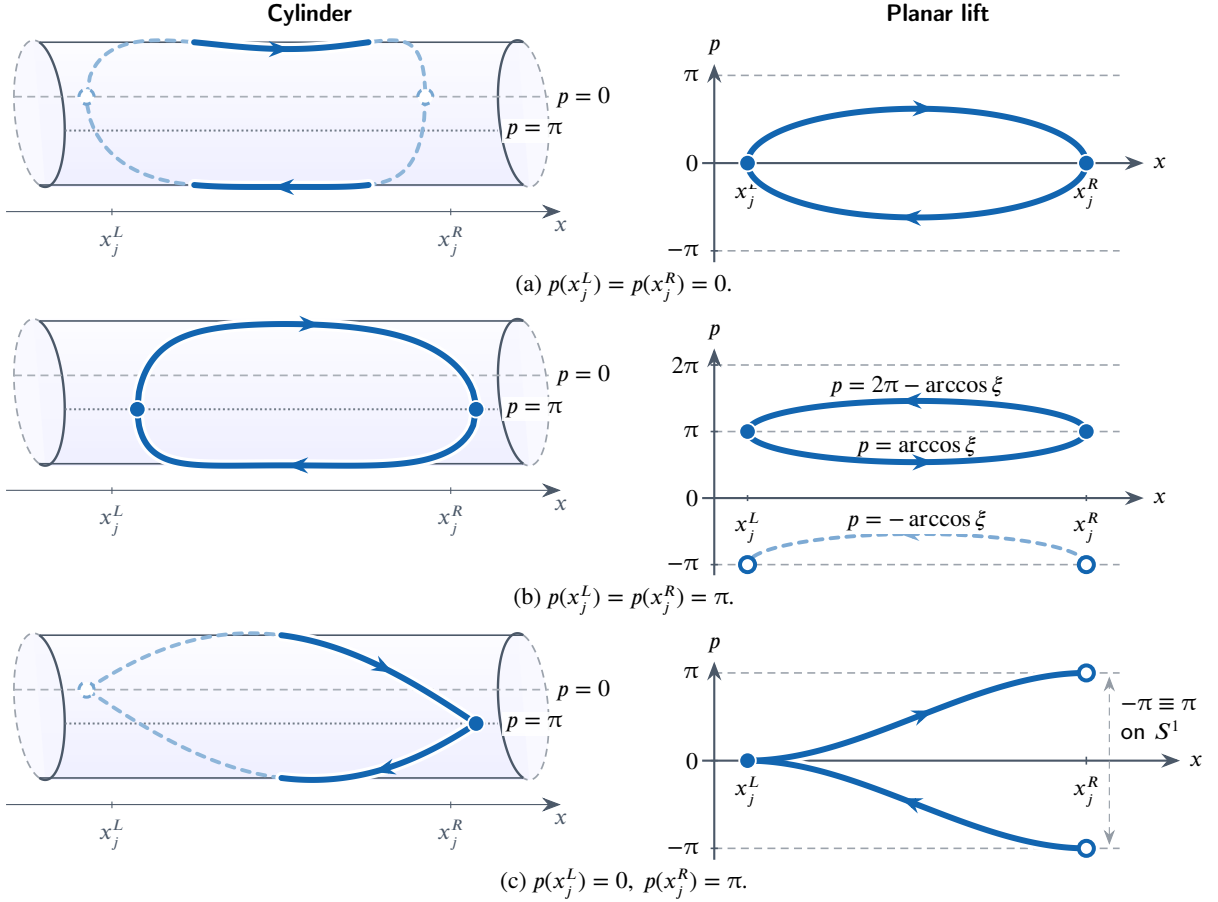

  \centering
  \captionsetup[subfigure]{font=small,skip=2pt}
  \newcommand{\WKBpanelscale}{0.8}

  \begin{subfigure}[t]{0.98\textwidth}
    \centering
    \begin{minipage}[t]{0.485\linewidth}
      \vspace{0pt}\centering
      \textbf{Cylinder}\par\smallskip
      \makebox[\linewidth][c]{\WKBCylinderCaseI[\WKBpanelscale]}
    \end{minipage}
    \hfill
    \begin{minipage}[t]{0.485\linewidth}
      \vspace{0pt}\centering
      \textbf{Planar lift}\par\smallskip
      \makebox[\linewidth][c]{\WKBPlaneCaseI[\WKBpanelscale]}
    \end{minipage}
    \caption{$p(x_j^L)=p(x_j^R)=0$.}
    \label{fig:wkb-inline-I}
  \end{subfigure}

  \vspace{0.45em}

  \begin{subfigure}[t]{0.98\textwidth}
    \centering
    \begin{minipage}[t]{0.485\linewidth}
      \vspace{0pt}\centering
      \makebox[\linewidth][c]{\WKBCylinderCaseII[\WKBpanelscale]}
    \end{minipage}
    \hfill
    \begin{minipage}[t]{0.485\linewidth}
      \vspace{0pt}\centering
      \makebox[\linewidth][c]{\WKBPlaneCaseII[\WKBpanelscale]}
    \end{minipage}
    \caption{$p(x_j^L)=p(x_j^R)=\pi$.}
    \label{fig:wkb-inline-II}
  \end{subfigure}

  \vspace{0.45em}

  \begin{subfigure}[t]{0.98\textwidth}
    \centering
    \begin{minipage}[t]{0.485\linewidth}
      \vspace{0pt}\centering
      \makebox[\linewidth][c]{\WKBCylinderCaseIII[\WKBpanelscale]}
    \end{minipage}
    \hfill
    \begin{minipage}[t]{0.485\linewidth}
      \vspace{0pt}\centering
      \makebox[\linewidth][c]{\WKBPlaneCaseIII[\WKBpanelscale]}
    \end{minipage}
    \caption{$p(x_j^L)=0,\ p(x_j^R)=\pi$.}
    \label{fig:wkb-inline-III}
  \end{subfigure}
  \caption{three types of closed classical trajectories $\Ga_j$ on $\mathbb R\times S^1$ (left
    column) and their planar lifts (right column) for a classically allowed interval
    $I_j\subset(0,\ell)$ (recall that at turning points $x_j^{L,R}(\vep)\ne0,\ell$ the
    quasimomentum $p$ is equal to either $0$ or $\pi$). Cases (a) and (b) have winding number
    zero; case (c) has winding number one. In case (b), the lower branch $p=-\arccos\xi$ can be
    replaced by the equivalent lift $p=2\pi-\arccos\xi$ to obtain a closed curve in $\RR^2$. By
    contrast, in case (c) there is no lift of $\Ga_j(\vep)$ to $\RR^2$ whose image is a closed
    curve.}
  \label{fig.wkb-inline-comparison}
\end{figure}

Assuming that, as is generically the case, the evanescent tails connecting different ``potential
wells'' $I_j(\vep)$ are exponentially small, we can apply semiclassical (Bohr--Sommerfeld)
quantization conditions to each well independently\footnote{More precisely, the following
  derivation applies to regular energies, for which the allowed intervals have simple turning
  endpoints or reflecting physical boundaries, the reciprocal-velocity integrals are finite, and
  distinct wells are separated by forbidden regions with nonzero tunneling action. This excludes
  band crossings, coalescing turning points, and localized edge or defect states, which require
  separate treatment.}. Taking into account that the lattice spacing $a$ plays the role of
$\hbar$, these conditions would normally be formulated as follows:
\begin{equation}\label{BS0}
  \frac1a\oint_{\Ga_j(\vep)}p\,\dd x+\ga_j(\vep)=2\pi n_j(\vep),\qquad \text{with}\quad
  n_j(\vep)\in\ZZ,\en j=1,\dots,g(\vep),
\end{equation}
where $\Ga_j(\vep)$ is the closed curve in $(x,p)$ phase space with equation
\[
  \cos p=\xi(x,\vep),
\]
and the phases $\ga_j(\vep)$ contain the turning-point, boundary, transport, and subprincipal
contributions that can only be determined using the appropriate matching conditions between the
different branches of the wave function. However, in contrast with the standard (continuous) WKB
theory, in the present context the phase space is \emph{not} $\RR^2$ but the cylinder
$\RR\times S^1$, as the momentum variable $p$ is now defined modulo $2\pi$. This makes the Cartan
one-form $p\,\dd x$ multivalued, i.e., defined up to integer multiples of $2\pi\,\dd x$. It may
superficially appear that this ambiguity does not affect eq.~\eqref{BS0}, since the integral of
$\dd x$ along a closed curve vanishes. This is, however, not the case in general, as the lift of
$\Ga_j(\vep)$ to $\RR^2$ is not always closed due to the multivaluedness of the quasimomentum $p$
(cf.~Fig.~\ref{fig.wkb-inline-comparison}). Thus due care must be taken in formulating the
Bohr--Sommerfeld quantization condition unambiguously.

To this end, we note that in $\RR^2$ the semiclassical quantization
formula~\eqref{BS0} can be equivalently written as
\begin{equation}\label{BSxdp}
  -\frac1a\oint_{\Ga_j(\vep)}x\,\dd p+\ga_j(\vep)=2\pi n_j(\vep),
\end{equation}
since the one-forms $p\dd x$ and $-x\dd p$ differ by the total differential $\dd(xp)$. The
advantage of this formulation is that, unlike eq.~\eqref{BS0}, it is also globally defined on the
cylinder, since
\[
  \dd\bigl(p+2\pi n\bigr)=\dd p,\qquad\all n\in\ZZ.
\]
Disregarding, to lowest order in $a$ (i.e., to $O(a^{-1})$), the phases $\ga_j$, we arrive at the
following formula for the counting function\footnote{As a matter of fact, $n_j(\vep)$ might differ
  from the counting function (level number) by a sign and a constant offset, neither of which
  affect the validity of eq.~\eqref{Djvep}.} $n_j(\vep)$ for each well $I_j(\vep)$:
\begin{equation}\label{njWKB}
  n_j(\vep)=-\frac1{2\pi a}\oint_{\Ga_j(\vep)}x\,\dd p.
\end{equation}
The contribution to the density of states of the $j$-th well is thus
\begin{equation}\label{Djvep}
  D_j(\vep)=\left|\frac{\dd n_j(\vep)}{\dd\vep}\right|=\frac1{2\pi a}
  \left|\frac{\dd}{\dd\vep}\oint_{\Ga_j(\vep)}x\,\dd p\right|.
\end{equation}
It is shown in Appendix~\ref{app.oint} that
\begin{equation}
  -\frac{\dd}{\dd\vep}\oint_{\Ga_j(\vep)}x\,\dd p =2\int_{I_j(\vep)}\dd
  x\,\frac{\pd p(x,\vep)}{\pd\vep}.
\label{ointpdx}
\end{equation}
Differentiating the identity
\[
  \cos p(x,\vep)=\xi(x,\vep)
\]
with respect to $\vep$ we obtain
\begin{equation}\label{dpdvep}
  -\sin p(x,\vep)\frac{\pd p(x,\vep)}{\pd\vep}=\frac{\pd \xi(x,\vep)}{\pd\vep}
  =\frac{\vep-\bmu(x)}{v(x)w(x)}\implies \frac{\pd p(x,\vep)}{\pd\vep}
  =\frac{\bmu(x)-\vep}{v(x)w(x)\sqrt{1-\xi(x,\vep)^2}}=\frac1{v_F(x)}
\end{equation}
(cf.~eq.~\eqref{vF}), so that
\begin{equation*}
  D_j(\vep)=
  \frac1{\pi a}\left|\int_{I_j(\vep)}\frac{\dd x}{v_F(x)}\right|.
\end{equation*}
Taking into account that $\vep-\bmu(x)$, and hence $v_F(x)$, does not change sign over the
interval $I_j(\vep)$ (indeed, $(\vep-\bmu(x))^2=\Emin^2(x)+2v(x)w(x)(1+\xi(x,\vep))>\Emin^2(x)\ge0$
on $\Om_\vep$) we can write the previous equation as
\begin{equation}
  \label{Djfinal}
  D_j(\vep)=\frac1{\pi a}\int_{I_j(\vep)}\frac{\dd x}{\abs{v_F(x,\vep)}}.
\end{equation}
By eq.~\eqref{etaAetaB}, with the rapidly oscillating factors
$\sin^2\left(\th_j\pm\frac\phi2\right)$ replaced by their local average $1/2$, it then follows
that\footnote{To be precise, the local average is taken over a spatial interval containing many
  oscillations but short compared with the scales of variation of the envelopes and $O(a^0)$ phase
  offsets.}
\begin{equation}
  \label{Djnorm}
  D_j(\vep)=\frac{2}{\pi a}\,\cN_j(\vep),
\end{equation}
with
\[
  \cN_j(\vep)=\norm\big{\Psi_j(\cdot,\vep)}^2=\int_0^\ell\dd x\,\abs*\big{\Psi_j(x,\vep)}^2.
\]
Although not strictly necessary for our purposes, the total density of states $D(\vep)$ is easily
obtained by summing the independent contributions $D_j(\vep)$ of each well $I_j(\vep)$, namely
\begin{equation}\label{Dvep}
  D(\vep)=\sum_{j=1}^{g(\vep)}D_j(\vep)=\frac1{\pi a}\sum_{j=1}^{g(\vep)}
  \int_{I_j(\vep)}\frac{\dd x}{\abs{v_F(x,\vep)}}
  =\frac1{\pi a}\int_0^\ell\dd x\,\frac{\Th\bigl(1-\xi(x,\vep)^2\bigr)}{\abs{v_F(x,\vep)}}.
\end{equation}

\begin{remark}
  Equation~\eqref{Dvep} can be alternatively justified starting from the dispersion relation of
  the bulk Hamiltonian of the RM chain (cf.~Appendix~\ref{app.bulk}), namely
  \[
    \vep_s(k)=\bmu+ sE(k),\qquad k=\frac{2\pi m}L,\quad m=0,\dots,L-1,\quad s=\pm\,,
  \]
  where $E(k)$ is given by eq.~\eqref{Ekbulk} and we have restored the sublattice-independent
  on-site potential $\bmu$. Since
  \[
    \De p=\frac{2\pi}L\en\implies\en \frac1{\ell\De p}=\frac{L}{2\pi\ell}=\frac1{2\pi a},
  \]
  in the thermodynamic limit the number of states with energy not exceeding $\vep$ is given by
  \[
    n(\vep)=\frac1{2\pi a}\int_0^\ell\dd x\sum_{s=\pm}\int_0^{2\pi}\dd
    k\,\Th\bigl(\vep-\vep_s(k)\bigr).
  \]
  If the chain's parameters are slowly varying, the local density approximation to the previous
  formula reads
  \[
    n(\vep)=\frac1{2\pi a}\int_0^\ell\dd x\sum_{s=\pm}\int_0^{2\pi}\dd
    k\,\Th\bigl(\vep-\bmu(x)-s E(x,k)\bigr),
  \]
  where $E(x,k)$ is defined in eq.~\eqref{Exq}. Differentiating this formula with respect to
  $\vep $we obtain the following expression for the total density of states:
  \[
    D(\vep)=\frac1{2\pi a}\int_0^\ell\dd x\sum_{s=\pm}\int_0^{2\pi}\dd
    k\,\de\bigl(\vep-\bmu(x)-s E(x,k)\bigr).
  \]
  We now perform the integration in $k$ taking into account that, by eq.~\eqref{cospx},
  \[
    \vep-\bmu(x)=s E(x,k)\en\iff\en \cos k=\xi(x,\vep),\en s=\sgn(\vep-\bmu(x))\,.
  \]
  For given $x\in[0,\ell]$ and $\vep\in\RR$, the equation $\cos k=\xi(x,\vep)$ has a solution
  $k\in[0,2\pi)$ if and only if $\abs{\xi(x,\vep)}\le1$. Thus we can restrict the $x$ integral to
  the set $\Om_\vep$, or equivalently write
  \[
    D(\vep)=\frac1{2\pi a}\int_0^\ell\dd
    x\,\Th\bigl(1-\xi(x,\vep)^2\bigr)\sum_{s=\pm}\int_0^{2\pi}\dd k\,\de\bigl(\vep-\bmu(x)-s
    E(x,k)\bigr).
  \]
  Secondly, for each $x\in\Om_\vep$ the equation $\cos k=\xi(x,\vep)$ has two solutions
  $k(x,\vep)$ and $2\pi-k(x,\vep)$. Using the well-known identity
  \[
    \de\bigl(f(k)\bigr)=\sum_i\frac{\de(k-k_i)}{\abs{f'(k_i)}},
  \]
  where the sum is extended to the roots of the equation $f(k)=0$, we readily obtain
  \begin{equation}\label{Dvep2}
    D(\vep)=\frac1{\pi a}\int_0^\ell\dd x\,\frac{\Th\bigl(1-\xi(x,\vep)^2\bigr)%
    }{\abs{\pd_kE(x,k)}}\bigg|_{k=\arccos\xi(x,\vep)}\,.
  \end{equation}
  Note that the factor of $2$ is due to the equal contributions of the two roots of the equation
  $\cos k=\xi(x,\vep)$ in $[0,2\pi)$, namely $k=\arccos\xi(x,\vep)$ and
  $k=2\pi-\arccos\xi(x,\vep)$. Equation~\eqref{Dvep2} is easily seen to coincide with
  eq.~\eqref{Dvep} on account of eq.~\eqref{vF}.\qed
\end{remark}

\subsection{Particle-density profiles}

Let us denote by $\wh\al$ and $\wh\be$ the components of the normalized WKB wave function
$\wh\Psi_j=\Psi_j/\norm*{\Psi_j}$, namely
\[
  \wh\al_j(x,\vep)=\frac{\al_j(x,\vep)}{\sqrt{\cN_j(\vep)}},\qquad
  \wh\be_j(x,\vep)=\frac{\be_j(x,\vep)}{\sqrt{\cN_j(\vep)}}.
\]
In the (semiclassical) independent-well approximation, the spectral densities of both sublattices
are given by
\begin{equation}\label{cDAB}
  \cD_A(x,\vep)=\sum_{j=1}^{g(\vep)}D_j(\vep)\abs{\wh\al_j(x,\vep)}^2,\qquad
  \cD_B(x,\vep)=\sum_{j=1}^{g(\vep)}D_j(\vep)\abs{\wh\be_j(x,\vep)}^2,
\end{equation}
where as usual we have neglected the exponentially small contributions coming from the classically
forbidden region. In the thermodynamic limit, the particle densities of sublattices A and B at a
Fermi energy $\vep_F$ are therefore respectively given by
\begin{align*}
  \rho_A(x,\vep_F)&=\int_{\vep_0}^{\vep_F}\dd\vep\,\cD_A(x,\vep)
  =\int_{\vep_0}^{\vep_F}\dd\vep\,\sum_{j=1}^{g(\vep)}
  D_j(\vep)\abs{\wh\al_j(x,\vep)}^2,\\
  \rho_B(x,\vep_F)&=\int_{\vep_0}^{\vep_F}\dd\vep\,\cD_B(x,\vep)
  =\int_{\vep_0}^{\vep_F}\dd\vep\,\sum_{j=1}^{g(\vep)}
  D_j(\vep)\abs{\wh\be_j(x,\vep)}^2,
\end{align*}
Here $\vep_0$ is the minimum single-particle energy,
which is given in this approximation by $\vep^-_{\mathrm{min}}$. Taking into account
eq.~\eqref{Djnorm}, the previous formulas simplify to
\begin{equation}\label{rhoAB}
  \rho_A(x,\vep_F)=\frac{2}{\pi a}\int_{\vep_0}^{\vep_F}\dd\vep\,
  \sum_{j=1}^{g(\vep)}\abs{\al_j(x,\vep)}^2,\qquad
  \rho_B(x,\vep_F)=\frac{2}{\pi a}\int_{\vep_0}^{\vep_F}\dd\vep\,
  \sum_{j=1}^{g(\vep)}\abs{\be_j(x,\vep)}^2.
\end{equation}
Introducing the \emph{effective} WKB wave function
\begin{equation}\label{Psieff}
  \Psi_{\text{eff}}(x,\vep)
  =\sum_{j=1}^{g(\vep)}\Psi_j(x,\vep)\eqqcolon\bigl(\al_{\eff}(x,\vep),\be_{\eff}(x,\vep)\bigr),
\end{equation}
we can rewrite the previous formulas more concisely as
\begin{equation}\label{rhoABeff}
  \rho_A(x,\vep_F)=\frac{2}{\pi a}\int_{\vep_0}^{\vep_F}\dd\vep\abs{\al_{\eff}(x,\vep)}^2,\qquad
  \rho_B(x,\vep_F)=\frac{2}{\pi a}\int_{\vep_0}^{\vep_F}\dd\vep\abs{\be_{\eff}(x,\vep)}^2,
\end{equation}
The average density and the sublattice imbalance are then given by
\begin{subequations}\label{rhosdef}
  \begin{align}\label{rhob}
    \bar\rho(x,\vep_F)
    &\coloneqq\frac12\left(\rho_A(x,\vep_F)+\rho_B(x,\vep_F)\right)=\frac1{\pi
      a}\int_{\bmu(x)-\Emax(x)}^{\vep_F}\dd\vep\,\abs{\Psi_{\eff}(x,\vep)}^2,
      \\
    \de\rho(x,\vep_F)
    &\coloneqq\frac12\left(\rho_B(x,\vep_F)-\rho_A(x,\vep_F)\right)
      =\frac1{\pi a}\int_{\bmu(x)-\Emax(x)}^{\vep_F}\dd\vep\,
      \left(\abs{\be_{\eff}(x,\vep)}^2-\abs{\al_\eff(x,\vep)}^2\right)\,.
      \label{derho}
  \end{align}
\end{subequations}
Note that in the previous formulas we have taken into account that
$\Psi_{\eff}(x,\vep)=0$ for $\vep<\bmu(x)-\Emax(x)$ to sharpen the lower integration limit.

\begin{remark}
  Let
  \[
    \cN(\vep)=\norm{\Psi_\eff(\cdot,\vep)}^2=\sum_{j=1}^{g(\vep)}\cN_j(\vep),
  \]
  where the last equality follows from eq.~\eqref{Psieff} taking into account that
  $\supp\Psi_j\cap\supp\Psi_k=\emptyset$ for $j\ne k$. Note that, by eqs.~\eqref{Djnorm}
  and~\eqref{Dvep}, we have
  \[
    \frac{D_j(\vep)}{\cN_j(\vep)}=\frac{D(\vep)}{\cN(\vep)}.
  \]
  Consider next the normalized effective WKB eigenfunction $\wh\Psi_\eff$, given by
  \[
    \wh\Psi_\eff(x,\vep)=\frac{\Psi_\eff(x,\vep)}{\sqrt{\cN(\vep)}}
    \eqqcolon\big(\wh\al_\eff(x,\vep),\wh\be_\eff(x,\vep)\big)\,.
  \]
  If $x\in I_k(\vep)$ we then have
  \[
    D(\vep)\abs{\wh\al_\eff(x,\vep)}^2
    =\frac{D(\vep)}{\cN(\vep)}\abs{\al_\eff(x,\vep)}^2
    =\frac{D(\vep)}{\cN(\vep)}\abs{\al_k(x,\vep)}^2
    =\frac{D_k(\vep)}{\cN_k(\vep)}\abs{\al_k(x,\vep)}^2
    =\sum_{j=1}^{g(\vep)}\frac{D_j(\vep)}{\cN_j(\vep)}\abs{\al_j(x,\vep)}^2,
   \]
   since $\al_j(x,\vep)=0$ for $j\ne k$, and similarly for the B sublattice. In other words,
   although $\wh\Psi_\eff$ is \emph{not} in general an eigenstate of $H$, it yields the correct
   (semiclassical) spectral densities through the formulas
  \[
    \cD_A(x,\vep)=D(\vep)\abs{\wh\al_\eff(x,\vep)}^2,\qquad
    \cD_B(x,\vep)=D(\vep)\abs{\wh\be_\eff(x,\vep)}^2.
  \]
  Put differently, the coefficients in the linear combination~\eqref{Psieff} of the
  individual WKB wave functions $\Psi_j$ are chosen so that the weight of $\Psi_\eff$ in the
  $j$-th well,
  \[
    \int_{I_j(\vep)}\dd x\,\abs{\Psi_\eff(x,\vep)}^2=\cN_j(\vep),
  \]
  is proportional---to leading semiclassical order---to the number
  \[
    D_j(\vep)\,\dd\vep=\frac{2}{\pi a}\cN_j(\vep)\,\dd\vep
  \]
  of exact eigenfunctions localized in that well with energies in a narrow range $\dd\vep$ around
  $\vep$.\qed
  
\end{remark}

It is also worth noting that, although eqs.~\eqref{cDAB} have been deduced by treating the
contributions from each well independently, they still hold in symmetric situations in which the
exact wave functions may be localized in several wells. Indeed, consider for the sake of
simplicity the case in which $g(\vep)=2$, and let $\wh\Psi_j(x,\vep)$ be the normalized WKB
quasi-mode localized in the $j$-th well (with $j=1,2$). For sufficiently deep and/or sufficiently
far away wells, the overlap and coupling between these states is exponentially small:
\[
  \ip*{\wh\Psi_1}{\wh\Psi_2}=O(\e^{-K/a}),\qquad
  t\coloneqq\mel*{\wh\Psi_1}{H}{\wh\Psi_2}=O(\e^{-K/a}),
\]
where $K=O(1)$ is a positive constant\footnote{Typically,
  \[
    K=\int_{\tilde I_{12}(\vep)}\dd x\,\arccosh\abs{\xi(x,\vep)},
  \]
  where $\tilde I_{12}(\vep)$ is the interval separating the two wells.}. If, by reasons
of symmetry, $\vep$ satisfies the Bohr--Sommerfeld quantization condition for both $\Psi_1$ and
$\Psi_2$, there are two exact eigenstates
\[
  \Psi_\pm\simeq\frac1{\sqrt 2}\left(\wh\Psi_1\pm\wh\Psi_2\right)\eqqcolon(\al_\pm,\be_\pm)
\]
with exponentially close energies\footnote{In this case the off-diagonal tunneling matrix $t$ is
  real, by symmetry.} $\vep\pm t$. Again by symmetry, the contribution to the density of states of
both wells is the same:
\[
  D_1(\vep)=D_2(\vep)\eqqcolon D_0(\vep).
\]
Furthermore, denoting by $D_\pm(\vep)$ the contribution to the density of states of the
symmetric/antisymmetric linear combinations $\Psi_\pm$, we have
\[
  D_+(\vep)=D_-(\vep)=D_0(\vep)+O(\e^{-K/a})\,.
\]
If both members of the doublet $\Psi_\pm$ are occupied, their contribution to the spectral density
of (say) the A sublattice is therefore
\begin{align*}
  D_+(\vep)&\abs{\al_+(x,\vep)}^2+D_-(\vep)\abs{\al_-(x,\vep)}^2\\
  &\simeq\frac{D_0(\vep)}2\left(\abs{\wh\al_1(x,\vep)}^2
    +\abs{\wh\al_2(x,\vep)}^2+2\Re\big(\wh\al_1(x,\vep)^*\wh\al_2(x,\vep)\big)\right)\\
  &\qquad{}+\frac{D_0(\vep)}2\left(\abs{\wh\al_1(x,\vep)}^2
    +\abs{\wh\al_2(x,\vep)}^2-2\Re\big(\wh\al_1(x,\vep)^*\wh\al_2(x,\vep)\big)\right)\\
           &=D_0(\vep)\left(\abs{\wh\al_1(x,\vep)}^2+\abs{\wh\al_2(x,\vep)}^2\right)
             =D_1(\vep)\abs{\wh\al_1(x,\vep)}^2+D_2(\vep)\abs{\wh\al_2(x,\vep)}^2=\cD_A(x,\vep),
\end{align*}
exactly the same as the contribution of the two WKB quasi-modes $\wh\Psi_{1,2}(x,\vep)$. More
generally, suppose that by symmetry reasons there are $r$ WKB localized quasi-modes
$\Psi_{j_1}(x,\vep),\dots,\Psi_{j_r}(x,\vep)$, corresponding to an $r$-plet of exact
eigenfunctions with energies differing from $\vep$ by exponentially small terms. After an
exponentially small orthonormalization of the normalized quasi-modes, if necessary, these
eigenfunctions can be written as
\[
  \Psi^{(\la)}\simeq\sum_{k=1}^rU_{\la k}\wh\Psi_{j_k},\qquad\la=1,\dots,r,
\]
where $(U_{\la k})_{1\le k,\la\le r}$ is a unitary matrix, whose rows are normalized eigenstates
of the Hermitian tunneling matrix with elements
\[
  \mel{\wh\Psi_{j_k}}{H}{\wh\Psi_{j_l}},\qquad k,l=1,\dots,r.
\]
Note that, up to the algebraic error of the WKB approximation, we have
\[
  \mel{\wh\Psi_{j_k}}{H}{\wh\Psi_{j_l}}\simeq\vep\de_{kl}+O(\e^{-K/a}),\qquad k,l=1,\dots,r.
\]
As before, by symmetry the contributions to the density of states of the corresponding WKB
quasi-modes $\Psi_{j_k}$ are all equal:
\[
  D_{j_1}(\vep)=\cdots=D_{j_r}(\vep)\eqqcolon D_0(\vep),
\]
and also coincide with the contribution $D^{(\la)}(\vep)$ of each element in the $r$-plet, up to
exponentially small terms. Therefore the contribution of the entire $r$-plet to the spectral
density of (for instance) the A sublattice is given by
\[
  \sum_{\la=1}^rD^{(\la)}(\vep)\abs{\al^{(\la)}(x,\vep)}^2\simeq
  D_0(\vep)\sum_{\la=1}^r\abs{\al^{(\la)}(x,\vep)}^2.
\]
But, by unitarity, we have
\[
  \sum_{\la=1}^r\abs{\al^{(\la)}(x,\vep)}^2=\sum_{\la=1}^r\abs{\sum_{k=1}^rU_{\la
      k}\wh\al_{j_k}(x,\vep)}^2 =\sum_{\la,k,l=1}^rU_{\la k}U_{\la
    l}^*\wh\al_{j_k}(x,\vep)\wh\al_{j_l}^*(x,\vep) =\sum_{k=1}^r\abs{\wh\al_{j_k}(x,\vep)}^2,
\]
and thus
\[
  \sum_{\la=1}^rD^{(\la)}(\vep)\abs{\al^{(\la)}(x,\vep)}^2 \simeq
  D_0(\vep)\sum_{k=1}^r\abs{\wh\al_{j_k}(x,\vep)}^2
  =\sum_{k=1}^rD_{j_k}(\vep)\abs{\wh\al_{j_k}(x,\vep)}^2,
\]
as for the $r=2$ case.

\subsubsection{Average density}

Let us start by computing the average density. Using eqs.~\eqref{dpdvep} and~\eqref{rhob}, and
replacing again $\sin^2\left(\th_j\pm\frac\phi2\right)$ by its average $1/2$, we obtain
\begin{equation}\label{brhomaster}
  \bar\rho(x,\vep_F)=\frac1{2\pi a}\int_{\bmu(x)-\Emax(x)}^{\vep_F}\dd\vep\,
  \left|\frac{\pd p(x,\vep)}{\pd\vep}\right|\,\Th\bigl(1-\xi(x,\vep)^2\bigr)\,.
\end{equation}
The appearance of the $\Th$ function in the last equation, which is of fundamental importance,
stems from the fact that $\Psi_\eff$ is supported on the set $\Om_\vep$ where
$\Th\bigl(1-\xi(x,\vep)^2\bigr)=1$. Note also that, by eq.~\eqref{dpdvep},
$\frac{\pd p(x,\vep)}{\pd\vep}$ has the sign of $\bmu(x)-\vep$. We must distinguish the following
cases (cf.~Fig.~\ref{fig.bands}):
\begin{figure}[pos=b]
  \centering
    \begin{tikzpicture}[
    scale=.7,
    x=1.05cm,
    y=0.88cm,
    line cap=round,
    line join=round,
    curve/.style={black, line width=1.15pt},
    guide/.style={black, densely dashed, line width=0.85pt},
    axis/.style={black, line width=1.15pt, -{Latex[length=3mm,width=2mm]}},
    every node/.style={font=\small}
    ]

    \def\xmark{3.40}
    \def\ellmark{8.65}
    \def\plotsamples{120}

    %
    %

    \pgfmathdeclarefunction{mufun}{1}{%
      \pgfmathparse{-0.80 + 0.41*#1 - 0.033*#1*#1}%
    }
    \pgfmathdeclarefunction{eminfun}{1}{%
      \pgfmathparse{1.05 + 0.04*#1 + 0.004*#1*#1}%
    }
    \pgfmathdeclarefunction{emaxfun}{1}{%
      \pgfmathparse{3.45 - 0.17*#1 + 0.025*#1*#1}%
    }

    \pgfmathsetmacro{\yupperx}
      {mufun(\xmark)+emaxfun(\xmark)}
    \pgfmathsetmacro{\yuppermiddlex}
      {mufun(\xmark)+eminfun(\xmark)}
    \pgfmathsetmacro{\ylowermiddlex}
      {mufun(\xmark)-eminfun(\xmark)}
    \pgfmathsetmacro{\ylowerx}
      {mufun(\xmark)-emaxfun(\xmark)}

    \pgfmathsetmacro{\yupperell}
      {mufun(\ellmark)+emaxfun(\ellmark)}
    \pgfmathsetmacro{\yuppermiddleell}
      {mufun(\ellmark)+eminfun(\ellmark)}
    \pgfmathsetmacro{\ylowermiddleell}
      {mufun(\ellmark)-eminfun(\ellmark)}
    \pgfmathsetmacro{\ylowerell}
      {mufun(\ellmark)-emaxfun(\ellmark)}


    \path[fill=regionred, fill opacity=0.62]
      plot[
        domain=0:\ellmark,
        samples=\plotsamples,
        variable=\t
      ]
      (\t,{mufun(\t)+emaxfun(\t)})
      --
      plot[
        domain=\ellmark:0,
        samples=\plotsamples,
        variable=\t
      ]
      (\t,{mufun(\t)+eminfun(\t)})
      -- cycle;

    \path[fill=regionviolet, fill opacity=0.65]
      plot[
        domain=0:\ellmark,
        samples=\plotsamples,
        variable=\t
      ]
      (\t,{mufun(\t)-eminfun(\t)})
      --
      plot[
        domain=\ellmark:0,
        samples=\plotsamples,
        variable=\t
      ]
      (\t,{mufun(\t)-emaxfun(\t)})
      -- cycle;


    \draw[axis] (-0.52,0) -- (10.40,0);

    \centering
    \draw[axis] (0,-4.58) -- (0,4.72)
      node[above, font=\normalsize] {$\varepsilon$};


    \draw[curve]
      plot[
        domain=0:\ellmark,
        samples=\plotsamples,
        variable=\t
      ]
      (\t,{mufun(\t)+emaxfun(\t)});

    \draw[curve]
      plot[
        domain=0:\ellmark,
        samples=\plotsamples,
        variable=\t
      ]
      (\t,{mufun(\t)+eminfun(\t)});

    \draw[curve]
      plot[
        domain=0:\ellmark,
        samples=\plotsamples,
        variable=\t
      ]
      (\t,{mufun(\t)-eminfun(\t)});

    \draw[curve]
      plot[
        domain=0:\ellmark,
        samples=\plotsamples,
        variable=\t
      ]
      (\t,{mufun(\t)-emaxfun(\t)});


    \draw[guide] (0,\yupperx) --
                 (\xmark,\yupperx);

    \draw[guide] (0,\yuppermiddlex) --
                 (\xmark,\yuppermiddlex);

    \draw[very thick]
      (\xmark,\yuppermiddlex) --
      (\xmark,\yupperx);

    \draw[guide] (0,\ylowermiddlex) --
                 (\xmark,\ylowermiddlex);

    \draw[guide] (0,\ylowerx) --
                 (\xmark,\ylowerx);

    \draw[guide]
      (\xmark,\ylowermiddlex) --
      (\xmark,\yuppermiddlex);

    \draw[very thick]
      (\xmark,\ylowerx) --
      (\xmark,\ylowermiddlex);


    \draw[guide]
      (\ellmark,0) --
      (\ellmark,\yupperell);

    \draw[guide]
      (\ellmark,\ylowerell) --
      (\ellmark,0);


    \node[below right=1pt] at (\xmark,0.1) {$x$};
    \node[below right=1pt] at (\ellmark,0.1) {$\ell$};


    \node[anchor=east] at (-0.55,\yupperx)
      {$\bar{\mu}(x)+E_{\max}(x)$};

    \node[anchor=east] at (-0.55,\yuppermiddlex)
      {$\bmu(x)+E_{\min}(x)$};

    \node[anchor=east] at (-0.55,\ylowermiddlex)
      {$\bmu(x)-E_{\min}(x)$};

    \node[anchor=east] at (-0.55,\ylowerx)
      {$\bar{\mu}(x)-E_{\max}(x)$};


    \node[anchor=west] at
      ({\ellmark+0.18},\yupperell)
      {$p=0,\ \xi=1$};

    \node[anchor=west] at
      ({\ellmark+0.18},\yuppermiddleell)
      {$p=\pi,\ \xi=-1$};

    \node[anchor=west] at
      ({\ellmark+0.18},\ylowermiddleell)
      {$p=\pi,\ \xi=-1$};

    \node[anchor=west] at
      ({\ellmark+0.18},\ylowerell)
      {$p=0,\ \xi=1$};


    \node[align=center] at (6.7,2.75)
      {$\displaystyle \abs{\xi}<1,\
        \frac{\partial p}{\partial \varepsilon}<0$};

    \node[align=center] at (6.7,-2.1)
      {$\displaystyle \abs{\xi}<1,\
        \frac{\partial p}{\partial \varepsilon}>0$};

    \node[align=center] at (4.2,2.75)
      {$\mathcal B_+$};

    \node[align=center] at (4.2,-1.6)
      {$\mathcal B_-$};

      \node[align=center] at (4.2,4.1)
      {$\xi>1$};

    \node[align=center] at (4.2,-3.5)
      {$\xi>1$};

    \node[align=center] at (4.2,.8)
      {$\xi<-1$};

\end{tikzpicture}
\caption{schematic representation of upper and lower bands $\cB_\pm$, as well as the corresponding
  local bands $\cB_\pm(x)$ at a generic point $x$ (thick vertical lines). The five energy ranges
  discussed in the text are limited by the boundaries $\bmu(x)\pm\Emin(x)$ and
  $\bmu(x)\pm\Emax(x)$ of the local bands at $x$.}
  \label{fig.bands}
\end{figure}
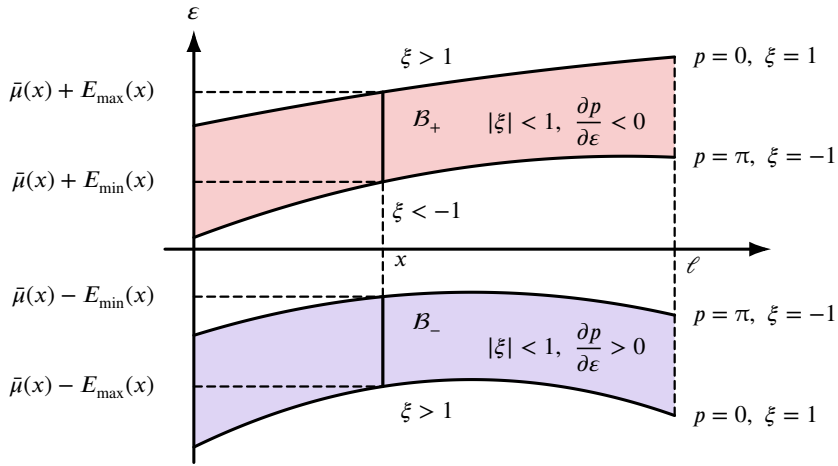
\begin{enumerate}[i.]
\item $\vep_F<\bmu(x)-\Emax(x)$.\en In this case $\bar\rho(x,\vep_F)=0$, due to the factor
  $\Th\bigl(1-\xi(x,\vep)^2\bigr)$ in eq.~\eqref{brhomaster}.

\item $\vep_F\in\cB_-(x)$.\en Now $\Th\bigl(1-\xi(x,\vep)^2\bigr)=1$ for
  $\bmu(x)-\Emax(x)<\vep<\vep_F$, $\frac{\pd p(x,\vep)}{\pd\vep}>0$, and
  $p\bigl(x,\bmu(x)-\Emax(x)\bigr)=0$. Hence
\[
  \bar\rho(x,\vep_F)=\frac1{2\pi a}\int_{\bmu(x)-\Emax(x)}^{\vep_F}\dd\vep\, \frac{\pd
    p(x,\vep)}{\pd\vep}=\frac{p(x,\vep_F)}{2\pi a}=\frac{p_F(x,\vep_F)}{2\pi a},
\]
since  $p=p_F$ for $\abs{\xi}\le1$ (cf.~eq.~\eqref{pF}).

\item $\bmu(x)-\Emin(x)<\vep_F<\bmu(x)+\Emin(x)$.\en In this case $\vep_F$ is inside the local
  insulator band at $x$, so that
  \[
    \Th\bigl(1-\xi(x,\vep)^2\bigr)=
    \begin{cases}
      1,\quad &\bmu(x)-\Emax(x)<\vep<\bmu(x)-\Emin(x),\\
      0, &\bmu(x)-\Emin(x)<\vep<\vep_F,
    \end{cases}
  \]
  with $p(x,\vep)=\pi$ at the top of the lower local band. We thus have
  \[
    \bar\rho(x,\vep_F)=\frac1{2\pi a}\int_{\bmu(x)-\Emax(x)}^{\bmu(x)-\Emin(x)}\dd\vep\, \frac{\pd
      p(x,\vep)}{\pd\vep}=\frac1{2\pi a}p\bigl(x,\bmu(x)-\Emin(x)\bigr)=\frac1{2a}.
  \]
\item $\vep_F\in\cB_+(x)$. Since $\frac{\pd p}{\pd\vep}$ is negative in the upper metallic band,
  we have
  \begin{align*}
    \bar\rho(x,\vep_F)
    &=\frac1{2a}-\frac1{2\pi a}\int_{\bmu(x)+\Emin(x)}^{\vep_F}\dd\vep\, \frac{\pd
      p(x,\vep)}{\pd\vep}=\frac1{2a}-\frac1{2\pi a}\left(p(x,\vep_F)-
      p\bigl(x,\bmu(x)+\Emin(x)\bigr)\right)\\
    &=\frac1{2\pi a}\big(2\pi-p_F(x,\vep_F)\big).
  \end{align*}

\item $\vep_F>\bmu(x)+\Emax(x)$.\en Now $\Th\bigl(1-\xi(x,\vep)^2\bigr)=0$
  for $\vep>\bmu(x)+\Emax(x)$, and therefore
  \[
    \bar\rho(x,\vep_F)=\frac1{a}-\frac1{2\pi a}\,p_F\bigl(x,\bmu(x)+\Emax(x)\bigr)
    =\frac1a.
  \]
\end{enumerate}
We conclude that $\bar\rho(x,\vep_F)$ is given by the LDA formula~\eqref{LDA1}.

\subsubsection{Filling fraction}

From the previous formulas it is also straightforward to derive a closed formula for the filling
fraction $\nu(\vep_F)$ as a function of the Fermi energy $\vep_F$. Indeed, by eqs.~\eqref{dpdvep}
and~\eqref{Dvep} we have
\[
  \nu(\vep_F)=\frac1{2L}\int_{\vep_0}^{\vep_F}\dd\vep D(\vep) =\frac1{2\pi
    \ell}\int_{\vep_0}^{\vep_F}\dd\vep\int_0^\ell\dd x\,
  \frac{\Th\bigl(1-\xi(x,\vep)^2\bigr)}{\abs{v_F(x,\vep)}} =\frac1{2\pi
    \ell}\int_{\vep_0}^{\vep_F}\dd\vep\int_0^\ell\dd x\, \abs{\frac{\pd
      p(x,\vep)}{\pd\vep}}\Th\bigl(1-\xi(x,\vep)^2\bigr).
\]
Reversing the order of integration and using eq.~\eqref{brhomaster}, we obtain the equivalent
formula
\[
  \nu(\vep_F)=\frac{a}{\ell}\int_0^{\ell}\dd x\,\bar\rho(x,\vep_F),
\]
as expected. This formula can be slightly simplified by noting that, by eq.~\eqref{LDA1}, we can
write
\[
  a\bar\rho(x,\vep_F)=\frac1{2}+\frac12\sgn\bigl(\vep_F-\bmu(x)\bigr)\left(
    1-\frac{p_F(x,\vep_F)}{\pi}\right),
\]
where we have taken into account that $p_F(x,\vep_F)=\pi$ for
$\bmu(x)-\Emin(x)<\vep_F<\bmu(x)+\Emin(x)$ (cf.~eq.~\eqref{pF}). We thus have
\begin{equation}\label{nuvepF}
  \nu(\vep_F)=\frac12+\frac1{2\ell}\int_0^\ell\dd x\,\sgn\bigl(\vep_F-\bmu(x)\bigr)\left(1
    -\frac{p_F(x,\vep_F)}{\pi}\right).
\end{equation}

\subsubsection{Sublattice imbalance}

The calculation of the sublattice imbalance $\de\rho(x,\vep_F)$ proceeds similarly. To begin with,
if $x\in I_j(\vep)$ and $\vep\in\cB_\pm(x)$, averaging eq.~\eqref{etaAetaB} over many cells we
have
\begin{equation}
  \abs{\be_\eff(x,\vep)}^2-\abs{\al_\eff(x,\vep)}^2
  =\frac{\abs{\vep-\mu^A(x)}-\abs{\vep-\mu^B(x)}}{4v(x)w(x)\sqrt{1-\xi(x,\vep)^2}}
    =\mp\frac{\dmu(x)}{2v(x)w(x)\sqrt{1-\xi(x,\vep)^2}}
  =\frac{\dmu(x)}{2\abs{\vep-\bmu(x)}}\,\frac{\pd p(x,\vep)}{\pd\vep},
    \label{bemal}
\end{equation}
and therefore
\begin{equation}
  \label{drhomaster}
  \de\rho(x,\vep_F)=\frac{\dmu(x)}{2\pi a}
  \int_{\bmu(x)-\Emax(x)}^{\vep_F}\frac{\dd\vep}{\abs{\vep-\bmu(x)}}\,
  \frac{\pd p(x,\vep)}{\pd\vep}\,\Th\bigl(1-\xi(x,\vep)^2\bigr).
\end{equation}
As before, we distinguish five different cases according to the position of the Fermi energy
$\vep_F$ relative to the boundaries of the local bands $\cB_\pm(x)$:
\begin{enumerate}[i.]
\item $\vep_F<\bmu(x)-\Emax(x)$.\en In this case $\de\rho(x,\vep_F)=0$, since
  $\Th\bigl(1-\xi(x,\vep)^2\bigr)=0$ throughout the whole integration range.
  
\item $\vep_F\in\cB_-(x)$.\en Changing the integration variable from $\vep$ to $p$ we obtain
  \begin{align}
    \de\rho(x,\vep_F)&=\frac{\dmu(x)}{2\pi a}\int_0^{p(x,\vep_F)}\frac{\dd  p}{\abs{\vep-\bmu(x)}}
    \notag\\
      &=\frac{\dmu(x)}{2\pi a}\int_0^{p(x,\vep_F)}\frac{\dd
      p}{\sqrt{\dmu(x)^2+v(x)^2+w(x)^2+2v(x)w(x)\cos p}}\notag\\
    &=
      \frac{\dmu(x)}{\pi a}\int_0^{\frac12p(x,\vep_F)}
      \frac{\dd s}{\sqrt{\dmu(x)^2+(v(x)+w(x))^2-4v(x)w(x)\sin^2s}}\notag\\
      &=\frac{\ka(x)\dmu(x)}{2\pi a\sqrt{v(x)w(x)}}\,
        F\Bigl(\frac12p(x,\vep_F),\ka(x)\Bigr),
        \label{derhoBm}
  \end{align}
  with $\ka(x)$ given by eq.~\eqref{kaLDA}. This coincides with the LDA result in
  eq.~\eqref{LDA2}, as $p(x,\vep_F)=p_F(x,\vep_F)$ for $\vep_F\in\cB_-(x)$.
  
\item $\bmu(x)-\Emin(x)<\vep_F<\bmu(x)+\Emin(x)$.\en In this case $\Th\bigl(1-\xi(x,\vep)^2\bigr)$
  vanishes for $\bmu(x)-\Emin(x)<\vep<\vep_F$. Taking into account that $p_F=\pi$ at the top of
  the lower local band $\cB_-(x)$ we obtain
  \begin{equation}\label{drhohalf}
    \de\rho(x,\vep_F)=\frac{\ka(x)\dmu(x)}{2\pi a\sqrt{v(x)w(x)}}\,
      F\Bigl(\frac\pi2,\ka(x)\Bigr)=\frac{\ka(x)\dmu(x)}{2\pi a\sqrt{v(x)w(x)}}\,
      K\bigl(\ka(x)\bigr).
    \end{equation}
    where $K(\ka)$ is the complete elliptic integral of the first kind with modulus $\ka$. This is
  again a particular case of eq.~\eqref{LDA2}, since $p_F\bigl(x,\bmu(x)-\Emin(x)\bigr)=\pi$.

\item $\vep_F\in\cB_+(x)$.\en We must now add to the contribution of the whole lower band computed
  above that of the interval $\bigl(\bmu(x)+\Emin(x),\vep_F\bigr)$. We thus have
  \begin{align*}
    \de\rho(x,\vep_F)&=\frac{\dmu(x)}{2\pi a}\left(\int_{\bmu(x)-\Emax(x)}^{\bmu(x)-\Emin(x)}
      +\int_{\bmu(x)+\Emin(x)}^{\vep_F}\right)\frac{\dd\vep}{\abs{\vep-\bmu(x)}}\,
      \frac{\pd p(x,\vep)}{\pd\vep}\\
      &= \frac{\dmu(x)}{2\pi a}\left(\int_{0}^{\pi}
        +\int_{\pi}^{p(x,\vep_F)}\right)\frac{\dd p}{\abs{\vep-\bmu(x)}}
        =
      \frac{\dmu(x)}{2\pi a}\int_{0}^{p(x,\vep_F)}\frac{\dd p}{\abs{\vep-\bmu(x)}}\,.
  \end{align*}
  Hence in this case $\de\rho(x,\vep_F)$ is still given by eq.~\eqref{derhoBm}, which coincides
  with the LDA formula~\eqref{LDA2} as $p=p_F$ for $\vep_F\in\cB_+(x)$.

\item $\vep_F>\bmu(x)+\Emax(x)$.\en Now $\Th\bigl(1-\xi(x,\vep)^2\bigr)=0$ for
  $\vep>\bmu(x)+\Emax(x)$, so that
  \[
    \de\rho(x,\vep_F)=\frac{\dmu(x)}{2\pi a}\int_{0}^{p_F\left(x,\bmu(x)+\Emax(x)\right)}\frac{\dd
      p}{\abs{\vep-\bmu(x)}}=0,
  \]
  since $p_F\bigl(x,\bmu(x)+\Emax(x)\bigr)=0$. Again, this is a particular case of
  eq.~\eqref{LDA2}.
\end{enumerate}
In summary, we have shown that the LDA formula~\eqref{LDA2} holds for all Fermi energies $\vep_F$.

\subsection{Discussion}\label{sec.disc}

It should be clear from the previous discussion---and, in particular, eqs.~\eqref{LDArhos}---that
the metal/insulator behavior of an inhomogeneous RM model at Fermi energy $\vep_F$ is dictated by
the topology of the classically allowed region---or, more precisely, of the open set
$\Om_{\vep_F}$ on which $\abs{\xi(x,\vep_F)}<1$, and hence $p(x,\vep_F)$ is real---i.e., the number
and location of the disjoint potential wells $I_j(\vep_F)$ that make up this region, as well as
the nature of the disjoint intervals $\tilde I_j(\vep_F)$ in its complement.

Indeed, inside each potential well $I_j(\vep_F)$, both $\bar\rho(x,\vep_F)$ and
$\de\rho(x,\vep_F)$ vary continuously as the Fermi quasimomentum $p_F(x,\vep_F)\in(0,\pi)$ traverses
its range. Therefore the system behaves as a metal on each of these intervals, which thus make up
a collection of disjoint metallic islands. On the other hand, the local behavior in a classically
forbidden interval $\tilde I_j(\vep_F)$ depends on whether $\xi(x,\vep_F)$ is greater than $1$ or
less than $-1$ (equivalently, $p_F(x,\vep_F)=0$ or $\pi$), and (in the former case) whether
$\vep_F-\bmu(x)$ is positive or negative\footnote{Note that, since $\abs{\xi(x,\vep_F)}>1$ on the
  interior of $\tilde I_j(\vep_F)$ and $\xi$ varies smoothly,  either $\xi(x,\vep_F)<-1$
  throughout this interval or $\xi(x,\vep_F)>1$.}. More precisely:
\begin{enumerate}[i)]
\item If $\xi(x,\vep_F)<-1$ on $x\in\tilde I_j(\vep_F)$, the whole lower band $\cB_-(x)$ is occupied
  and the local upper band $\cB_+(x)$ is empty for all $x\in\tilde I_j(\vep_F)$. In other words,
  $\vep_F$ lies in the insulator gap at every point of $\tilde I_j(\vep_F)$. Hence in this case the
  interval $\tilde I_j(\vep_F)$ is insulating, and at each of its points the chain is half-filled.
  The last statement also follows from eq.~\eqref{LDA1}, since in this case
  \[
    p_F(x,\vep_F)=\pi\enspace\implies\enspace a\bar\rho(x,\vep_F)=\frac{1}{2}\,.
  \]
  Note, however, that by eq.~\eqref{LDA2} the sublattice imbalance $\de\rho(x,\vep_F)$ need not be
  constant.
  
\item If $\xi(x,\vep_F)>1$ and $\vep_F-\bmu(x)<0$ inside $\tilde I_j(\vep_F)$, by eq.~\eqref{LDA1}
  the average density $\bar\rho(x,\vep_F)$ (and, hence, $\de\rho(x,\vep_F)$) vanishes on this
  interval. Hence the interval $\tilde I_j(\vep_F)$ contains no fermions.
  
\item Finally, if $\xi(x,\vep_F)>1$ and $\vep_F-\bmu(x)>0$ inside $\tilde I_j(\vep_F)$, by
  eq.~\eqref{LDA1} $a\bar\rho(x,\vep_F)=1$ (and, consequently, $\de\rho(x,\vep_F)=0$). Hence in
  this case every site of the interval $\tilde I_j(\vep_F)$ is occupied.
\end{enumerate}

The chain therefore can be effectively divided into a sequence of insulator, metallic, or frozen
(fully occupied/empty) regions, analogous to a semiconductor heterostructure, although here the
structure is generated by the spatial variation of the Rice--Mele parameters rather than by
joining distinct materials. Moreover, as explained above, the location and nature of these regions
are determined by the function $\xi(x,\vep_F)$ (and, when $\xi(x,\vep_F)>1$, by the sign of
$\vep_F-\bmu(x)$). Note, in particular, that in the inhomogeneous case a fixed Fermi energy
$\vep_F$ can be in either of the conduction bands, or in the insulator gap, depending on the
region considered.

As remarked in Section~\ref{sec.consWKB}, in the inhomogeneous case the energy spectrum will not
have a global insulator gap whenever the global bands $\cB_\pm$ overlap, i.e., when
\[
  \max_{0\le x\le \ell}(\bmu(x)-\Emin(x))\ge\min_{0\le x\le \ell}(\bmu(x)+\Emin(x)).
\]
Note that this can only occur when the on-site potential $\bmu(x)$ is not identically zero.
However, even in this case the system could have one or several local insulating intervals,
determined by the condition $\xi(x,\vep_F)<-1$. This fact will be discussed in detail in the
following examples.

\section{Examples}\label{sec.examples}

In this section we shall present a few examples illustrating the previous results.

\subsection{Homogeneous RM chain in a harmonic trap}

Perhaps the simplest nontrivial example where the previous results can be fully tested is provided
by a homogeneous RM chain subject to the on-site harmonic potential
\begin{equation}\label{munharm}
  \bmu_n=\mu_0\left(1-\frac{2n}L\right)^2,
\end{equation}
with $\mu_0>0$, and $\dmu_n=\dmu$ constant. Models of this kind have been considered in the
context of Thouless pumping, for instance in refs.~\cite{WTD13,NTTI16} (see also~\cite{ZGWVE24}).
Since
\[
  \Emin(x)=\sqrt{\dmu^2+(v-w)^2},\qquad \Emax(x)=\sqrt{\dmu^2+(v+w)^2}
\]
are both constant, it is convenient to parametrize $v$ and $w$ in terms of $\Emin$ and $\Emax$ as
\[
  v=\frac12\left(\sqrt{\Emax^2-\dmu^2}+\si\sqrt{\Emin^2-\dmu^2}\right),\qquad
  w=\frac12\left(\sqrt{\Emax^2-\dmu^2}-\si\sqrt{\Emin^2-\dmu^2}\right),\qquad
\]
where $\si=\pm1$ is the sign of $v-w$. The necessary and sufficient condition for the existence of
a global energy gap is that
\[
  \max_{0\le x\le \ell}(\bmu(x)-\Emin)<\min_{0\le x\le \ell}(\bmu(x)+\Emin),
\]
where
\[
  \bmu(x)=\mu_0\left(1-\frac{2x}\ell\right)^2
\]
is the continuum version of $\bmu_n$. Since $\Emin$ and $\Emax$ are constant, and
\[
  \max_{0\le x\le\ell}\bmu(x)=\mu_0,\qquad \min_{0\le x\le\ell}\bmu(x)=0,
\]
the gap will not close provided that $\mu_0<2\Emin$. We shall therefore discuss separately the
cases $\mu_0<2\Emin$ and $\mu_0>2\Emin$.

\subsubsection{$\mu_0<2\Emin$}
In this case, the local lower band $\cB_-(x)$ is full and the local upper band $\cB_+(x)$ is empty
throughout the chain. Hence the system behaves as an insulator for Fermi energies
$\vep_F\in(\mu_0-\Emin,\Emin)$, with global energy gap of width $2\Emin-\mu_0$. Note that for
these Fermi energies $\xi(x,\vep_F)<-1$, and consequently $p_F(x,\vep_F)=\pi$ for all
$x\in[0,\ell]$. By eqs.~\eqref{LDArhos}, the average occupation number $a\bar\rho(x)$ is $1/2$ (as
expected), whereas the sublattice imbalance $\de\rho$ is also constant but nonzero, given by
\[
  \de\rho=\frac{\de\mu}{\pi a\Emax}\,K(\ka),\qquad
  \text{with}\quad\ka=\sqrt{1-\frac{\Emin^2}{\Emax^2}}\,.
\]
\begin{figure}[pos=t]
  \centering
  \begin{subfigure}{0.49\textwidth}
    \centering
    \begin{minipage}[c][\BandPanelHeight][c]{\linewidth}
      \begin{tikzpicture}
        \begin{axis}[
          width=\BandPlotWidth,
          height=\BandPlotHeight,
          xmin=0,
          xmax=1040,
          ymin=-3.25,
          ymax=5.45,
          axis lines=middle,
          axis line style={-Latex},
          axis on top,
          clip=false,
          xlabel={$x$},
          ylabel={$\vep_F$},
          every axis x label/.style={
            at={(current axis.right of origin)},
            anchor=west
          },
          every axis y label/.style={
            at={(current axis.above origin)},
            anchor=south
          },
          xtick={500,1000},
          xticklabels={$\ell/2$,$\ell$},
          ytick={-3,-2,-1,0,2,3,4,5},
          yticklabels={
            {$-\Emax$},
            {$-\Emin$},
            {$\mu_0-\Emax$},
            {$\mu_0-\Emin$},
            {$\Emin$},
            {$\Emax$},
            {$\mu_0+\Emin$},
            {$\mu_0+\Emax$}
          },
          tick align=outside,
          xticklabel style={font=\small, yshift=2pt},
          yticklabel style={font=\small},
          tick style={black, line width=0.2pt},
          major tick length=4pt,
          samples=201,
          domain=0:1000,
          no markers
          ]

          \draw[black, line width=0.2pt]
          ([xshift=-4pt]axis cs:0,0) -- ([xshift=3pt]axis cs:0,0);
          
          \path[fill=gray!70, fill opacity=0.20]
          (axis cs:0,-3) rectangle (axis cs:1000,-2);
          \path[fill=blue!65, fill opacity=0.20]
          (axis cs:0,-2) rectangle (axis cs:1000,-1);
          \path[fill=gray!70, fill opacity=0.20]
          (axis cs:0,-1) rectangle (axis cs:1000,0);
          \path[fill=blue!65, fill opacity=0.20]
          (axis cs:0,0) rectangle (axis cs:1000,2);
          \path[fill=gray!70, fill opacity=0.20]
          (axis cs:0,2) rectangle (axis cs:1000,3);
          \path[fill=blue!65, fill opacity=0.20]
          (axis cs:0,3) rectangle (axis cs:1000,4);
          \path[fill=gray!70, fill opacity=0.20]
          (axis cs:0,4) rectangle (axis cs:1000,5);
          
          \addplot[red,  thick] {3 + 2*(1 - 2*x/1000)^2};
          \addplot[blue, thick] {2 + 2*(1 - 2*x/1000)^2};
          \addplot[black, dashed, thick] {2*(1 - 2*x/1000)^2};
          \addplot[blue, thick] {-2 + 2*(1 - 2*x/1000)^2};
          \addplot[red,  thick] {-3 + 2*(1 - 2*x/1000)^2};
          
          \addplot[black, very thick] {2};
          \addplot[black, very thick] {0};
          
          \node[anchor=east, font=\small, xshift=-4pt]
          at (axis cs:0,0) {$\mu_0-\Emin$};
          
          \draw[black, line width=1pt]
          (axis cs:500,-0.09) -- (axis cs:500,0.09);
          
        \end{axis}
      \end{tikzpicture}
    \end{minipage}
    \caption{$\Emax-\Emin<\mu_0<2\Emin$.}
    \label{fig:bands-gap}
  \end{subfigure}
  \hfill
  \begin{subfigure}{0.49\textwidth}
    \centering
    \begin{minipage}[c][\BandPanelHeight][c]{\linewidth}
      \begin{tikzpicture}
        \begin{axis}[
          width=\BandPlotWidth,
          height=\BandPlotHeight,
          xmin=0,
          xmax=1040,
          ymin=-1.75,
          ymax=5.9,
          axis lines=middle,
          axis line style={-Latex},
          axis on top,
          clip=false,
          xlabel={$x$},
          ylabel={$\vep_F$},
          every axis x label/.style={
            at={(current axis.right of origin)},
            anchor=west
          },
          every axis y label/.style={
            at={(current axis.above origin)},
            anchor=south
          },
          xtick={500,1000},
          xticklabels={$\ell/2$,$\ell$},
          ytick={-1.5,-1,1,1.5,2.5,3,5,5.5},
          yticklabels={
            {$-\Emax$},
            {$-\Emin$},
            {$\Emin$},
            {$\Emax$},
            {$\mu_0-\Emax$},
            {$\mu_0-\Emin$},
            {$\mu_0+\Emin$},
            {$\mu_0+\Emax$}
          },
          tick align=outside,
          xticklabel style={font=\small, yshift=2pt},
          yticklabel style={font=\small},
          samples=201,
          domain=0:1000,
          no markers
          ]
          
          \path[fill=gray!70, fill opacity=0.20]
          (axis cs:0,-1.5) rectangle (axis cs:1000,-1);
          \path[fill=blue!65, fill opacity=0.20]
          (axis cs:0,-1) rectangle (axis cs:1000,1);
          \path[fill=gray!70, fill opacity=0.20]
          (axis cs:0,1) rectangle (axis cs:1000,1.5);
          \path[fill=blue!65, fill opacity=0.20]
          (axis cs:0,1.5) rectangle (axis cs:1000,2.5);
          \path[fill=gray!70, fill opacity=0.20]
          (axis cs:0,2.5) rectangle (axis cs:1000,3);
          \path[fill=blue!65, fill opacity=0.20]
          (axis cs:0,3) rectangle (axis cs:1000,5);
          \path[fill=gray!70, fill opacity=0.20]
          (axis cs:0,5) rectangle (axis cs:1000,5.5);
          
          \addplot[red,  thick] {1.5 + 4*(1 - 2*x/1000)^2};
          \addplot[blue, thick] {1   + 4*(1 - 2*x/1000)^2};
          \addplot[black, dashed, thick] {4*(1 - 2*x/1000)^2};
          \addplot[blue, thick] {-1   + 4*(1 - 2*x/1000)^2};
          \addplot[red,  thick] {-1.5 + 4*(1 - 2*x/1000)^2};
          
          \addplot[black, very thick] {1};
          \addplot[black, very thick] {3};
          
          \draw[black, line width=1.1pt]
          (axis cs:500,-0.09) -- (axis cs:500,0.09);
          
        \end{axis}
      \end{tikzpicture}
    \end{minipage}
    \caption{$\mu_0>2\Emax$.}
    \label{fig:bands-overlap}
  \end{subfigure}
  \caption{Energy intervals governing the metal/insulator behavior of the homogeneous RM model in
    a harmonic potential trap, represented by alternating light gray and light blue horizontal
    bands, for examples with a global gap (panel (a)) and overlapping global bands (panel (b)). In
    both panels, the outer boundaries $\bmu(x)\pm\Emax$ and the inner boundaries $\bmu(x)\pm\Emin$
    of the local bands $\cB_\pm(x)$ are shown respectively in red and blue, while the dashed black
    curve is the graph of the sublattice-independent on-site potential $\bmu(x)$. The thick black
    lines are the boundaries of the global energy gap $(\mu_0-\Emin,\Emin)$
    (panel~\subref{fig:bands-gap}) or the overlap $(\Emin,\mu_0-\Emin)$ between the two global
    bands $\cB_\pm$ (panel~\subref{fig:bands-overlap}).}
  \label{fig:bands-trap}
\end{figure}
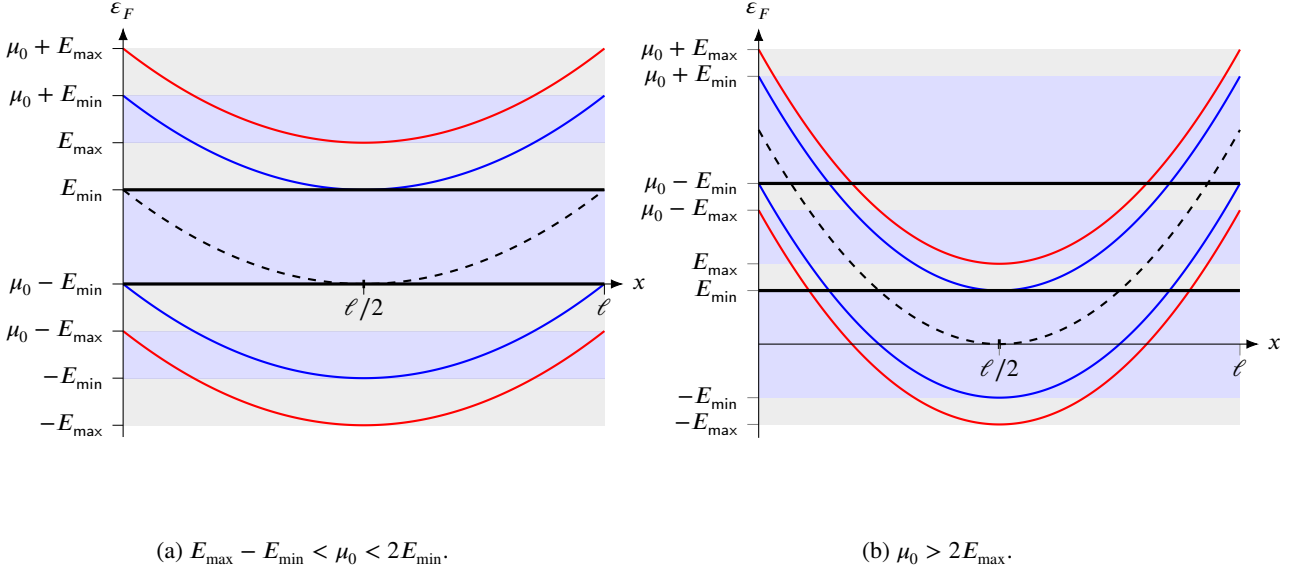%
Thus, in this range of Fermi energies, the particle densities behave as in the homogeneous case.
Note, however, that the eigenfunctions and their Wannier centers depend on the on-site
potential~$\bmu(x)$, and therefore differ from those of the homogeneous model.

On the other hand, for Fermi energies outside the gap $(\mu_0-\Emin,\Emin)$ the situation is far
more interesting since, as explained in Section~\ref{sec.disc}, the metal/insulator behavior
varies along the chain. Assuming, for definiteness, that $\mu_0>\Emax-\Emin$ (the case
$\mu_0<\Emax-\Emin$ is slightly simpler), there are nine different energy intervals characterizing
this behavior, as summarized in Table~\ref{tab.bands1} (see also Fig.~\ref{fig:bands-gap}).

Consider, for instance, the case $\Emax<\vep_F<\mu_0+\Emin$. In this case the two upper local band
boundaries, $\vep=\bmu(x)+\Emin$ and $\vep=\bmu(x)+\Emax$, each intersect the horizontal line
$\vep=\vep_F$ at two points lying in the interval $[0,\ell]$, respectively given by
\[
  x_{\text{min},\pm}=\frac{\ell}2\,\left(1\pm\sqrt{\frac{\vep_F-\Emin}{\mu_0}}\right),\qquad
  x_{\text{max},\pm}=\frac{\ell}2\,\left(1\pm\sqrt{\frac{\vep_F-\Emax}{\mu_0}}\right),
\]
with
\[
  0<x_{\text{min},-}<x_{\text{max},-}<x_{\text{max},+}<x_{\text{min},+}<\ell.
\]
On the outer intervals $(0,x_{\text{min,-}})$ and $(x_{\text{min},+},\ell)$ we have
\[
  \xi(x,\vep_F)<-1\implies p_F(x,\vep_F)=\pi,
\]
since the Fermi energy lies in the local insulator gap at every point in these intervals.
Moreover, by eq.~\eqref{LDA1} the average occupation is given by
\[
  a\bar\rho(x,\vep_F)=\frac12.
\]
Therefore the chain is insulating on these intervals. On the other hand, the Fermi energy lies in
the upper metallic band at every point of the inner intervals
$(x_{\text{min,-}},x_{\text{max},-})$ and $(x_{\text{max},+},x_{\text{min},+})$. Hence the chain
is metallic in these intervals. Finally, at every point of the central interval
$(x_{\text{max},-},x_{\text{max},+})$ we have
\[
  \xi(x,\vep_F)>1\implies p_F(x,\vep_F)=0,
\]
which, by eq.~\eqref{LDA1}, implies that $a\bar\rho(x,\vep_F)=1$. Thus, the chain is fully occupied
for $x\in(x_{\text{max},-},x_{\text{max},+})$.

\begin{table}[pos=t]
  \centering

  \begin{subtable}[t]{0.48\textwidth}
    \centering
    \caption{$\Emax-\Emin<\mu_0<2\Emin$}
    \label{tab.bands1}
    \begin{tabular*}{\dimexpr .85\linewidth}{@{}LL@{}}
      \toprule
      Interval & Behavior  \\ 
      \midrule
      $\vep_F<-\Emax$& (0)\\
      $-\Emax<\vep_F<-\Emin$& (0m0) \\
      $-\Emin<\vep_F<\mu_0-\Emax$& (0mim0) \\
      $\mu_0-\Emax<\vep_F<\mu_0-\Emin$& (mim) \\
      $\mu_0-\Emin<\vep_F<\Emin$& (i)\\
      $\Emin<\vep_F<\Emax$& (imi)\\
      $\Emax<\vep_F<\mu_0+\Emin$& (im1mi)\\
      $\mu_0+\Emin<\vep_F<\mu_0+\Emax$& (m1m)\\
      $\vep_F>\mu_0+\Emax$& (1)\\
      \bottomrule
    \end{tabular*}
  \end{subtable}
  \begin{subtable}[t]{0.48\textwidth}
    \centering
    \caption{$\mu_0>2\Emax$}
    \label{tab.bands2}
    \begin{tabular*}{\dimexpr .85\linewidth}{@{}LL@{}}
      \toprule
      Interval & Behavior  \\ 
      \midrule
      $\vep_F<-\Emax$& (0)\\
      $-\Emax<\vep_F<-\Emin$& (0m0) \\
      $-\Emin<\vep_F<\Emin$& (0mim0) \\
      $\Emin<\vep_F<\Emax$& (0mimim0) \\
      $\Emax<\vep_F<\mu_0-\Emax$& (0mim1mim0) \\
      $\mu_0-\Emax<\vep_F<\mu_0-\Emin$& (mim1mim)\\
      $\mu_0-\Emin<\vep_F<\mu_0+\Emin$& (im1mi)\\
      $\mu_0+\Emin<\vep_F<\mu_0+\Emax$& (m1m)\\
      $\vep_F>\mu_0+\Emax$& (1)\\
      \bottomrule
    \end{tabular*}
  \end{subtable}
  \caption{Local metal/insulator behavior of the homogeneous RM model in a harmonic potential trap
    for the two examples in Fig.~\ref{fig:bands-trap}. The symbols in the right column have the
    following meaning: $0=\text{zero occupation}$, $1=\text{full occupation}$,
    $\text{i}=\text{insulator (half filling)}$, $\text{m}=\text{metal}$.}
  \label{tab.allbands}
\end{table}

\begin{figure}[pos=t,width=0.65\textwidth,abovecap=0pt]
  
\centering

\begin{minipage}{0.32\textwidth}
  \centering
  \includegraphics[width=\linewidth]{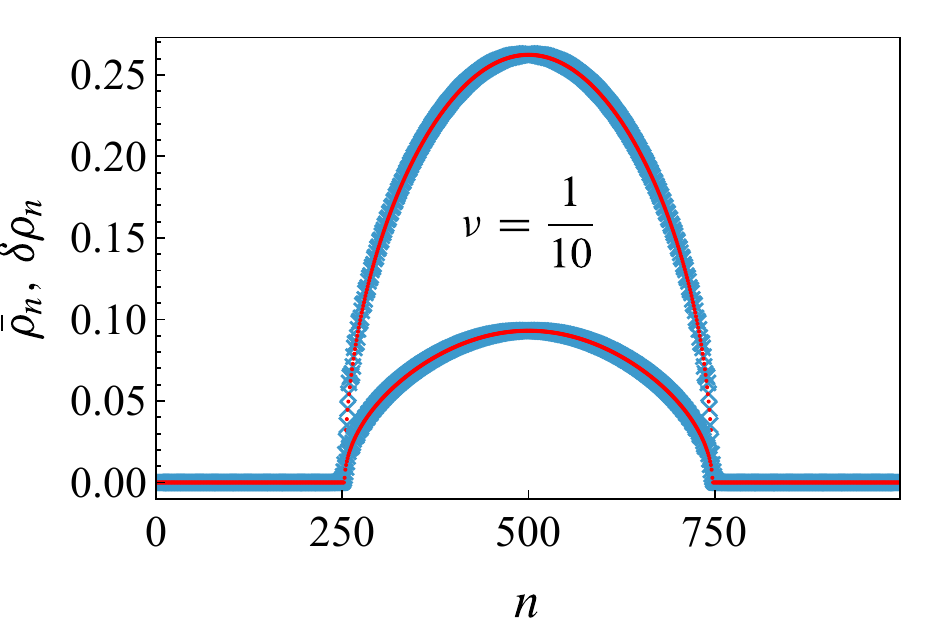}
\end{minipage}\hfill
\begin{minipage}{0.32\textwidth}
  \centering
  \includegraphics[width=\linewidth]{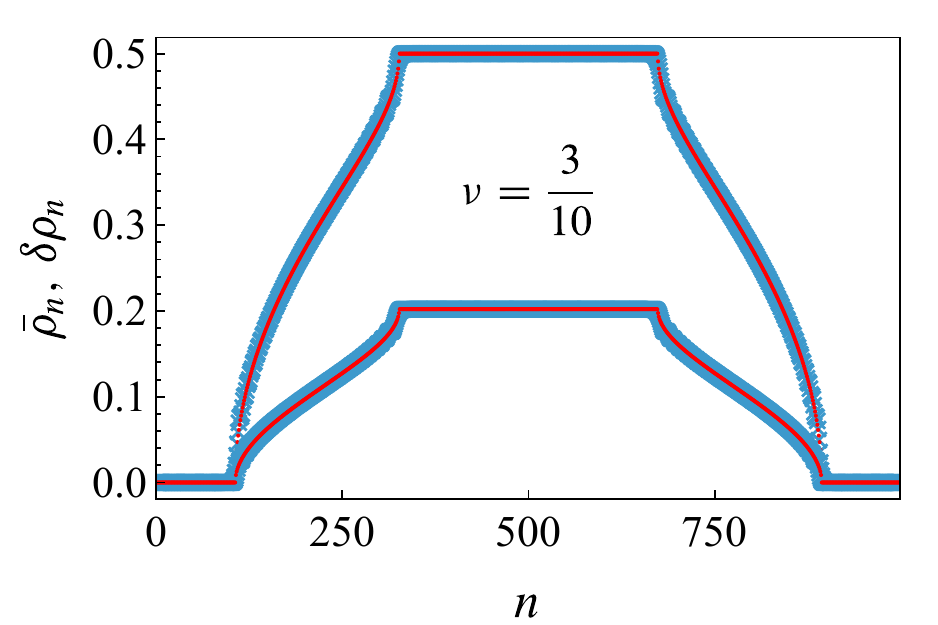}
\end{minipage}\hfill
\begin{minipage}{0.32\textwidth}
  \centering
  \includegraphics[width=\linewidth]{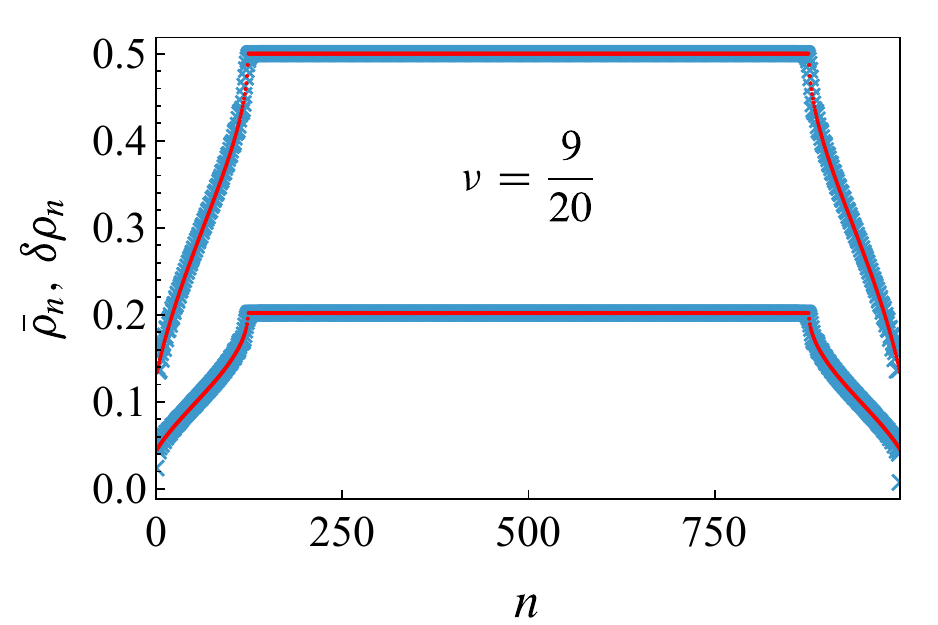}
\end{minipage}

\smallskip

\begin{minipage}{0.32\textwidth}
  \centering
  \includegraphics[width=\linewidth]{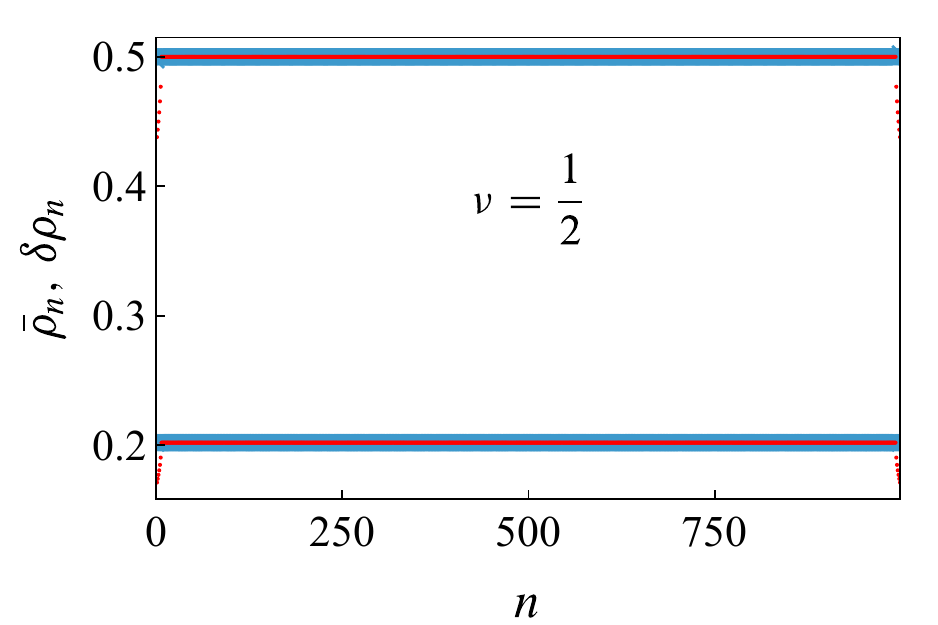}
\end{minipage}\hfill
\begin{minipage}{0.32\textwidth}
  \centering
  \includegraphics[width=\linewidth]{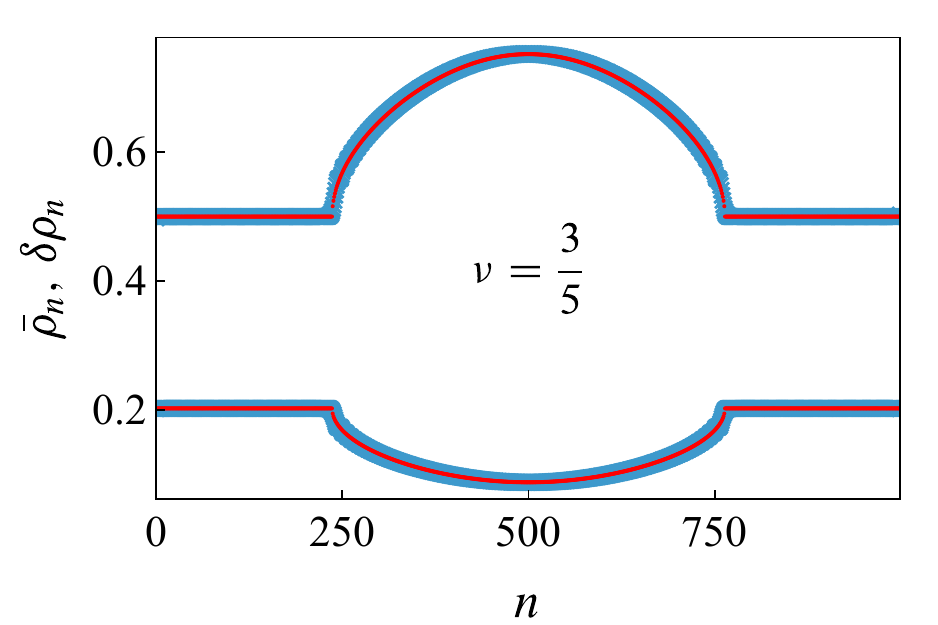}
\end{minipage}\hfill
\begin{minipage}{0.32\textwidth}
  \centering
  \includegraphics[width=\linewidth]{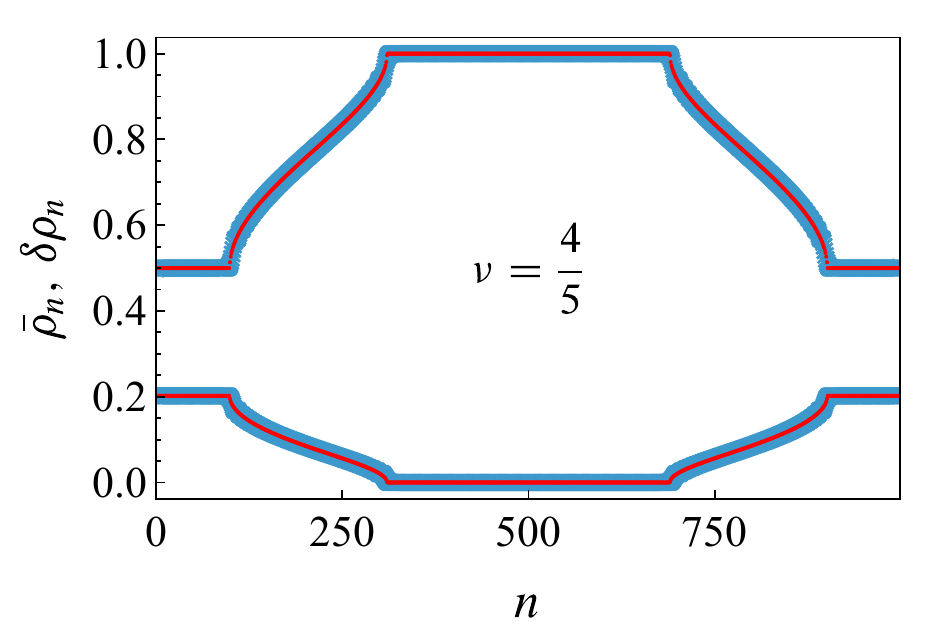}
\end{minipage}

\smallskip

\begin{minipage}[t]{0.32\textwidth}
  \vspace{0pt}
  \centering
  \includegraphics[width=\linewidth]{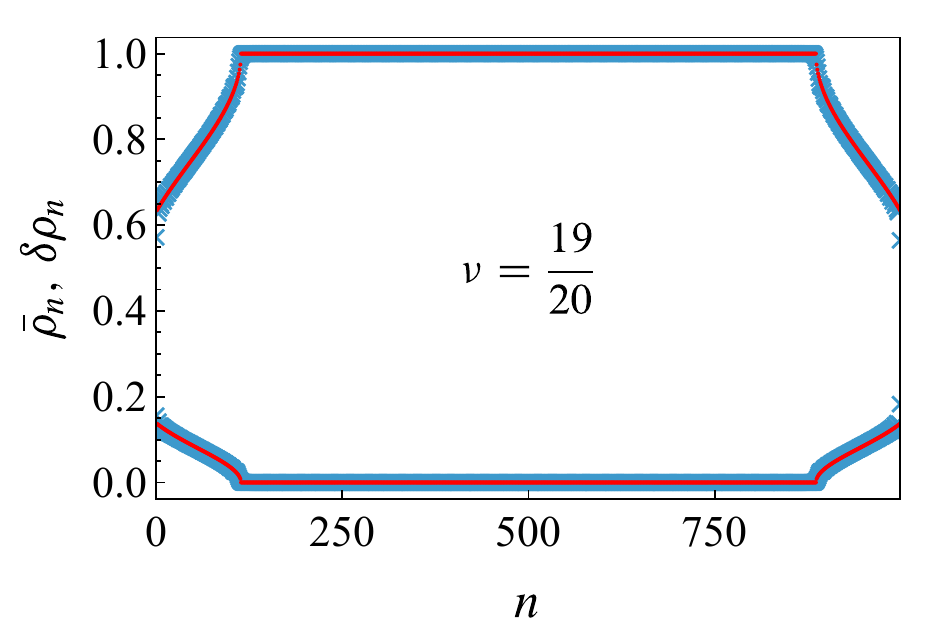}
\end{minipage}\hfill
\begin{minipage}[t]{0.65\textwidth}
  \vspace{6pt}
  \caption{Density profiles of the homogeneous RM model with harmonic on-site
    potential~\eqref{munharm} (blue marks), compared to their WKB approximation~\eqref{LDArhos}
    (red line) for $L=1000$ cells and $a=1$. The values of the chain parameters are $\Emin = 2$,
    $\Emax = 3$, $\mu_0=2$, $\de\mu=1$, and $\si=1$, so that $\Emax-\Emin<\mu_0<2\Emin$. The
    values of the filling fraction $\nu$, which correspond to Fermi energies in rows 2--8 of
    Table~\ref{tab.bands1}, are indicated inside each panel.}
  \label{fig:densities1}
\end{minipage}

\end{figure}

In Fig.~\ref{fig:densities1} we plot the density profiles $\bar\rho$ and $\de\rho$ for several
filling fractions $\nu$, corresponding to Fermi energies in rows 2--8 of Table~\ref{tab.bands1}.
The agreement between the numerically computed density profiles and their WKB
approximations~\eqref{LDArhos} is remarkable in all cases, even near the turning points where the
semiclassical WKB wave function diverges. Of particular interest are the consequences of the WKB
approximation on intervals where the system is insulating. Indeed, in these intervals
$p(x,\vep_F)=\pi$ is constant and does not depend on $\vep_F$, so that $\bar\rho(x,\vep_F)=1/(2a)$
(i.e., the chain is half full). On the other hand, the sublattice imbalance $\de\rho(x,\vep_F)$ is
in general a function of $x$. Since in our case the parameters $v$, $w$, and $\de\mu$ are
constant,
\begin{equation}\label{drhoins}
  \de\rho=\frac{\de\mu}{\pi a\Emax}K\biggl(\textstyle\sqrt{1-\frac{\Emin^2}{\Emax^2}}\,\biggr)
\end{equation}
is also constant. For instance, for the values of the chain parameters in Fig.~\ref{fig:densities1}
we have
\begin{equation}\label{drhoinsnum}
  \de\rho=\frac{1}{3\pi}K\Bigl({\textstyle\frac{\sqrt{5}}3}\Bigr)=0.202046,
\end{equation}
regardless of the value of $\vep_F$ (or, equivalently, the filling fraction $\nu$) and $x$. This
result is in excellent agreement with the numerical value, as can be seen from
Fig.~\ref{fig:densities1}.

\subsubsection{$\mu_0>2\Emin$}
Now
\[
  \cB_-\cap\cB_+=\bigl[\Emin,\mu_0-\Emin\bigr]\ne\emptyset,
\]
so that the upper and lower bands overlap, and the system is no longer a global insulator. In this
case there are several slightly different behaviors, depending on the position of $\mu_0$ relative
to $\Emax\pm\Emin$ and $2\Emax$. We shall restrict ourselves, for definiteness, to the case
$\mu_0>2\Emax$ portrayed in Fig.~\ref{fig:bands-overlap}. As in the case $\mu_0<2\Emin$ discussed
above, the local metal/insulator behavior changes qualitatively as the Fermi energy $\vep_F$
traverses the nine energy intervals listed in Table~\ref{tab.bands2}
(cf.~Fig.~\ref{fig:bands-overlap}). For instance, for $\Emax<\vep_F<\mu_0-\Emax$ the interval
$[0,\ell]$ is divided into 9 disjoint intervals, limited by the points~$(0,x_1,\dots,x_8,\ell)$
with
\begin{align*}
  x_{1,8}&=\frac{\ell}2\,\left(1\mp\sqrt{\frac{\vep_F+\Emax}{\mu_0}}\right),\qquad
           x_{2,7}=\frac{\ell}2\,\left(1\mp\sqrt{\frac{\vep_F+\Emin}{\mu_0}}\right),\\
  x_{3,6}&=\frac{\ell}2\,\left(1\mp\sqrt{\frac{\vep_F-\Emin}{\mu_0}}\right),\qquad
           x_{4,5}=\frac{\ell}2\,\left(1\mp\sqrt{\frac{\vep_F-\Emax}{\mu_0}}\right).
\end{align*}
In each of the intervals $(x_i,x_{i+1})$, the system's metal/insulator behavior is as listed in
Table~\ref{tab.bands2}. Indeed, throughout the central interval $(x_4,x_5)$ we have
$\xi(x,\vep_F)>1$ and $\vep_F>\bmu(x)$, so that $p_F(x,\vep_F)=0$ and $\bar\rho(x,\vep_F)=1$ by
eq.~\eqref{LDA1}. Hence the chain is fully occupied in this interval. On the other hand, on the
symmetric intervals $(x_3,x_4)$ and $(x_5,x_6)$ we have $\abs{\xi(x,\vep_F)}<1$, and consequently
the system behaves as a metal. The same is true for the intervals $(x_1,x_2)$ and $(x_7,x_8)$.
Likewise, on the intervals $(x_2,x_3)$ and $(x_6,x_7)$ we have $\xi(x,\vep_F)<-1$, so that on
these intervals the system behaves as an insulator. Finally, on the outer intervals $(0,x_1)$ and
$(x_8,\ell)$ we have $\xi(x,\vep_F)>1$ and $\vep_F<\bmu(x)$, and consequently $\bar\rho(x,\vep_F)=0$
by eq.~\eqref{LDA1}.

\begin{figure}[pos=t,width=0.65\textwidth,abovecap=0pt]
  
\centering

\begin{minipage}{0.32\textwidth}
  \centering
  \includegraphics[width=\linewidth]{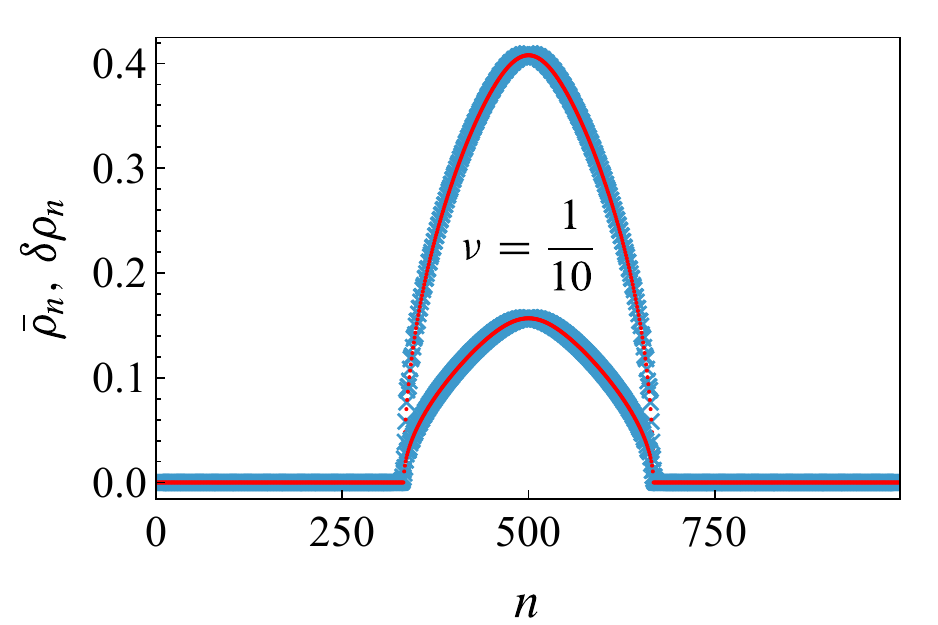}
\end{minipage}\hfill
\begin{minipage}{0.32\textwidth}
  \centering
  \includegraphics[width=\linewidth]{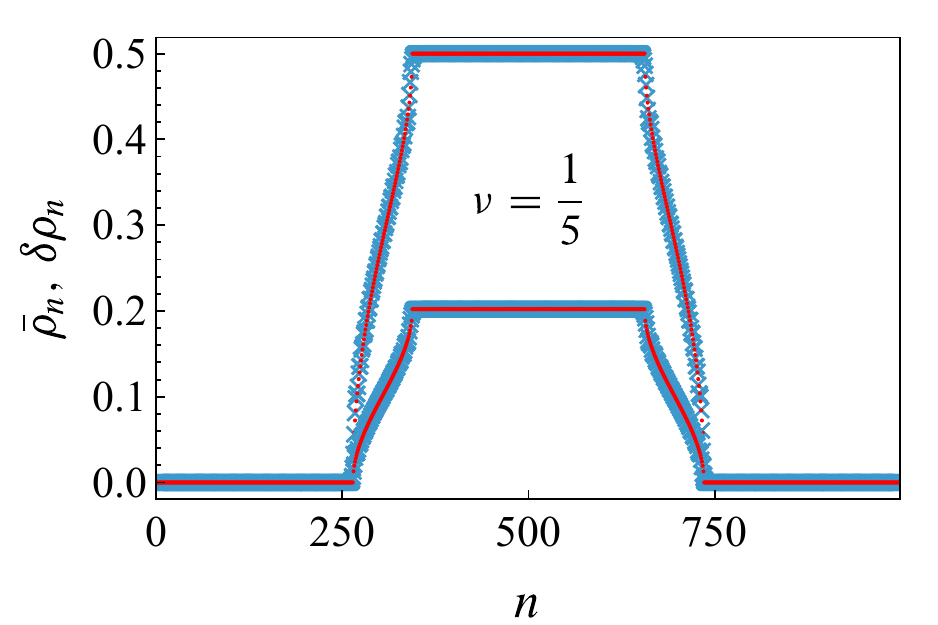}
\end{minipage}\hfill
\begin{minipage}{0.32\textwidth}
  \centering
  \includegraphics[width=\linewidth]{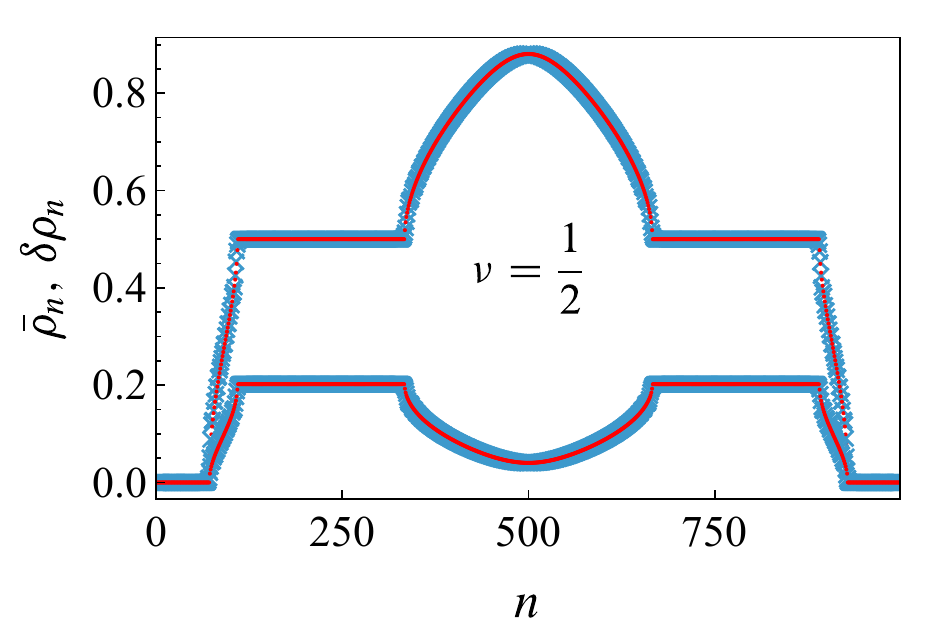}
\end{minipage}

\smallskip

\begin{minipage}{0.32\textwidth}
  \centering
  \includegraphics[width=\linewidth]{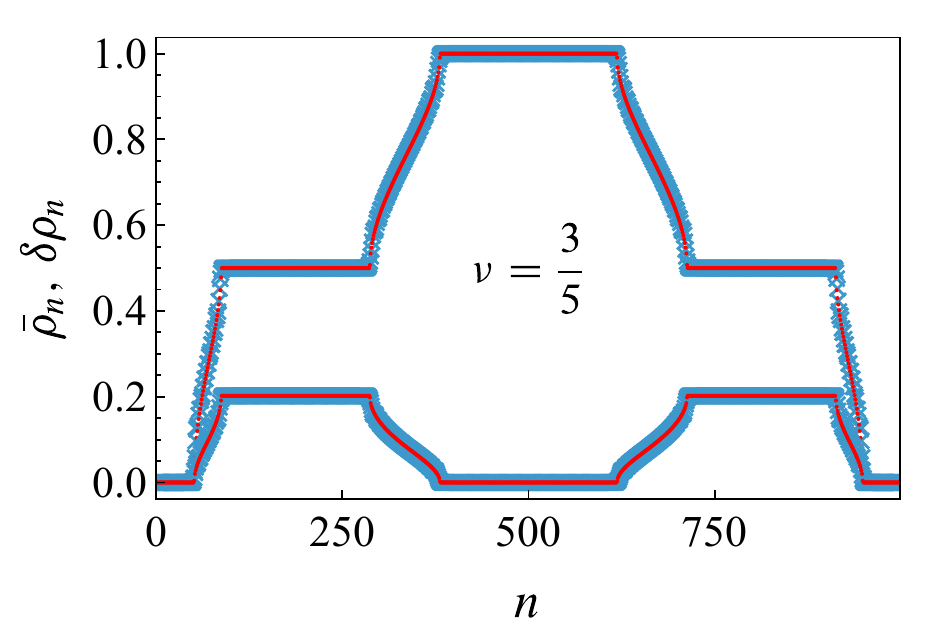}
\end{minipage}\hfill
\begin{minipage}{0.32\textwidth}
  \centering
  \includegraphics[width=\linewidth]{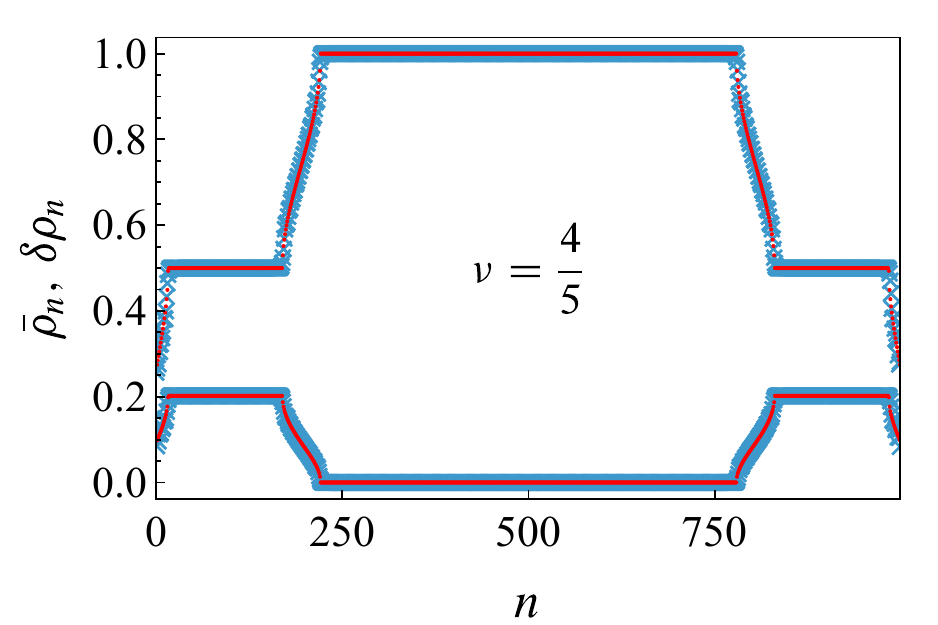}
\end{minipage}\hfill
\begin{minipage}{0.32\textwidth}
  \centering
  \includegraphics[width=\linewidth]{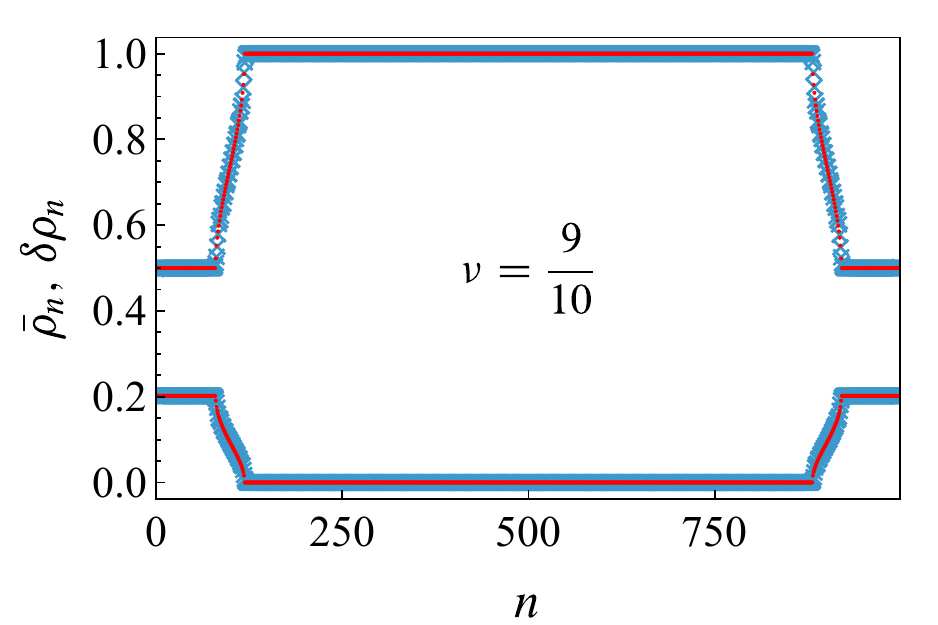}
\end{minipage}

\smallskip

\begin{minipage}[t]{0.32\textwidth}
  \vspace{0pt}
  \centering
  \includegraphics[width=\linewidth]{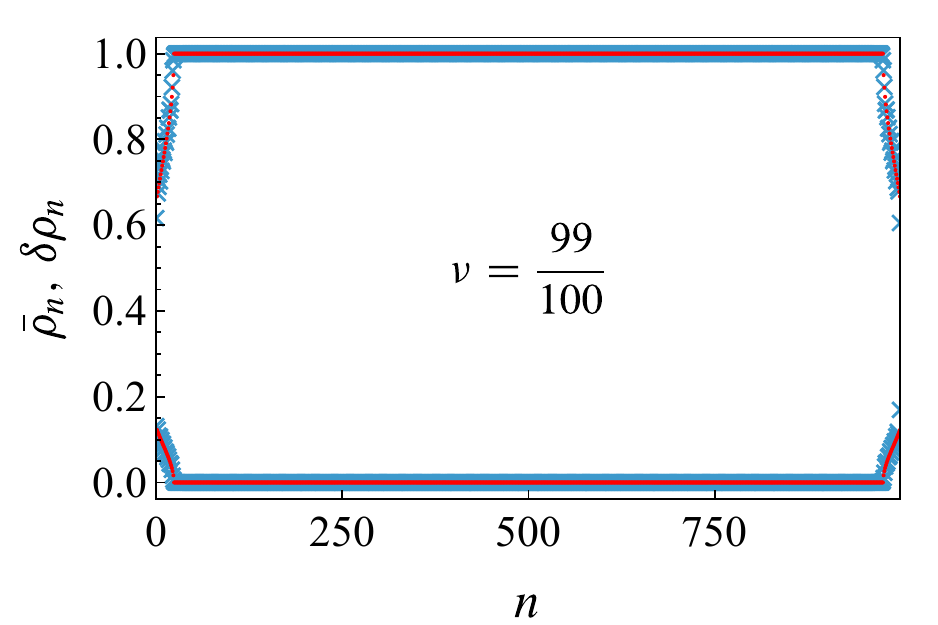}
\end{minipage}\hfill
\begin{minipage}[t]{0.65\textwidth}
  \vspace{6pt}
  \caption{Density profiles of the homogeneous RM model with harmonic on-site
    potential~\eqref{munharm} (blue marks), compared to their WKB approximation~\eqref{LDArhos}
    (red line) for $L=1000$ cells and $a=1$. The chain parameters are given by $\Emin = 1$,
    $\Emax = 3/2$, $\mu_0=4$, $\de\mu=1/2$, and $\si=1$, so that $\mu_0>2\Emax$. The values of the
    filling fraction $\nu$, which correspond to Fermi energies in rows 2--8 of
    Table~\ref{tab.bands2}, are indicated inside each panel.}
  \label{fig:densities2}
\end{minipage}
\end{figure}

Again, the agreement between the numerical density profiles for each of the seven nontrivial
energy intervals (rows 2--8) in Table~\ref{tab.bands2} and their corresponding WKB approximations
is excellent (cf.~Fig.~\ref{fig:densities2}). In particular, the location of the insulator,
metallic, and frozen (empty or fully occupied) intervals deduced from this approximation, and
listed in the table, is borne out by the numerical results. Another nontrivial prediction of the
WKB approximation is that the value of $\de\rho$ in the insulator intervals is the same as for the
example presented in Fig.~\ref{fig:densities1}. Indeed, by eq.~\eqref{drhoins} this value depends
only on the ratios $\de\mu/\Emax$ and $\Emin/\Emax$, which coincide in both examples.

\subsection{RM chain with a domain wall}\label{sec.DWexa}

Consider next the RM chain with variable hopping amplitudes
\begin{subequations}\label{ex2params}
  \begin{equation}\label{ex2vw}
    v_n=v_0\frac{n+1}{L},\qquad w_n=v_0\left(1-\frac{n+1}L\right)
  \end{equation}
  and on-site potentials
  \begin{equation}\label{ex2mus}
    \bmu_n=\mu_0\left(1-\frac{2n+2}L\right)^2,\qquad \de\mu_n=\de_0\left(1-\frac{2n+2}L\right),
  \end{equation}
\end{subequations}
with $v_0$, $\mu_0$, and $\de_0$ positive parameters. Their corresponding continuous versions
$v(x)$, $w(x)$, $\bmu(x)$, and $\dmu(x)$ are obtained by replacing $(n+1)/L$ with $x/\ell$. Thus in
this case
\[
  \Emin(x)=M\abs{1-\frac{2x}\ell},\qquad
  \Emax(x)=\sqrt{\de_0^2\left(1-\frac{2x}\ell\right)^2+v_0^2},
\]
with
\[
  M=\sqrt{v_0^2+\de_0^2}.
\]
The parameters in eqs.~\eqref{ex2params} favor intercell hopping and occupation of the B
sublattice in the chain's left half, whereas in the right half the situation is reversed. At the
midpoint, both the dimerization $v-w$ and the staggered potential $\dmu$ vanish, so that $\Emin=0$
and the local spectral gap closes. The chain's midpoint, therefore, constitutes a domain wall
separating the two opposite dimerization and sublattice-staggering patterns~\cite{SSH79,TLM80}.

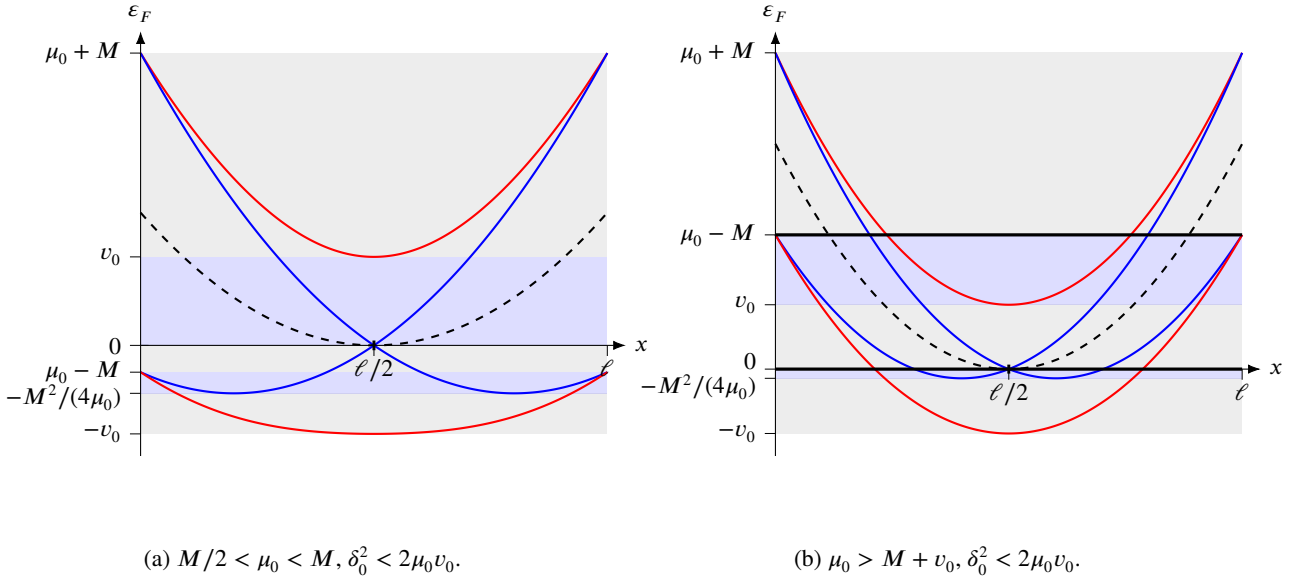
\begin{figure}[pos=t]
  \centering
  \newcommand{\mywidth}{8cm}
  \newcommand{\myheight}{7.2cm}
  
  \centering
  
  \begin{subfigure}{0.49\textwidth}
    \centering
    \begin{minipage}[c][\BandPanelHeight][c]{\linewidth}
      
      \begin{tikzpicture}
        \begin{axis}[
          width=\mywidth,
          height=\myheight,
          xmin=0,
          xmax=1040,
          ymin=-1.25,
          ymax=3.55,
          axis lines=middle,
          axis line style={-Latex},
          axis on top,
          clip=false,
          xlabel={$x$},
          ylabel={$\vep_F$},
          every axis x label/.style={
            at={(current axis.right of origin)},
            anchor=west
          },
          every axis y label/.style={
            at={(current axis.above origin)},
            anchor=south
          },
          xtick={500,1000},
          xticklabels={$\ell/2$,$\ell$},
          ytick={
            -1,
            {-13/24},
            {(3-sqrt(13))/2},
            0,
            1,
            {(3+sqrt(13))/2}
          },
          yticklabels={,,,,,},
          tick align=outside,
          xticklabel style={font=\small,yshift=2pt},
          yticklabel style={font=\small},
          tick style={black,line width=0.2pt},
          major tick length=4pt,
          samples=301,
          domain=0:1000,
          no markers
          ]
          
          \draw[black,line width=0.2pt]
          ([xshift=-4pt]axis cs:0,0)
          --
          ([xshift=3pt]axis cs:0,0);
          
          
          \path[fill=gray!70,fill opacity=0.20]
          (axis cs:0,-1)
          rectangle
          (axis cs:1000,{-13/24});
          
          \path[fill=blue!65,fill opacity=0.20]
          (axis cs:0,{-13/24})
          rectangle
          (axis cs:1000,{(3-sqrt(13))/2});
          
          \path[fill=gray!70,fill opacity=0.20]
          (axis cs:0,{(3-sqrt(13))/2})
          rectangle
          (axis cs:1000,0);
          
          \path[fill=blue!65,fill opacity=0.20]
          (axis cs:0,0)
          rectangle
          (axis cs:1000,1);
          
          \path[fill=gray!70,fill opacity=0.20]
          (axis cs:0,1)
          rectangle
          (axis cs:1000,{(3+sqrt(13))/2});
          
          %
          %
          %
          
          \addplot[red,thick]
          {1.5*(1-2*x/1000)^2
            + sqrt(1+2.25*(1-2*x/1000)^2)};
          
          \addplot[blue,thick]
          {1.5*(1-2*x/1000)^2
            + sqrt(13)/2*abs(1-2*x/1000)};
          
          \addplot[black,dashed,thick]
          {1.5*(1-2*x/1000)^2};
          
          \addplot[blue,thick]
          {1.5*(1-2*x/1000)^2
            - sqrt(13)/2*abs(1-2*x/1000)};
          
          \addplot[red,thick]
          {1.5*(1-2*x/1000)^2
            - sqrt(1+2.25*(1-2*x/1000)^2)};
          
          
          \node[anchor=east,font=\small,xshift=-4pt]
          at (axis cs:0,-1)
          {$-v_0$};
          
          \node[anchor=east,font=\small,xshift=-4pt]
          at (axis cs:0,{-7/12})
          {$-M^2/(4\mu_0)$};
          
          \node[anchor=east,font=\small,xshift=-4pt]
          at (axis cs:0,{(3-sqrt(13))/2})
          {$\mu_0-M$};
          
          \node[anchor=east,font=\small,xshift=-4pt]
          at (axis cs:0,0)
          {$0$};
          
          \node[anchor=east,font=\small,xshift=-4pt]
          at (axis cs:0,1)
          {$v_0$};
          
          \node[anchor=east,font=\small,xshift=-4pt]
          at (axis cs:0,{(3+sqrt(13))/2})
          {$\mu_0+M$};
          
          \draw[black,line width=1.1pt]
          (axis cs:500,-0.055)
          --
          (axis cs:500,0.055);
          
        \end{axis}
      \end{tikzpicture}

    \end{minipage}
    \caption{$M/2<\mu_0<M$, $\de_0^2<2\mu_0v_0$.}
    \label{fig:bands-ex2-nooverlap}
  \end{subfigure}\hfill
  \begin{subfigure}{0.49\textwidth}
    \centering
    \begin{minipage}[c][\BandPanelHeight][c]{\linewidth}
      \begin{tikzpicture}
        \begin{axis}[
          width=\mywidth,
          height=\myheight,
          xmin=0,
          xmax=1040,
          ymin=-1.35,
          ymax=5.25,
          axis lines=middle,
          axis line style={-Latex},
          axis on top,
          clip=false,
          xlabel={$x$},
          ylabel={$\vep_F$},
          every axis x label/.style={
            at={(current axis.right of origin)},
            anchor=west
          },
          every axis y label/.style={
            at={(current axis.above origin)},
            anchor=south
          },
          xtick={500,1000},
          xticklabels={$\ell/2$,$\ell$},
          ytick={
            -1,
            -0.142857142857,
            0,
            1,
            2.085786437627,
            4.914213562373
          },
          yticklabels={,,,,,},
          tick align=outside,
          xticklabel style={font=\small,yshift=2pt},
          yticklabel style={font=\small},
          tick style={black, line width=0.2pt},
          major tick length=4pt,
          samples=301,
          domain=0:1000,
          no markers
          ]
          
          \draw[black, line width=0.2pt]
          ([xshift=-4pt]axis cs:0,0) -- ([xshift=3pt]axis cs:0,0);
          
          
          \path[fill=gray!70, fill opacity=0.20]
          (axis cs:0,-1)
          rectangle
          (axis cs:1000,-0.142857142857);
          
          \path[fill=blue!65, fill opacity=0.20]
          (axis cs:0,-0.142857142857)
          rectangle
          (axis cs:1000,0);
          
          \path[fill=gray!70, fill opacity=0.20]
          (axis cs:0,0)
          rectangle
          (axis cs:1000,1);
          
          \path[fill=blue!65, fill opacity=0.20]
          (axis cs:0,1)
          rectangle
          (axis cs:1000,2.085786437627);
          
          \path[fill=gray!70, fill opacity=0.20]
          (axis cs:0,2.085786437627)
          rectangle
          (axis cs:1000,4.914213562373);
          
          %
          
          \addplot[red, thick]
          {3.5*(1-2*x/1000)^2
            + sqrt(1+(1-2*x/1000)^2)};
          
          \addplot[blue, thick]
          {3.5*(1-2*x/1000)^2
            + sqrt(2)*abs(1-2*x/1000)};
          
          \addplot[black, dashed, thick]
          {3.5*(1-2*x/1000)^2};
          
          \addplot[blue, thick]
          {3.5*(1-2*x/1000)^2
            - sqrt(2)*abs(1-2*x/1000)};
          
          \addplot[red, thick]
          {3.5*(1-2*x/1000)^2
            - sqrt(1+(1-2*x/1000)^2)};
          
          
          \addplot[black, very thick] {0};
          
          \addplot[black, very thick] {3.5-sqrt(2)};
          
          
          \node[anchor=east,font=\small,xshift=-4pt]
          at (axis cs:0,-1)
          {$-v_0$};
          
          \node[anchor=east,font=\small,xshift=-4pt,yshift=-3pt]
          at (axis cs:0,-0.142857142857)
          {$-M^2/(4\mu_0)$};
          
          \node[anchor=east,font=\small,xshift=-4pt,yshift=3pt]
          at (axis cs:0,0)
          {$0$};
          
          \node[anchor=east,font=\small,xshift=-4pt]
          at (axis cs:0,1)
          {$v_0$};
          
          \node[anchor=east,font=\small,xshift=-4pt]
          at (axis cs:0,2.085786437627)
          {$\mu_0-M$};
          
          \node[anchor=east,font=\small,xshift=-4pt]
          at (axis cs:0,4.914213562373)
          {$\mu_0+M$};
          
          \draw[black,line width=1.1pt]
          (axis cs:500,-0.07) -- (axis cs:500,0.07);
          
        \end{axis}
      \end{tikzpicture}
    \end{minipage}
    \caption{$\mu_0>M+v_0$, $\de_0^2<2\mu_0v_0$.}
    \label{fig:bands-ex2-overlap}
  \end{subfigure}
  
  \caption{Energy intervals governing the metal/insulator behavior of the inhomogeneous RM model
    with parameters~\eqref{ex2params}, shown as alternating light gray and light blue horizontal
    bands, for two examples without (panel (a)) and with (panel (b)) overlapping global bands. The
    red and blue curves respectively represent the outer boundaries $\bmu(x)\pm\Emax(x)$ and the
    inner boundaries $\bmu(x)\pm\Emin(x)$ of the local bands $\cB_\pm(x)$, while the dashed black
    curve is the graph of the uniform on-site potential $\bmu(x)$. In panel (b), the thick black
    lines are the boundaries of the overlap $(0,\mu_0-M)$ between the two global bands $\cB_\pm$.
  }
  \label{fig:bands-trap-2}
\end{figure}%
As in the previous example, the system's local metal/insulator behavior is determined by the
position of the Fermi energy $\vep_F$ relative to the extrema of the four local band-edge curves
$\bmu(x)\pm\Emin(x)$ and $\bmu(x)\pm\Emax(x)$. The resulting sequence of metallic, insulator, and
frozen regions is controlled by the dimensionless ratios $\mu_0/v_0$ and $\de_0/v_0$.

To begin with, since
\[
  \max\left(\bmu(x)-\Emin(x)\right)=\max(0,\mu_0-M)\ge0=\min\left(\bmu(x)+\Emin(x)\right),
\]
in this case there is no global insulating gap. In fact, there are two main qualitative behaviors
depending on the sign of $\mu_0-M$. More precisely, if $\mu_0<M$ we have
\[
  \bmu(x)-\Emin(x)\le0=\bmu\left(\tfrac\ell2\right)-\Emin\left(\tfrac\ell2\right),
\]
so that $\cB_-\cap\cB_+=\{0\}$ and the global gap closes at $x=\ell/2$. On the other hand, if
$\mu_0>M$ then
\[
\max\left(\bmu(x)-\Emin(x)\right)=\mu_0-M>0,
\]
and therefore the overlap of the global metallic bands $\cB_{\pm}$ is the interval $[0,\mu_0-M]$.

\begin{figure}[pos=b]
  \includegraphics[width=.32\linewidth]{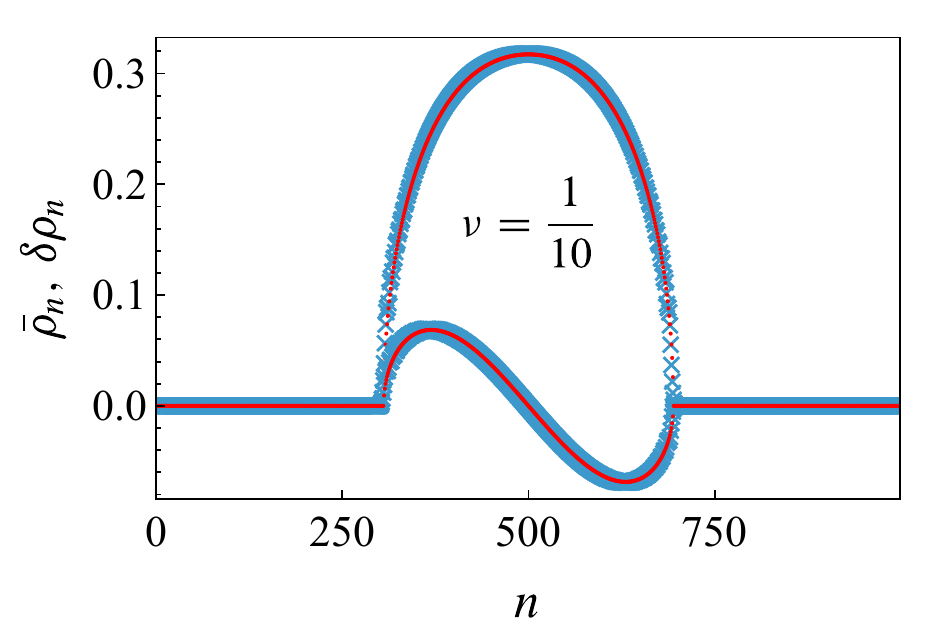}\hfill
  \includegraphics[width=.32\linewidth]{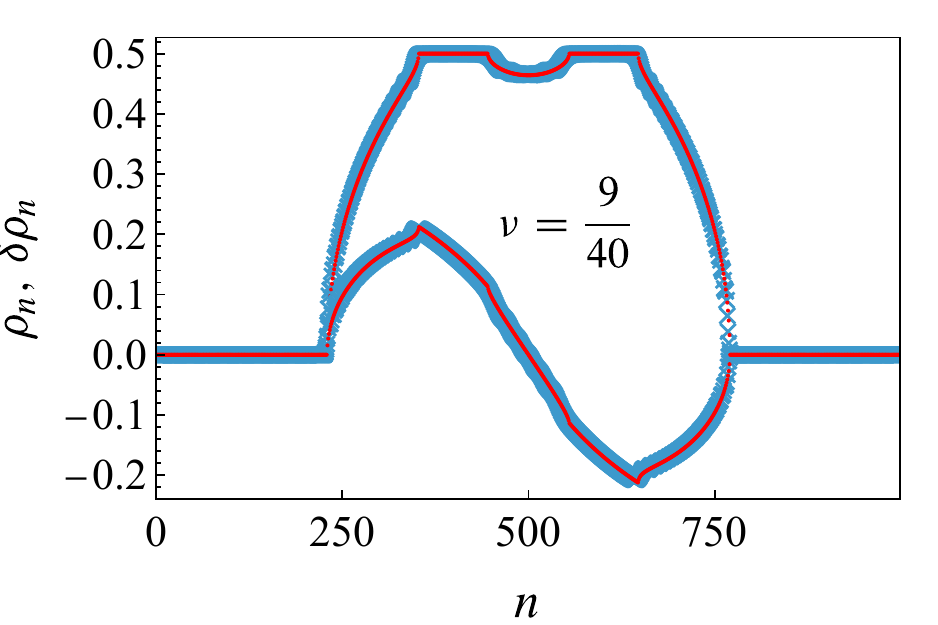}\hfill
  \includegraphics[width=.32\linewidth]{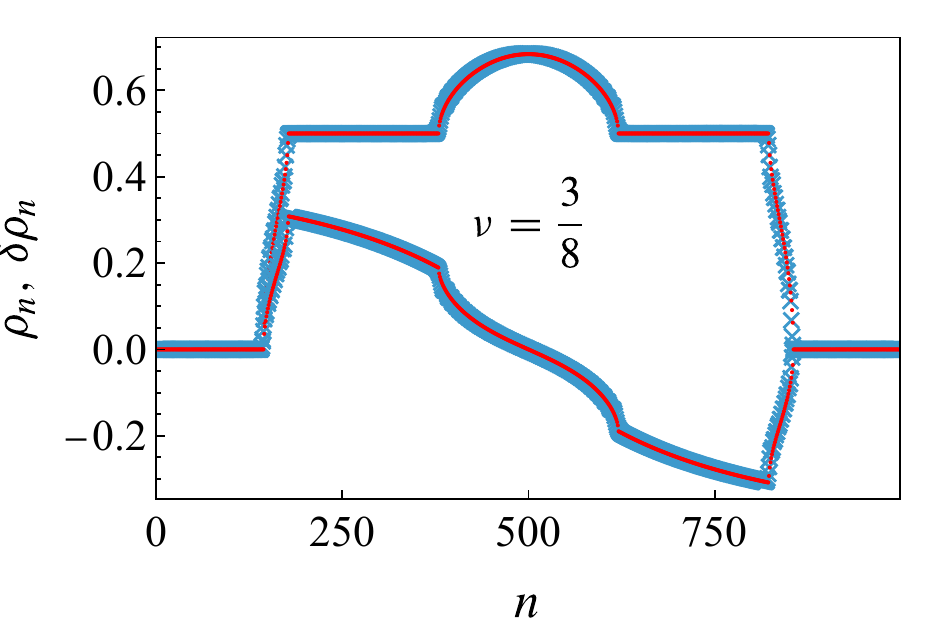}\\
  \begin{center}
    \includegraphics[width=.32\linewidth]{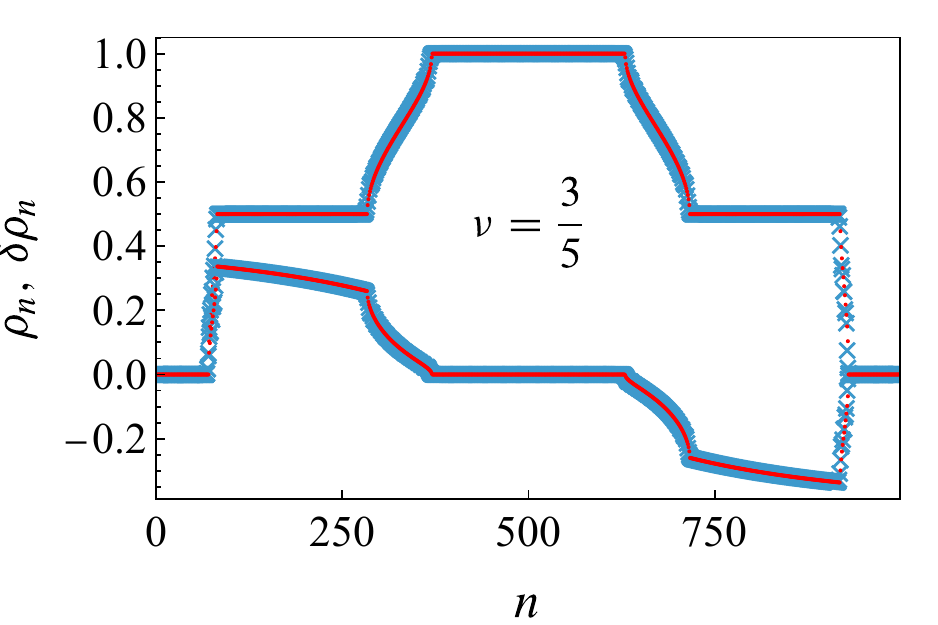}\quad
    \includegraphics[width=.32\linewidth]{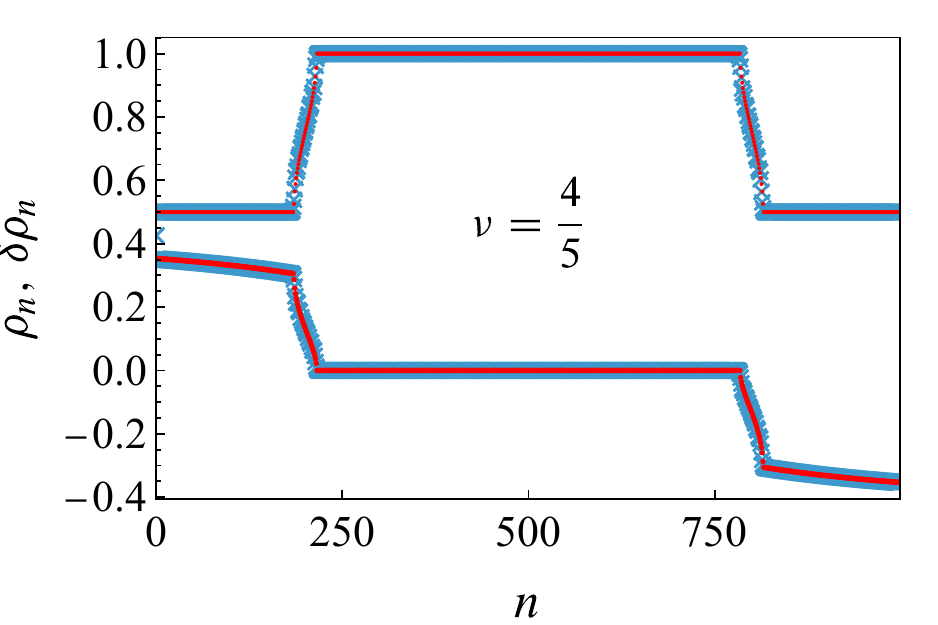}
  \end{center}
  \caption{Density profiles of the inhomogeneous RM model in eq.~\eqref{ex2params} for $L=1000$
    cells, $a=1$, and parameters $v_0=\de_0=1$, $\mu_0=7/2$ (blue marks), compared to their WKB
    approximation~\eqref{LDArhos} (red line). The values of the filling fraction $\nu$,
    corresponding to Fermi energies in rows 2--6 of Table~\ref{tab.bands2-2}, are indicated inside
    each panel.}
  \label{fig.exa22}
\end{figure}

In Fig.~\ref{fig:bands-trap-2} we present two examples illustrating these different qualitative
behaviors; see also Table~\ref{tab.allbands-2} for the corresponding layout of local metallic,
insulator, and frozen intervals depending on the Fermi energy $\vep_F$. In both cases, the WKB
formulas~\eqref{LDArhos} provide an excellent approximation to the average density and the
sublattice imbalance for all values of the Fermi energy, even close to the turning points; see,
e.g., Fig.~\ref{fig.exa22} for the case $\mu_0>M+v_0$, $\de_0^2<2\mu_0v_0$ represented in
Fig.~\ref{fig:bands-ex2-overlap}.

\begin{table}[pos=t]
  \centering

  \begin{subtable}[t]{0.48\textwidth}
    \centering
    \caption{$M/2<\mu_0<M$, $\de_0^2<2\mu_0v_0$}
    \label{tab.bands1-2}
    \begin{tabular*}{\dimexpr .85\linewidth}{@{}LL@{}}
      \toprule
      Interval & Behavior  \\ 
      \midrule
      $\vep_F<-v_0$& (0)\\
      $-v_0<\vep_F<-M^2/(4\mu_0)$& (0m0) \\
      $-M^2/(4\mu_0)<\vep_F<\mu_0-M$& (0mimim0) \\
      $\mu_0-M<\vep_F<0$& (imi) \\
      $0<\vep_F<v_0$& (imi)\\
      $v_0<\vep_F<\mu_0+M$& (im1mi)\\
      $\vep_F>\mu_0+M$& (1)\\
      \bottomrule
    \end{tabular*}
  \end{subtable}
  \begin{subtable}[t]{0.48\textwidth}
    \centering
    \caption{$\mu_0>M+v_0$, $\de_0^2<2\mu_0v_0$}
    \label{tab.bands2-2}
    \begin{tabular*}{\dimexpr .85\linewidth}{@{}LL@{}}
      \toprule
      Interval & Behavior  \\ 
      \midrule
      $\vep_F<-v_0$& (0)\\
      $-v_0<\vep_F<-M^2/(4\mu_0)$& (0m0) \\
      $-M^2/(4\mu_0)<\vep_F<0$& (0mimim0) \\
      $0<\vep_F<v_0$& (0mimim0)\\
      $v_0<\vep_F<\mu_0-M$& (0mim1mim0)\\
      $\mu_0-M<\vep_F<\mu_0+M$& (im1mi)\\
      $\vep_F>\mu_0+M$& (1)\\
      \bottomrule
    \end{tabular*}
  \end{subtable}
  \caption{Local metal/insulator behavior of the inhomogeneous RM model with
    parameters~\eqref{ex2params} for the two examples in Fig.~\ref{fig:bands-trap-2}. The meaning
    of the symbols in the right column is explained in Table~\ref{tab.allbands}. Rows 4-5 of panel
    (a), or (3)-(4) of panel (b), differ in the filling of the central metallic region: for
    $\vep_F<0$ the lower local band is partially filled and the upper one is empty, whereas for
    $\vep_F>0$ the lower band is full and the upper one is partially filled.}
  \label{tab.allbands-2}
\end{table}

We shall concentrate in what follows on the example represented in
Fig.~\ref{fig:bands-ex2-overlap}, which exhibits greater complexity than its partner in
Fig.~\ref{fig:bands-ex2-nooverlap} due to the overlap of the global metallic bands. In particular,
at zero Fermi energy the classically allowed region features an isolated point ($x=\ell/2$)
separated from two intervals symmetrically placed with respect to this point. In the continuum
limit, this suggests the existence of a Jackiw--Rebbi mode~\cite{JR76} with energy close to zero,
localized near the domain wall $x=\ell/2$, together with neighboring states supported on the
adjacent intervals. In fact, the existence of the Jackiw--Rebbi mode can be deduced from the
continuum version of the recurrence relation~\eqref{contsysdisc}. In fact, due to the highly
oscillatory behavior of the eigenfunctions of the discrete Hamiltonian, it is convenient to first
perform the change of variable
\[
  \Phi_n=(-1)^n\begin{pmatrix}
    \al_n\\ \be_n
  \end{pmatrix}\eqqcolon\begin{pmatrix}
    \phi_{A,n}\\ \phi_{B,n}
  \end{pmatrix}.
\]
The system of equations satisfied by the continuum version $\Phi(x)=(\phi_A(x)\en\phi_B(x))^\T$ of
$\Phi_n$ is then
\begin{align*}
  \big(\vep-\bmu(x)-\dmu(x)\big)\phi_A(x)&=v(x)\phi_B(x)-w\Bigl(x-\frac{a}2\Bigr)\phi_B(x-a),\\
  \big(\vep-\bmu(x)+\dmu(x)\big)\phi_B(x)&=v(x)\phi_A(x)-w\Bigl(x+\frac{a}2\Bigr)\phi_A(x+a).
\end{align*}
Expanding the terms in $x\pm a$ and $x\pm\frac{a}2$ to first order in the lattice spacing $a$, we
obtain the system of first-order differential equations
\begin{equation}\label{dweigv}
  \mathcal H(x)\Phi(x)=\vep\Phi(x),
\end{equation}
where the Dirac Hamiltonian $\mathcal H(x)$ is given by
\[
  \mathcal H(x)=\iu a
  \si^y\left(w(x)\pd_x+\frac{w'(x)}2\right)+m(x)\si^x+\dmu(x)\si^z+\bmu(x),\qquad\text{with}\quad
  m(x):=v(x)-w(x).
\]
Note that $\cH$ is Hermitian, since
\[
  \iu\left(w(x)\pd_x+\frac{w'(x)}2\right)=\frac12\left\{\iu\pd_x,w(x)\right\}.
\]
In our case,
\begin{equation}\label{paramst}
  w(x)=\frac{v_0}2(1-t),\qquad m(x)=v_0t,\qquad \bmu(t)=\mu_0t^2,\qquad \de\mu=-\de_0t,
\end{equation}
where
\[
  t=\frac{2x}\ell-1\in[-1,1]
\]
is the distance to the chain's midpoint in units of $\ell/2$. By the previous remarks, we look for
solutions of the eigenvalue eq.~\eqref{dweigv} with energy close to zero concentrated near $t=0$.
Since $\bmu(x)$ is much smaller near $t=0$ than the other coefficients $m(x)$, $w(x)$, and
$\dmu(x)$, we treat this term as a small perturbation and solve the zero-th order eigenvalue
problem
\begin{equation}\label{cH0}
  \mathcal H_0\Phi_0=0,
\end{equation}
where
\[
  \mathcal H_0=\iu a w(x)\si^y\pd_x+m(x)\si^x+\frac{\iu a}2 w'(x)\si^y+\dmu(x)\si^z.
\]
Multiplying the left-hand side of eq.~\eqref{cH0} by $\si^y$ we obtain
\[
  \frac{1-t}t\,\pd_t(w^{1/2}\Phi_0)=\frac{LM}{v_0}\Ga(w^{1/2}\Phi_0),
  \qquad\text{with}\quad\Ga:=\frac1M(\de_0\si^x+v_0\si^z),
\]
or equivalently
\[
  \pd_\xi(w^{1/2}\Phi_0)=\frac{LM}{v_0}\Ga(w^{1/2}\Phi_0),\qquad
  \text{with}\quad \xi\coloneqq-t-\log(1-t).
\]
The matrix $\Ga$, which plays the role of the chirality operator $\si^z$ when $\de_0\ne0$, has
eigenvalues $\pm 1$. Its corresponding normalized eigenvectors can be taken as
\begin{equation}\label{upm}
  u_+=\frac1{\sqrt{2M(M+v_0)}}\begin{pmatrix}
    M+v_0\\ \de_0
  \end{pmatrix},\qquad
  u_-=\frac1{\sqrt{2M(M+v_0)}}\begin{pmatrix}
    \de_0\\ -(M+v_0)
  \end{pmatrix}.
\end{equation}
The general solution of eq.~\eqref{cH0} is thus a linear combination of the two independent
solutions of opposite ``chiralities''
\[
  \Phi_0^{(\pm)}\propto(1-t)^{-\frac12}\e^{\pm\frac{LM}{v_0}\xi}u_\pm=(1-t)^{\mp\frac{LM}{v_0}-\frac12}
  \e^{\mp\frac{LM}{v_0}t}u_\pm\eqqcolon
  f_\pm(t)u_\pm.
\]
It is straightforward to show that $f_+(t)$ decreases from $2^{-1/2}(\e/2)^{\frac{LM}{v_0}}\gg 1$
at $t=-1$ to $1$ at $t=-1/(2\be)$ and increases without bound as $t$ runs from $0$ to $1$,
diverging at $t=1$ as $(1-t)^{-LM/v_0-\frac12}$. Thus the positive chirality solution $\Phi_0^{(+)}$,
being unbounded (and not normalizable, for $L$ large enough), is unphysical and should be
discarded. Note, by contrast, that in the models with $\bmu=\dmu=0$ (and an even number of sites)
studied in Ref.~\cite{PCLMV26}, in the continuum limit there are two chiral zero modes, one
supported on each sublattice (i.e., with opposite chiralities). We have verified that the
chain~\eqref{ex2vw} with $\bmu=\dmu=0$ likewise has a pair of eigenvalues whose splitting is
exponentially small in $L$. Suitable linear combinations of the corresponding eigenstates are
supported on opposite sublattices and thus have opposite chiralities, becoming zero modes in the
thermodynamic limit.

On the other hand, the function $f_-(t)$ is monotonically increasing for $-1<t<1/(2\be)$ (i.e.,
$0<x<\ell/2+\ell/(4\be)$) and decreasing for $1/(2\be)<t<1$ (i.e., $\ell/2+\ell/(4\be)<x<\ell$).
To check that the corresponding negative chirality solution $\Phi_0^{(-)}$ is localized near $t=0$
(i.e., $x=\ell/2$), and hence is a Jackiw--Rebbi mode, it suffices to note that
\[
  -t-\log(1-t)=\frac{t^2}2+O(t^3).
\]
Hence we can take with very good accuracy
\begin{equation}\label{Phi0m}
  \Phi_0^{(-)}(x)=\sqrt{\frac2a}\,\left(\frac{M}{\pi L v_0}\right)^{1/4}\e^{-\frac{LM}{2v_0}t^2}u_-,
\end{equation}
where the factor in front of the exponential is due to normalization. Note that
\[
  \mel{\Phi_0^{(-)}}{t^2}{\Phi_0^{(-)}}=\frac{v_0}{2LM},
\]
and therefore
\[
  \mel{\Phi_0^{(-)}}{\Big(x-\frac\ell2\Big)^2}{\Phi_0^{(-)}}=
  \frac{\ell^2}4\mel{\Phi_0^{(-)}}{t^2}{\Phi_0^{(-)}}
  =\frac{v_0a\ell}{8M}.
\]
Thus the width of the Jackiw--Rebbi mode is proportional to $\sqrt{a\ell}=a\sqrt{L}$, in agreement
with the general results of Ref.~\cite{PCLMV26} for the case $\bmu=\dmu=0$. Note that this is much
smaller (in the thermodynamic limit) than the length of the adjacent intervals on which
neighboring eigenstates are supported, which is $O(aL)$. Finally, by eq.~\eqref{paramst} the
correction to the energy of the Jackiw--Rebbi mode, to first order in perturbation theory, is
given by\footnote{The following estimate does not include all lattice and sampling corrections of
  order $ L^{-1}$, and is therefore not a complete leading-order prediction for the discrete
  eigenvalue.}
\begin{equation}\label{dvep}
  \de\vep=\mel{\Phi_0^{(-)}}{\bmu(x)}{\Phi_0^{(-)}}=\mu_0\mel{\Phi_0^{(-)}}{t^2}{\Phi_0^{(-)}}=\frac{\mu_0v_0}{2LM}=O(L^{-1})\,.
\end{equation}
It is also worth mentioning that for $\de_0\ne0$ the Jackiw--Rebbi mode has non-zero weight both
in the A and B sublattices, unlike what happens for the RM chains with $\bmu=\dmu=0$ studied in
Ref.~\cite{PCLMV26}; in fact, by eq.~\eqref{upm} the quotient of the A and B components of this
mode is $-\de_0/(M+v_0)$.

\begin{figure}[pos=t]
  \includegraphics[width=8cm]{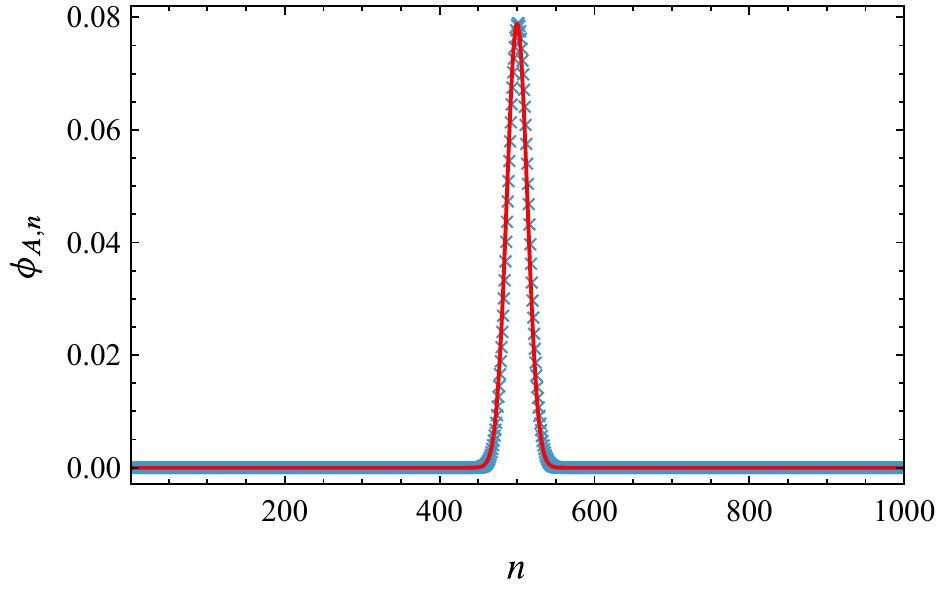}\hfill \includegraphics[width=8cm]{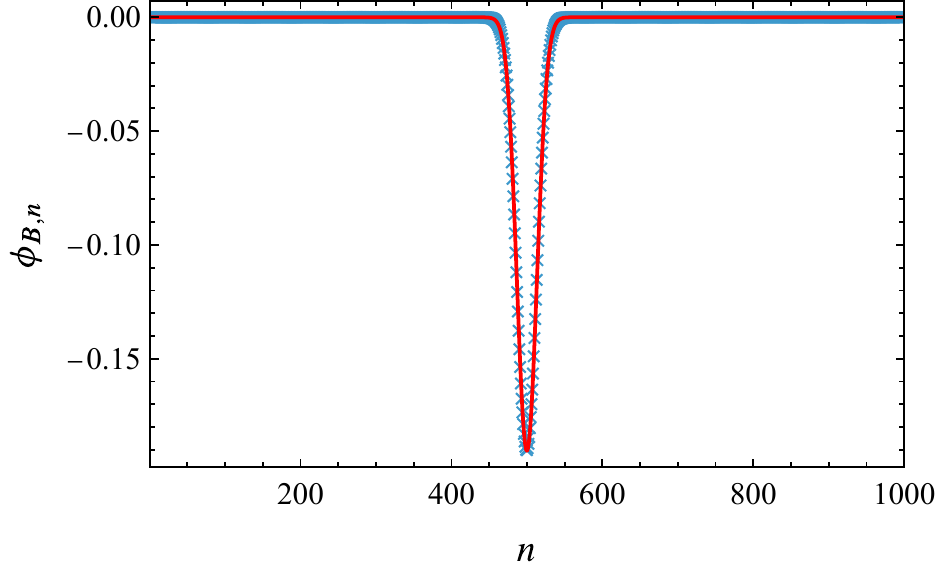}
  \caption{Plot of $\phi_{A,n}=(-1)^n\al_n$ and $\phi_{B,n}=(-1)^n\be_n$ (blue marks) compared to
    the corresponding A and B components of the continuum approximation $\Phi_0^{(-)}(na)$ in
    eq.~\eqref{Phi0m} (red curve) for the Jackiw--Rebbi mode of the inhomogeneous RM
    chain~\eqref{ex2params} with $L=1000$, $a=v_0=\de_0=1$, $\mu_0=7/2$.}
  \label{fig:JR}
\end{figure}

\begin{figure}[pos=b]
  \includegraphics[width=8cm]{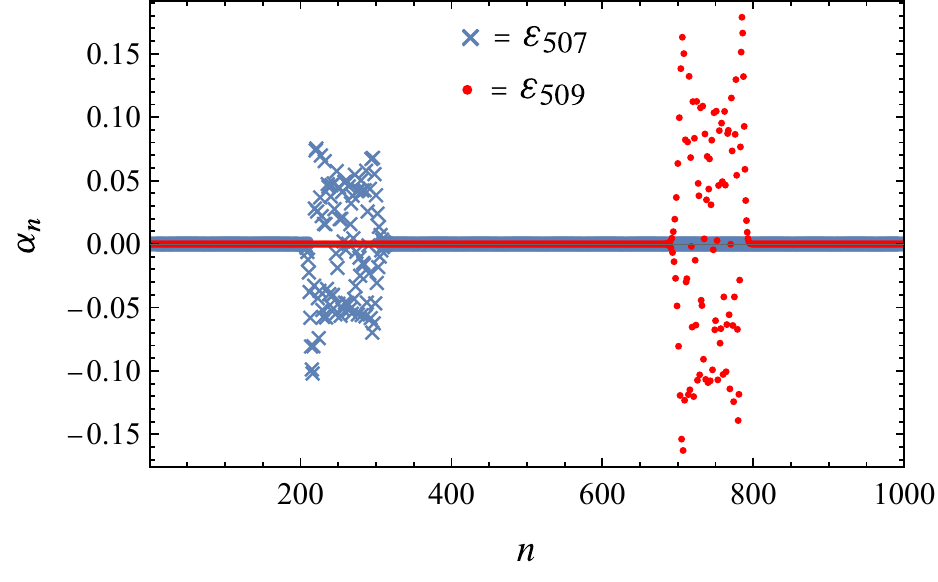}\hfill
  \includegraphics[width=8cm]{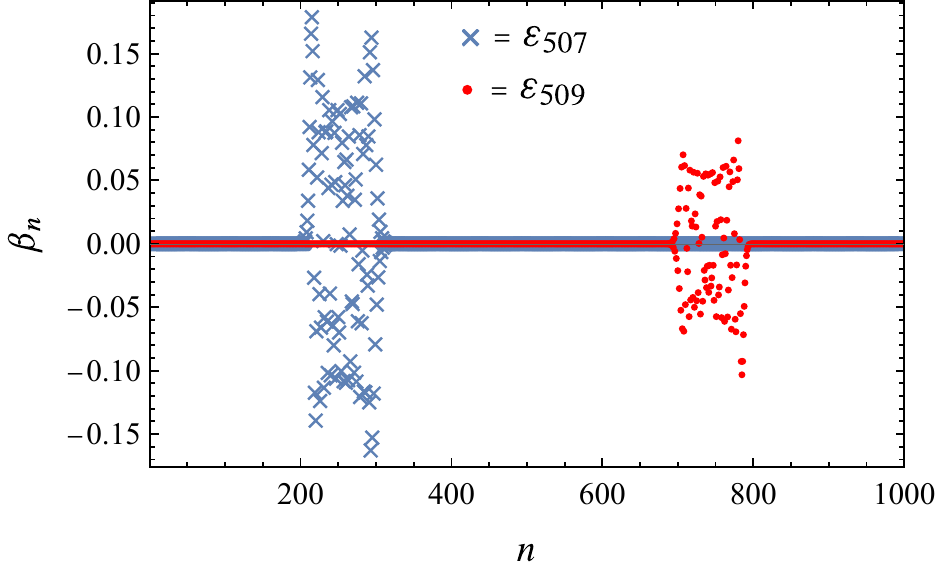}
  \caption{Plot of the components $\al_n$ and $\be_n$ of the eigenfunctions associated with the
    two energy levels adjacent to the Jackiw--Rebbi mode shown in Fig.~\ref{fig:JR}.}
  \label{fig:albe}
\end{figure}

The existence of a Jackiw--Rebbi mode with energy near zero, localized at the chain's midpoint, is
confirmed by our numerical calculations, for both $\mu_0<M$ and $\mu_0>M$. For instance, for
$L=1000$ cells with unit lattice spacing and parameters $v_0=\de_0=1$, $\mu_0=7/2>M+v_0=\sqrt2+1$,
the WKB approximation~\eqref{nuvepF} to the filling fraction at Fermi energy $\vep_F=0$ yields
$\nu\simeq0.25362$. Since $2L\nu\simeq 507.24$, the Jackiw--Rebbi mode is expected to lie near the
$507$-th energy level. In fact, it corresponds to the $508$-th level, with energy
$\vep=1.09495\cdot 10^{-3}$ of the same order of magnitude as the rough estimate
$\de\vep=1.23744\cdot 10^{-3}$ in eq.~\eqref{dvep}. The corresponding eigenfunction is
approximated with remarkable accuracy by the continuum wave function $\Phi_0^{(-)}$ in
eq.~\eqref{Phi0m}, as is apparent from Fig.~\ref{fig:JR}. By contrast, the eigenfunctions
associated with energy levels near the Jackiw--Rebbi mode are not exponentially localized, but are
instead alternately supported on two disjoint intervals, one on either side of the midpoint
$x=\ell/2$ (see, e.g., Fig.~\ref{fig:albe}). As is apparent from Fig.~\ref{fig:bands-ex2-overlap},
the endpoints of these intervals are the intersections of the horizontal axis with the borders of
the lower local band, i.e., the four non-central roots of the algebraic equations
$\bmu(x)-\Emin(x)=0$ and $\bmu(x)-\Emax(x)=0$.

\section{Conclusions and outlook}\label{sec.conc}

In this work we have developed a discrete WKB approach to the single-particle wave functions of
two-band free-fermion chains with parameters varying slowly on the scale of the lattice spacing,
and applied it to the inhomogeneous Rice--Mele model. This extends our previous treatment of
scalar recurrence relations to two-band fermionic systems, providing a semiclassical description
of their single-particle states throughout the local energy bands. Starting from the
single-particle WKB wave functions, we have derived closed-form expressions for both the particle
density and the sublattice imbalance at arbitrary filling. We have shown that these expressions
coincide with those obtained from the LDA, and are also in excellent agreement with the numerical
data. Our results thus provide a semiclassical description of the particle-density profiles, and
establish the validity of the LDA at leading order in the slowly varying limit.

The spatial variation of the local band edges in inhomogeneous chains gives rise to different
configurations of metallic, insulating, empty, and fully occupied regions, depending on the Fermi
energy and the model parameters. We have analyzed this behavior in two representative examples: a
homogeneous Rice--Mele chain subject to a harmonic trapping potential, which is directly relevant
to the experimental setups of refs.~\cite{WTD13,NTTI16}, and a fully inhomogeneous chain
containing a domain wall supporting a Jackiw--Rebbi mode. In particular, these examples illustrate
the applicability of the present approach to systems with spatially varying gaps, including
situations in which the local gap closes at a domain wall.

While the homogeneous RM model and its extensions with spatially varying on-site potentials have
been experimentally realized~\cite{NTTI16,NTKKMT21}, the hopping amplitudes in the corresponding
tight-binding Hamiltonians are taken to be spatially uniform. To complement our analytic
treatment, we have also proposed several experimental realizations of fully inhomogeneous RM
chains that should be feasible with current ultracold-atom optical superlattice techniques.

Several extensions of the present work are worth exploring. A natural generalization is to
multiband free-fermion chains, such as generalized Rice--Mele or Aubry--André--Harper models,
which would require extending the discrete WKB formalism to wave functions with more than two
components. Another interesting direction is a more detailed analysis of the WKB solutions near
classical turning points, including the derivation of appropriate connection formulas. Such an
analysis would extend the validity of the semiclassical approximation to the boundaries of the
classically allowed regions and could provide further insight into the behavior of localized
states. Another potential line of research could focus on defects in inhomogeneous free-fermion
systems, exploiting their closed-form solvability in the thermodynamic limit. This would allow us
to analytically explore the interplay between spatial inhomogeneity and quench
physics~\cite{VSDH16} or the bulk-boundary correspondence~\cite{PKKLS20}.

Finally, the WKB wave functions could be used to investigate other observables beyond the local
particle densities. In particular, it would be of interest to study the polarization of
inhomogeneous Rice--Mele chains and its evolution under adiabatic parameter variations, with a
view toward applications to Thouless pumping. Another promising direction is the semiclassical
evaluation of the ground-state correlation matrix, which could in turn provide an approach to the
entanglement entropy of spatial subsystems. These extensions would broaden the scope of the
discrete WKB framework as a tool for studying the spectral, local, and nonlocal properties of
inhomogeneous free-fermion systems.

\appendix                     
\numberwithin{equation}{section}

\section{Gauge-fixing the open chain}\label{app.gauge}

In general, the Hamiltonian of a fermionic chain with $L$ cells with an arbitrary number of sites
$Z$ can be taken as
\[
  H=\sum_{n=0}^{L-1}\sum_{\si=0}^{Z-2}\left(v_{n\si}a_{n\si}^\dagger
    a_{n,\si+1}+\Hc\right)
  +\sum_{n=0}^{L-2}\left(w_{n}a_{n,Z-1}^\dagger a_{n+1,0}+\Hc\right)
  +\sum_{n=0}^{L-1}\sum_{\si=0}^{Z-1}\mu_{n\si}a_{n\si}^\dagger a_{n\si},
\]
or in more abbreviated fashion
\[
  H=\sum_{n,\si}\left(v_{n\si}a_{n\si}^\dagger a_{n,\si+1}+\Hc\right)
  +\sum_{n}\left(w_{n}a_{n,Z-1}^\dagger a_{n+1,0}+\Hc\right)
  +\sum_{n,\si}\mu_{n\si}a_{n\si}^\dagger a_{n\si}.
\]
We shall next show that, if $v_{n\si},w_n\ne0$ for all $n,\sigma$, the hopping amplitudes
$v_{n\si},w_n$ can be taken without loss of generality as real and positive. Indeed, write
\[
  v_{n\si}=\abs{v_{n\si}}\e^{\iu\,\al_{n,\si}},\qquad w_n=\abs{w_n}\e^{\iu\,\be_n},
\]
with $\al_{n\si}$, $\be_{n}$ real, and introduce the new set of fermionic operators
\[
  \tilde a_{n\si}=\e^{\iu\,\ga_{n\si}}a_{n\si},
\]
with $\ga_{n\si}\in\RR$. We then have
\[
  H=\sum_{n,\si}\abs{v_{n\si}}\left(\tilde a_{n\si}^\dagger \tilde a_{n,\si+1}+\Hc\right)
  +\sum_{n}\abs{w_{n}}\left(\tilde a_{n,Z-1}^\dagger
    \tilde a_{n+1,0} +\Hc\right)
  +\sum_{n,\si}\mu_{n\si}\tilde a_{n\si}^\dagger \tilde a_{n\si}
\]
provided that
\[
  \ga_{n,\si+1}=\ga_{n\si}+\al_{n\si},\qquad \ga_{n+1,0}=\ga_{n,Z-1}+\be_n,
\]
for all $n,\si$. The solution of this system is straightforward:
\[
  \ga_{n\si}=\ga_{00}+\sum_{i=0}^{n-1}\sum_{s=0}^{Z-2}\al_{is}+\sum_{s=0}^{\si-1}\al_{ns}
  +\sum_{i=0}^{n-1}\be_i,
\]
with $\ga_{00}\in\RR$ arbitrary.

\section{Realization of inhomogeneous RM models in optical superlattices}\label{app.AAH}

In this Appendix we shall briefly outline how to obtain eq.~\eqref{hij} for the matrix elements of
the tight-binding Hamiltonian obtained from the optical potential $V(x)$ in eq.~\eqref{Voptlatt}.

To begin with, we can write the potential $V(x)$ in eq.~\eqref{Voptlatt} as
\[
  V(x)=V_0\sin^2\left(\frac{2\pi x}a\right)-C(x)\cos\left(\frac{2\pi
      x}{a}\right)-S(x)\sin\left(\frac{2\pi x}{a}\right) +\frac{V_1(x)}2,
\]
with $C(x)$ and $S(x)$ defined in eq.~\eqref{CSx}. Since the main optical potential
$V_0\sin^2\bigl(\frac{2\pi x}a\bigr)$ in eq.~\eqref{h0} is $a/2$-periodic and real, we can choose
its lowest-band Wannier functions $W_m(x)$ to be real, symmetric about their centers $x_m=ma/2$,
and translation-covariant, i.e., satisfying
\[
  W_m(x)=W(x-x_m),\qquad \text{with}\quad W\equiv W_0.
\]

Taking all of this into account, the contribution to the off-diagonal element $h_{m,m+1}$ of the
lattice Hamiltonian $h_0$ is given by
\begin{align*}
h_{m,m+1}^{(0)}
&=\int\dd x\,W\left(x-\frac{ma}2\right)
\left(-\frac{\hbar^2}{2M}\pd_x^2+V_0\sin^2\left(\frac{2\pi x}{a}\right)\right)
  W\left(x-\frac a2(m+1)\right)\\
&=\int\dd s\,W(s)
\left(-\frac{\hbar^2}{2M}\pd_s^2+V_0\sin^2\left(\frac{2\pi s}{a}\right)\right)
  W\left(s-\frac a2\right)=A_1+V_0B_1,
\end{align*}
with
\[
A_1=-\frac{\hbar^2}{2M}\int\dd s\,W(s)W''\left(s-\frac a2\right),\qquad
B_1=\int\dd s\,\sin^2\left(\frac{2\pi s}{a}\right)W(s)W\left(s-\frac a2\right).
\]
Similarly, the contribution of $h_0$ to the on-site potential is
\begin{align*}
  h_{mm}^{(0)}
    &=\int\dd s\,
    W(s)\left(-\frac{\hbar^2}{2M}\pd_s^2+V_0\sin^2\left(\frac{2\pi s}{a}\right)\right)
    W(s)
  =A_0+V_0B_0,
\end{align*}
with
\[
A_0=-\frac{\hbar^2}{2M}\int\dd s\,W(s)W''(s),\qquad
B_0=\int\dd s\,\sin^2\left(\frac{2\pi s}{a}\right)\abs{W(s)}^2.
\]
Likewise, the contribution of the term $-C(x)\cos(2\pi x/a)$ to the diagonal matrix element
$h_{mm}$ is
\begin{align*}
  h_{mm}^{(C)}
  &=-\int\dd x\,
    C(x)\cos\left(\frac{2\pi x}{a}\right)
    \abs{W\left(x-\frac{ma}2\right)}^2\\
  &=-(-1)^m\int\dd s\,
    C\left(s+\frac{ma}2\right)\cos\left(\frac{2\pi s}{a}\right)\abs{W(s)}^2
  \simeq -(-1)^mC\left(\frac{ma}2\right)t_C,
\end{align*}
with
\[
  t_C=\int\dd s\,\cos\left(\frac{2\pi s}{a}\right)\abs{W(s)}^2.
\]
It is straightforward to show that the analogous contribution to the off-diagonal element
$h_{m,m+1}$ of this term vanishes:
\begin{align*}
  h_{m,m+1}^{(C)}
  &=-\int\dd x\,
    C(x)\cos\left(\frac{2\pi x}{a}\right)
    W\left(x-\frac{ma}2\right)W\left(x-\frac a2(m+1)\right)\\
  &=-(-1)^m\int\dd s\,
    C\left(s+\frac{ma}2\right)\cos\left(\frac{2\pi s}{a}\right)W(s)
    W\left(s-\frac{a}2\right)\\
  &\simeq-(-1)^mC\left(\frac{ma}2+\frac{a}4\right)\int\dd s\,
    \cos\left(\frac{2\pi s}{a}\right)W(s)W\left(s-\frac{a}2\right)=0,
\end{align*}
since the last integral vanishes by parity (recall that $W(x)=W_0(x)$ is symmetric about its
center $x_0=0$). An analogous calculation shows that the term
$-S(x)\sin\left(\frac{2\pi x}a\right)$ does not contribute to $h_{mm}$, while its contribution to
$h_{m,m+1}$ is given by
\[
  h^{(S)}_{m,m+1}\simeq -(-1)^mS\left(\frac{ma}2+\frac{a}4\right)t_S,\qquad
  t_S\coloneqq
  \int\dd s\,\sin\left(\frac{2\pi s}{a}\right)W(s)W\left(s-\frac a2\right).
\]
Finally, the contributions of the non-oscillating term $V_1(x)/2$ to the matrix
element of the tight-binding Hamiltonian are
\[
  h^{(1)}_{mm}\simeq\frac12V_1\left(\frac{ma}2\right)\braket{W_m}{W_m}
  =\frac12V_1\left(\frac{ma}2\right),\qquad
  h^{(1)}_{m,m+1}\simeq\frac12V_1\left(\frac{ma}2+\frac{a}4\right)\braket{W_m}{W_{m+1}}=0.
\]
Summing all these contributions, and taking into account that
$S\left(\frac{ma}2+\frac{a}4\right)\simeq S\left(\frac{ma}2\right)$ due to the slow variation of
$S$, yields eq.~\eqref{hij} in Section~\ref{sec.model}.

\section{Bulk Hamiltonian of the homogeneous RM chain}\label{app.bulk}

For convenience, in this appendix we summarize the main properties of the bulk Hamiltonian of the
homogeneous RM model, usually defined by
  \[
    \Hb=v\sum_{n=0}^{L-1}\left(a_n^\dagger b_n+b_n^\dagger a_n\right)+w\sum_{n=0}^{L-1}\left(b_{n}^\dagger
    a_{n+1}+a_{n+1}^\dagger b_{n}\right)+\dmu\sum_{n=0}^{L-1}\left(a_n^\dagger a_n-b_n^\dagger
    b_n\right),\qquad
  \text{with}\quad a_L\equiv a_0.
  \]
  The translation invariance of $\Hb$ makes it possible to diagonalize it through the Fourier
  transform
  \[
    a_n=\frac1{\sqrt L}\sum_k\e^{\iu kn}\ha_k,\qquad b_n=\frac1{\sqrt L}\sum_k\e^{\iu kn}\hb_k,
  \]
  where the momenta $k$ range over the set
  \[
    \left\{\frac{2\pi m}L:m=0,\dots,L-1\right\}.
  \]
  In terms of the momentum creation/annihilation operators $\ha_k$ and $\hb_k$, the Hamiltonian
  becomes
  \[
    \Hb=\sum_k\begin{pmatrix}
    \ha_k^\dagger&\hb_k^\dagger
    \end{pmatrix}h(k)\begin{pmatrix}
    \ha_k\\\hb_k
  \end{pmatrix},
  \]
  with
  \[
    h(k)=\begin{pmatrix}
    \dmu & z(k)\\
    z^*(k)& -\dmu
  \end{pmatrix},
    \qquad z(k)=v+w\e^{-\iu k}.
  \]
  The Bloch matrix $h(k)$ has eigenvalues $\pm E(k)$, with
  \begin{equation}\label{Ekbulk}
    E(k)\coloneqq\sqrt{\dmu^2+\abs{z(k)}^2}=\sqrt{\dmu^2+v^2+w^2+2vw\cos k},
  \end{equation}
  whose corresponding normalized eigenvectors can be taken as
  \begin{equation}\label{uAB}
    \bu_\pm(k)=\frac1{\sqrt{2E(k)(E(k)\mp\dmu)}}
    \begin{pmatrix}
      z(k)\\[2.5ex] \pm(E(k)\mp\dmu)\,
    \end{pmatrix}\eqqcolon
    \begin{pmatrix}
      u^A_\pm(k)\\[1ex] u^B_\pm(k)
    \end{pmatrix}
    .
  \end{equation}
  Note that, since $\abs{z(k)}\ge|v-w|>0$, the system is gapped, with energy gap
  \[
    2\min_kE(k)=\Delta=2\sqrt{\dmu^2+|v-w|^2}>0.
  \]
  Setting
  \[
    \hc^\dagger_\pm(k)=
    \begin{pmatrix}
      \ha_k^\dagger&\hb_k^\dagger
    \end{pmatrix}
    \cdot\bu_\pm(k)=u^A_\pm(k)\,\ha^\dagger_k+u^B_\pm(k)\,\hb^\dagger_k
  \]
  we have
  \[
    \Hb=\sum_k E(k)\left(\hc_+^\dagger(k)\hc_+(k)-\hc_-^\dagger(k)\hc_-(k)\right).
  \]
  Thus $\hc_\pm^\dagger(k)$ creates a fermion with energy $\pm E(k)$ and momentum $k$ (modulo
  $2\pi$), whose wave function can be written as
  \[
    \ket{\psi_\pm(k)}=\hc_\pm^\dagger(k)\ket0=u^A_\pm(k)\,\ha_k^\dagger\ket0+u^B_\pm(k)\,
    \hb^\dagger(k)\ket0.
  \]
  The projection of $\ket{\psi_\pm(k)}$ onto the $n$-th A-type (respectively B-type) basis
  state $\ket n_{\!A}$ (resp.~$\ket n_{\!B}$) is given by
  \[
    {}_A\!\braket{n}{\psi_\pm(k)}=\frac{u^A_\pm(k)\e^{\iu kn}}{\sqrt L}\,,
    \qquad
    {}_B\!\braket{n}{\psi_\pm(k)}=\frac{u^B_\pm(k)\e^{\iu kn}}{\sqrt L}\,.
  \]
  We therefore have
  \begin{subequations}\label{nuderhob}
    \begin{align}
      \left|{}_A\!\braket{n}{\psi_\pm(k)}\right|^2+\left|{}_B\!\braket{n}{\psi_\pm(k)}\right|^2
      &=\frac{\left|\bu_\pm(k)\right|^2}L=\frac1L,
      \label{nubulk}\\
      \left|{}_B\!\braket{n}{\psi_\pm(k)}\right|^2-\left|{}_A\!\braket{n}{\psi_\pm(k)}\right|^2
      &=\frac1{2LE(k)}\left[E(k)\mp\dmu-\frac{\abs{z(k)}^2}{E(k)\mp\dmu}\right]
        =\mp\frac{\dmu}{LE(k)},
        \label{excbulk}
    \end{align}
  \end{subequations}
  both independent of the cell index $n$.

  \section{Sublattice-resolved particle densities of the homogeneous RM model}%
  \label{sec:homdens}

  In this appendix we compute the sublattice-resolved particle densities of the homogeneous RM
  model in the thermodynamic limit. We shall work with the model with an odd number of sites
  since, as shown in Section~\ref{sec.homogeneous}, its single-particle spectrum is obtainable in
  closed form (cf.~eqs.~\eqref{sp-odd} and~\eqref{disprel-ens}-\eqref{qkeven}).

Suppose that we fill the first $M$ levels in the lower band, 
\[
  \vep_{-}(q_1)<\cdots<\vep_-(q_M),\qquad 1\le M\le L,
\]
so that the filling is
\[
  \nu=\frac{M}{2L+1}.
\]
Consider first the sites in the B sublattice, for which
\[
  \langle b_n^\dagger b_n\rangle=\frac1{L+1}\sum_{k=1}^M\frac{E(q_k)+\dmu}{E(q_k)}\,
  \sin^2\bigl((n+1)q_k\bigr).
\]
Since
\[
  \de q_k\coloneq q_{k+1}-q_k=\frac{\pi}{L+1},
\]
in the thermodynamic limit, and at a fixed cell $n\ll L$, we have
\begin{equation}\label{rhoAnTD}
  \langle b_n^\dagger
  b_n\rangle=\frac1{\pi}\int_0^{2\pi\nu}\dd q\,\frac{E(q)+\dmu}{E(q)}\,\sin^2\bigl((n+1)q\bigr).
\end{equation}
This result is \emph{exact}. We now apply the Riemann--Lebesgue lemma, according to which
\[
  \lim_{\la\to\infty}\int_a^b\dd q\,f(q)\sin(\la q)=\lim_{\la\to\infty}\int_a^b\dd q\,f(q)\cos(\la
  q)=0,\qquad \all f\in L^1([a,b]),
\]
which implies that\footnote{In fact, if $f'\in L^1([a,b])$ (or, more generally, if $f$ is of
  bounded variation in $[a,b]$), it can be shown that the error is $O(|\la|^{-1})$.}
\[
  \lim_{\la\to\infty}\int_a^b\dd q\,f(q)\sin^2(\la q)=\lim_{\la\to\infty}\int_a^b\dd
  q\,f(q)\cos^2(\la q)=\frac12\int_a^b\dd q\,f(q),\qquad \all f\in L^1([a,b]).
\]
From the Riemann--Lebesgue lemma and eq.~\eqref{rhoAnTD} we deduce that when $n\gg 1$ (i.e., in
the bulk)
\begin{equation}\label{bndbnint}
  \langle b_n^\dagger b_n\rangle\simeq\frac1{2\pi}\int_0^{2\pi\nu}\dd
  q\,\frac{E(q)+\dmu}{E(q)}=\nu+\frac{\dmu}{2\pi}\int_0^{2\pi\nu}\frac{\dd q}{E(q)}.
\end{equation}
Likewise, in the thermodynamic limit we have
\[
  \expval*{a_n^\dagger a_n}=\frac1{\pi}\int_0^{2\pi\nu}\dd q\,
  \frac{\big(v\sin\bigl((n+1)q\bigr)+w\sin(nq)\big)^2}{E(q)\big(E(q)+\dmu\big)}\,.
\]
Noting that
\begin{align}
  \big(v\sin\bigl((n+1)q\bigr)+w\sin(nq)\big)^2
  &=v^2\sin^2\bigl((n+1)q\bigr)
    +w^2\sin^2(nq)+vw\big(\cos q-\cos\bigl((2n+1)q\bigr)\big)\notag\\
  &=\frac12(E(q)^2-\dmu^2)-\frac12\big(v^2\cos\bigl(2(n+1)q\bigr)
    +w^2\cos(2nq)+2vw\cos\bigl((2n+1)q\bigr)\big)
                                                 \label{sinsinRL}
\end{align}
and applying the Riemann--Lebesgue lemma again we obtain
\begin{equation}\label{andanint}
  \expval*{a_n^\dagger a_n}\simeq\frac1{2\pi}\int_0^{2\pi\nu}\dd q\,\frac{E(q)-\dmu}{E(q)}
  =\nu-\frac{\dmu}{2\pi}\int_0^{2\pi\nu}\frac{\dd q}{E(q)}\,.
\end{equation}
The last integral can be readily expressed in terms of the incomplete elliptic integral of the
first kind
\[
  F(\vp,\ka)=\int_0^\vp\frac{\dd s}{\sqrt{1-\ka^2\sin^2s}},\qquad 0\le \ka\le 1.
\]
Indeed,
\begin{align*}
  \int_0^{2\pi\nu}\frac{\dd q}{E(q)}
  &=\int_0^{2\pi\nu}\frac{\dd q}{\sqrt{\dmu^2+v^2+w^2+2vw\cos
    q}} =\int_0^{2\pi\nu}\frac{\dd q}{\sqrt{\dmu^2+(v+w)^2-4vw\sin^2(q/2)}}\\
  &=\frac{\ka}{\sqrt{vw}}F(\pi\nu,\ka),
\end{align*}
with
\[
  \ka=\sqrt{\frac{4vw}{\dmu^2+(v+w)^2}}\le\sqrt{\frac{4vw}{(v+w)^2}}<1,
\]
and therefore
\begin{equation}
  \label{rhoA}
  \langle a_n^\dagger a_n\rangle\simeq\nu-\frac{\ka\dmu}{2\pi\sqrt{vw}}\,F(\pi\nu,\ka),
  \qquad
  \expval*{b_n^\dagger b_n}\simeq\nu+\frac{\ka\dmu}{2\pi\sqrt{vw}}\,F(\pi\nu,\ka)\,.
\end{equation}
Thus for fillings $\nu\le1/2$, i.e., in the lower band, the average density and the sublattice
imbalance are \emph{constant}, and given respectively by
\begin{equation}\label{barrhoderho}
  \bar\rho=\frac1{2a}\big(\expval*{a_n^\dagger a_n}+\expval*{b_n^\dagger
    b_n}\big)=\frac\nu{a},
  \quad
  \de\rho=\frac1{2a}\big(\expval*{b_n^\dagger b_n}-\expval*{a_n^\dagger
    a_n}\big)=\frac{\ka\dmu}{2\pi a\sqrt{vw}}\,F(\pi\nu,\ka),\qquad 0\le\nu\le\frac12\,,
\end{equation}
where $a$ is the lattice spacing (i.e., the width of the unit cell).

Suppose next that $1/2<\nu\le 1$, i.e., $L+1<M\le 2L+1$. In other words, the whole lower band plus
the next $M-L$ levels are filled. These $M-L$ levels have energies
\begin{equation}\label{enupband}
  \dmu<\vep_+(q_{L})<\cdots<\vep_+(q_{2L+2-M}),
\end{equation}
although the eigenfunction with eigenvalue $\dmu$ only contributes to the average occupation of
the A sublattice. The contribution to the average occupation $\expval*{b_n^\dagger b_n}$ of the
levels~\eqref{enupband} is therefore
\[
  \frac1{L+1}\sum^{k=L}_{2L+2-M}\frac{E(q_k)-\dmu}{E(q_k)}\, \sin^2\bigl((n+1)q_k\bigr),
\]
which in the thermodynamic limit becomes
\[
  \frac1\pi\int^\pi_{2\pi(1-\nu)}\dd q\,\frac{E(q)-\dmu}{E(q)}\,\sin^2\bigl((n+1)q\bigr)
  \simeq\frac1{2\pi}\int^\pi_{2\pi(1-\nu)}\dd q\,\frac{E(q)-\dmu}{E(q)}
  =\nu-\frac12-\frac{\dmu}{2\pi}\int^\pi_{2\pi(1-\nu)}\frac{\dd q}{E(q)}\,,
\]
where in the first approximation we have applied the Riemann--Lebesgue lemma. Adding the
contribution of the first band (see eq.~\eqref{bndbnint} with $\nu=1/2$) we finally obtain
\[
  \expval*{b_n^\dagger b_n}\simeq\nu+\frac{\dmu}{2\pi}\int_0^\pi\frac{\dd q}{E(q)}
  -\frac{\dmu}{2\pi}\int^\pi_{2\pi(1-\nu)}\frac{\dd q}{E(q)}
  =\nu+\frac{\dmu}{2\pi}\int_0^{2\pi(1-\nu)}\frac{\dd q}{E(q)}
  =\nu+\frac{\ka\dmu}{2\pi\sqrt{vw}}\,F\bigl(\pi(1-\nu),\ka\bigr)\,.
\]
Similarly, in the thermodynamic limit the contribution of the $M-L$ levels~\eqref{enupband} to the
average occupation of the A sublattice is given by\footnote{It should be noted that the edge mode
  does not contribute to the bulk density in the thermodynamic limit.}
\[
  \frac1{\pi}\int_{2\pi(1-\nu)}^{\pi}\dd q\,
  \frac{\big(v\sin\bigl((n+1)q\bigr)+w\sin(nq)\big)^2}{E(q)\big(E(q)-\dmu\big)}
  \simeq\frac1{2\pi}\int_{2\pi(1-\nu)}^{\pi}\dd q\,\frac{E(q)+\dmu}{E(q)}
  =\nu-\frac12+\frac{\dmu}{2\pi}\int_{2\pi(1-\nu)}^{\pi}\frac{\dd q}{E(q)}\,,
\]
where we have again applied the Riemann--Lebesgue lemma (cf.~eq.~\eqref{sinsinRL}). Adding the
contribution of the filled lower band (cf.~eq.~\eqref{andanint} with $\nu=1/2$) finally yields
\[
  \expval*{a_n^\dagger a_n}\simeq\nu-\frac{\dmu}{2\pi}\int_0^\pi\frac{\dd q}{E(q)}
  +\frac{\dmu}{2\pi}\int^\pi_{2\pi(1-\nu)}\frac{\dd q}{E(q)}
  =\nu-\frac{\dmu}{2\pi}\int_0^{2\pi(1-\nu)}\frac{\dd q}{E(q)}
  =\nu-\frac{\ka\dmu}{2\pi\sqrt{vw}}\,F\bigl(\pi(1-\nu),\ka\bigr)\,.
\]
Thus for $1/2\le\nu\le 1$ the average density and the sublattice imbalance are given by
\begin{equation}\label{barrhoderhoup}
  \bar\rho=\frac\nu{a},\quad
  \de\rho=\frac{\ka\dmu}{2\pi a\sqrt{vw}}\,F\bigl(\pi(1-\nu),\ka\bigr),\qquad \frac12\le\nu\le1\,.
\end{equation}

In summary, the average density and sublattice imbalance of the homogeneous Rice--Mele model with
positive parameters $(v,w)$ (with $v\ne w$) are given by
\begin{subequations}\label{homdens}
  \begin{equation}\label{rhoderhohom}
    \bar\rho=\frac\nu{a},\qquad
    \de\rho=\frac{\ka\dmu}{2\pi a\sqrt{vw}}\,F\bigl(\pi\min\bigl(\nu,1-\nu\bigr),\ka\bigr),
  \end{equation}
  with
  \begin{equation}
    \label{ka}
    \ka=\sqrt{\frac{4vw}{\dmu^2+(v+w)^2}}\,.
  \end{equation}
\end{subequations}

\begin{remark}
  The previous formulas could also have been obtained starting from the bulk Hamiltonian. Indeed,
  from eq.~\eqref{nubulk} it immediately follows that the average occupation $a\bar\rho$ is equal
  to the filling fraction $\nu$. The equation for the sublattice imbalance $\de\rho$ is similarly
  obtained from eq.~\eqref{excbulk}, taking into account that in this case $\De k=2\pi/L$ but, as
  $E(k)=E(2\pi-k)$, each level with $k\ne0,\pi$ is twice degenerate.

  Note also that, at half-filling (i.e., for $\nu=1/2$), eqs.~\eqref{rhoderhohom} essentially
  coincide with the corresponding formulas derived by Lin et al.~\cite{LKPWS20}, although these
  authors leave the relevant integral unevaluated rather than expressing it in closed form in
  terms of an elliptic integral.\qed
\end{remark}

\section{Parametrization of the classical action on the cylinder}\label{app.oint}

In this appendix we will justify eq.~\eqref{ointpdx}, used in the derivation of the contribution
to the density of states of the $j$-th potential well $I_j(\vep)$.

As mentioned in the text, the main difference with the standard (continuous) WKB theory is that
the phase space is the cylinder $\RR\times S^1$, since the momentum variable $p$ is defined modulo
$2\pi$. This makes the one-form $p\,\dd x$ multivalued, as it is defined up to integer multiples
of $2\pi\,\dd x$. On the other hand, the one-form $-x\dd p$ is well defined, so that the
Bohr--Sommerfeld semiclassical quantization condition~\eqref{BSxdp} is unambiguous. Moreover, the
identity
\begin{equation}\label{intparts}
  -x\,\dd p=p\,\dd x-\dd(x p)
\end{equation}
holds independently of the lift of phase space $\RR\times S^1$ to $\RR^2$, i.e., under the
replacement $p\mapsto p+2\pi n$ (with $n\in\ZZ$). As can be seen from
Fig.~\ref{fig.wkb-inline-comparison}, the lift of the classical trajectory $\Gamma_j(\vep)$
consists of two branches $(x,p_\pm(x,\vep))$, with $p_\pm(x,\vep)$ defined uniquely by
\[
  p_\pm(x,\vep)=\pm p(x,\vep)=\pm\arccos\xi(x,\vep),
\]
which have common endpoints $\bigl(x_j^{L,R}(\vep),p\bigl(x_j^{L,R}(\vep),\vep\bigr)\bigr)$ modulo
$2\pi$.

Suppose first that both endpoints of the classically allowed interval $I_j(\vep)$ lie in the
interior of the chain, so that $0<x_j^L(\vep)<x_j^R(\vep)<\ell$. Then
$p\bigl(x_j^{L,R}(\vep),\vep\bigr)\in\{0,\pi\}$ (cf.~Fig.~\ref{fig.wkb-inline-comparison}), and
the line integral in eq.~\eqref{BSxdp} is taken along the two branches $p=\pm\arccos\xi(x,\vep)$.
At an endpoint where $p=\pi$, the values $+\pi$ and $-\pi$ represent the same point on the
momentum circle, so the two branches already meet on the cylinder. Thus the vertical segment
joining these values in a planar representation is \emph{not} included in the integration contour.
Integrating by parts---which is allowed, by eq.~\eqref{intparts}---we therefore obtain the
standard result~\cite{Br93}
\begin{equation}
  -\oint_{\Ga_j(\vep)}x\,\dd p=-2\int_g{x_j^{L}(\vep)}^{x_j^{R}(\vep)}x\,\dd p(x,\vep)
  =2\int_{x_j^{L}(\vep)}^{x_j^{R}(\vep)}\dd x\,p(x,\vep)
  - 2x p(x,\vep)\biggr|_{x=x_j^L(\vep)}^{x=x_j^R(\vep)}.
  \label{eq:WKB_action}
\end{equation}
Differentiating both sides of this equation with respect to $\vep$, and taking into account that
$p\bigl(x_j^{L,R}(\vep),\vep\bigr)$ is locally constant (either $0$ or $\pi$), we immediately
obtain eq.~\eqref{ointpdx} used in the text.

Consider, next, the case in which one or both endpoints $x_j^{b}(\vep)$ ($b=L,R$) coincide with
one of the chain's boundaries. In this case we must complete the integration contour on the
cylinder by adding a vertical reflection segment joining the points
$(x_j^b(\vep),\pm p(x_j^{b}(\vep),\vep))$ (from $(\ell,p)$ to $(\ell,-p)$ at the right boundary
and from $(0,-p)$ to $(0,p)$ at the left boundary). The segment at $x=0$ does not contribute to
$-\oint_{\Ga_j(\vep)}x\,\dd p$, whereas the segment at $x=\ell$ contributes $2\ell p(\ell,\vep)$.
Thus, when $x_j^R(\vep)=\ell$, eq.~\eqref{eq:WKB_action} becomes
\[
  -\oint_{\Ga_j(\vep)}x\,\dd p
  =2\int_{x_j^{L}(\vep)}^{\ell}\dd x\,p(x,\vep)+2x_j^L(\vep)p\bigl(x_j^{L}(\vep),\vep\bigr).
\]
If $x_j^L(\vep)\ne0$, the endpoint momentum $p(x_j^L(\vep),\vep)$ is locally constant, equal to
$0$ or $\pi$; if $x_j^L(\vep)=0$, the last term vanishes identically. Differentiating with respect
to $\vep$ therefore yields eq.~\eqref{ointpdx} in either case. Finally, when only the left
endpoint coincides with a chain boundary, eq.~\eqref{eq:WKB_action} remains unchanged, and its
differentiation likewise gives eq.~\eqref{ointpdx}.




\end{document}